%% file: SAMpulley.tex
\documentclass[1p]{elsarticle}
\usepackage{pstricks}
\usepackage{graphicx}
\usepackage{amssymb,amsmath,amsthm,amsfonts,latexsym}
\usepackage{epsfig}
\usepackage{psfrag}
\usepackage{fancyhdr}

\newcommand{\ut}{\,\mathrm}
\newcommand{\dd}{\mathrm{d}}
\newcommand{\rmd}{\mathrm{d}}

\newcommand{\E}{{\cal E}}

\renewcommand{\H}{{\cal H}}

\renewcommand{\L}{{\cal L}}

\renewcommand{\P}{{\cal P}}

\newcommand{\R}{{\cal R}}
\newcommand{\calcS}{{\cal S}}

\newtheorem{thm}{Theorem}[section]
\newtheorem{lem}[thm]{Lemma}

\newtheorem{ur:remark}{Remark}[section]

\newtheorem{ur:notation}{Notation}[section]
\newtheorem{ur:remarks}{Remarks}[section]
\newenvironment{rems}{\begin{ur:remarks}\rm}{\end{ur:remarks}}
\newtheorem{ur:exemple}[thm]{Example}

\newtheorem{ur:example}{Example}[section]

\newcommand{\bm}[1]{\mbox{\boldmath ${#1}$\unboldmath}}

\newcommand{\brr}[1]{\left\{{#1}\right\}}
\newcommand{\p}[1]{\left({#1}\right)}

\def\nc{{\mathbb C}}
\def\nz{{\mathbb Z}}

\def\ack{\bigskip\noindent\ignorespaces\section*{Acknowledgments}}

\begin{document}

\title{Swinging Atwood's Machine: experimental and theoretical studies}

\author[a1]{O Pujol\corref{cora}}
\ead{olivier.pujol@loa.univ-lille1.fr}

\author[a2]{J P P\'erez}
\ead{perez@ast.obs-mip.fr}

\author[a3]{J P Ramis}
\ead{ramis.jean-pierre@wanadoo.fr}

\author[a4]{C Sim\'o}
\ead{carles@maia.ub.es}

\author[a5]{S Simon}
\ead{Sergi.Simon@port.ac.uk}

\author[a6]{J A Weil}
\ead{Jacques-Arthur.Weil@unilim.fr}

\address[a1]{Universit\'e de Lille, UFR de Physique fondamentale, Laboratoire
d'Optique Atmosph\'erique, 59655 Villeneuve d'Ascq cedex, France}
\cortext[cora]{Corresponding author}

\address[a2]{Universit\'e de Toulouse, Laboratoire d'Astrophysique
Toulouse-Tarbes, CNRS, 14 avenue \'Edouard Belin, 31400 Toulouse, France}

\address[a3]{Universit\'e de Toulouse, Laboratoire \'Emile Picard,
118 route de Narbonne 31062 Toulouse, France}

\address[a4]{Universitat de Barcelona, Departament de Matem\`atica Aplicada i
An\`alisi, Gran Via de les Corts Catalanes, 585, 08007 Barcelona, Spain}

\address[a5]{University of Portsmouth, Department of Mathematics, Lion Terrace,
Portsmouth, Hampshire PO1 3HF, United Kingdom}

\address[a6]{Universit\'e de Limoges, D\'epartement Math\'ematiques
Informatique, XLIM - UMR CNRS n. 6172, 123 avenue Albert Thomas, 87060 Limoges 
cedex, France}

\begin{abstract}
A \emph{Swinging Atwood Machine} (\emph{SAM}) is built and some
experimental results concerning its dynamic behaviour are
presented. Experiments clearly show that pulleys play a role in
the motion of the pendulum, since they can rotate and have
non-negligible radii and masses. Equations of motion must
therefore take into account the inertial momentum of the pulleys,
as well as the winding of the rope around them. Their influence is
compared to previous studies. A preliminary discussion of the role
of dissipation is included. The theoretical behaviour of the system
with pulleys is illustrated numerically, and the relevance of
different parameters is highlighted. Finally, the integrability of
the dynamic system is studied, the main result being that the
Machine with pulleys is non-integrable. The status of the
results on integrability of the pulley-less Machine is also
recalled.
\end{abstract}

\begin{keyword} 
Swinging Atwood's Machine (\emph{SAM}) \sep 
Chaotic system \sep Nonlinear dynamics \sep Integrability \sep 
Experimental \emph{SAM} apparatus.


\end{keyword}

\maketitle

\newpage
\tableofcontents
\thispagestyle{empty}
\newpage

\pagestyle{fancy}
\fancyhf{}
\lhead{}
\rhead{\thepage}

\section{Introduction}

This paper deals with the \emph{Swinging Atwood Machine} (\emph{SAM}), a
non-linear two-degrees-of-freedom system derived from the
well-known simple \emph{Atwood machine}. The latter was devised in
$1784$ by George Atwood, a London Physics lecturer who built his
own apparata as a means of practical illustration, in order to
experimentally demonstrate the uniformly accelerated motion of a
system falling under the earth gravity field $\bm{g}$ with
mass dependence \cite{Atwood-1784}. In Atwood's original machine,
two masses are mechanically linked by an inextensible thread wound
round a pulley. In \emph{SAM}, one of the masses $(m)$ is allowed
to swing in a plane while the other mass $(M)$ plays the role of a
counterweight; hence \emph{SAM} can be seen as a parametric
pendulum whose length is varying as a function of the parameter
$\mu = M/m$.

\medskip

For about twenty-five years, many studies have been conducted
concerning the mechanical behaviour of \emph{SAM}.
Said studies were conducted exclusively on a simplified model for
\emph{SAM} neglecting any influence from a massive set of pulleys. 
Through numerical investigations, \cite{Tufillaro-1984} inferred the
pulley-less \emph{SAM} to be an extremely intricate system
exhibiting significant changes in the qualitative behaviour of
trajectories, depending on $\mu$-values. Assuming $\mu>1$, motion
is limited in space and two types of trajectories can be
distinguished owing to the initial conditions: singular ones for
which pendulum length is initially zero, and non-singular ones
where the pendulum is initially released from rest with a non-zero
length. For the former, it appears that $\mu=3$ is a particular
condition corresponding to terminating trajectories, i.e. those
for which pendulum length becomes zero after a given duration,
regardless of the initial conditions \cite{Tufillaro-1994}. The
latter is divided 
into periodic, quasi-periodic and what can be conjectured to be
ergodic trajectories in some domain. \emph{SAM} without massive
pulleys was also studied by means of Poincar\'e sections wherein
chaotic dynamic behaviour becomes prominent as $\mu$ is increased
\cite{Tufillaro-1985}. An interesting and surprising result is the
integrability of the pulley-less \emph{SAM} for $\mu=3$, a
conclusion which is also supported using Hamilton-Jacobi theory
\cite{Tufillaro-1986}
and Noether symmetries \cite{AlmeidaMoreira}. For $\mu > 3$,
\cite{Casasayas-1990} proved that \emph{SAM} without massive
pulleys is not integrable, contrary to what was speculated by
\cite{Tufillaro-1985}. The belonging of $\mu=M/m$ to a special set of parameters $\left\{\mu_p: p\in\nz\right\}$ was established as
a necessary condition for integrability; this result was proven independently in \cite{Casasayas-1990},
\cite{Yehia} and \cite{AlmeidaMoreiraSantos}, and is proven in Remark \ref{finalrems}(2) of the present paper as well.
Moreover, unbounded trajectories $(\mu\leq 1)$ have been studied
via energetic considerations \cite{Tufillaro-1988};
\cite{Nunes-1995} identified and classified all periodic
trajectories in the pulley-less \emph{SAM} for $\mu=3$.
Finally, a very recent result co-written by one of the authors of the present
article (\cite{MS2009}) proved the non-integrability of this pulley-less
model for \emph{SAM} for the exceptional values $\left\{\mu_p: p\in\nz\right\}$; this had been an open problem, at least,
since \cite{Casasayas-1990} explicitly tackled the issue for the
first time. It is worth noting that all of these studies are
theoretical, albeit for the most part strongly supported by
massive numerical simulations.

\medskip

In this paper, we intend to describe a useful physical
construction of \emph{SAM} in detail, as well as present further
experimental and theoretical results. In addition, a theoretical
premise is introduced which stands as a novelty all its own: to
wit, as suggested by experiments, pulleys are no longer neglected,
in order to take account of non-zero radii and rotation around
their axes of revolution. 
When dealing with $N$-degree-of-freedom non-linear systems, 
the modern researcher's tendency to restrict adjectives such as ``complex'' to $N\gg 1$ 
should not divert us from the fact that even the dynamics of apparently elementary cases such as $N=2$ are 
often very difficult to determine (\cite{Birkhoff-1927}), and thus a source of interest in their own right. 
As shown in this paper, such is the case for SAM. 
A schematic representation of \emph{SAM}
is featured in Section~\ref{repsam} partly aimed at the derivation
of the equations of motion with pulleys in Section~\ref{eqmotion}.
The constructed apparatus is then described in detail in
Section~\ref{expappar}, and some experimental results are
presented in Section~\ref{resexp}. A comparison with the
theoretical model is performed in Section~\ref{comptheo} through
numerical simulations of the general equation of motion obtained
in Section~\ref{eqmotion}. Section~\ref{integ} is devoted to prove
the non-integrability for \emph{SAM}. 
The rigorous proof shown therein is requisite to establish definitively the non-integrability of SAM, 
hence incumbent upon any proper completion of experimental and numerical results. 
Indeed, although for some systems non-integrability is somehow suggested by a thorough Poincar\'e section analysis, 
chaotic zone detection may require extremely careful numerics and will at times become laborious -- we might call this \emph{shy chaos}. 
Furthermore the lack of integrability of some systems cannot be discovered by looking at the real phase space. 
There exist non-integrable systems without any recurrent motion in the real phase space and such that chaos is confined away from the real domain. 
For an example, see \cite{mosi96}. 
Finally, Section~\ref{conc}
concludes with perspectives on further experiments and some
comments on results concerning integrability of the pulley-less
case.

\medskip

\cite{Tufillaro-1984} suggested a \emph{SAM} physical demonstration model using
a vertically mounted air table, and alleged a successful
experimental demonstration of the system's motions. However, there
is no experimental result in the aforementioned reference and its
proposed model for \emph{SAM} is unequivocally not as close to the theoretical system
as the model described herein (cf.~Section~\ref{repsam}). To our present knowledge, detailed
experimental studies of \emph{SAM}, let alone comparisons of any
such experiments with the theory, do not exist in the literature prior to our work.

\medskip

Therefore, our work arguably completes the above
theoretical and experimental research on \emph{SAM}, and, at the same time, opens
new problems and sets a starting point for further experimental and theoretical studies.

\section{Schematic representation of \emph{SAM}}\label{repsam}

\begin{figure}[!ht]
\includegraphics{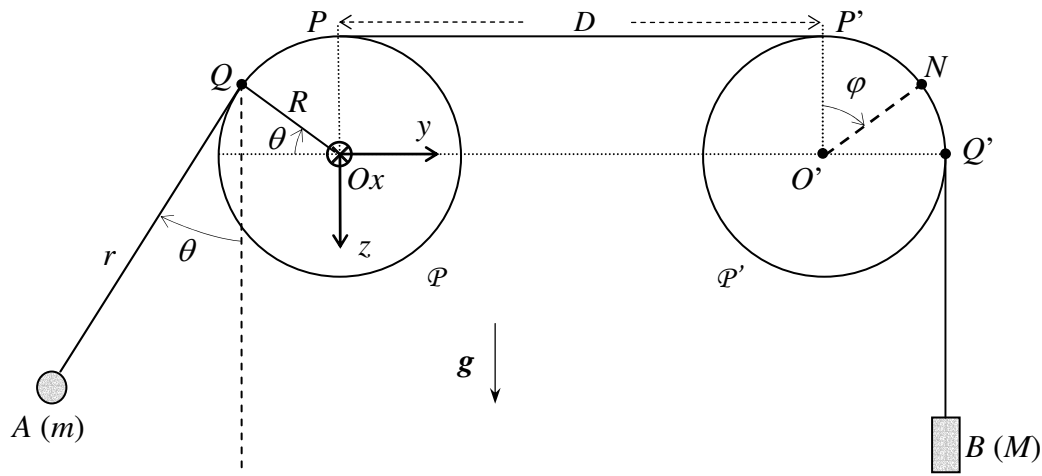}
\caption{Schematisation of the system $\calcS = $ \{pendulum-thread-pulleys-counterweight\} representing \emph{SAM}.
The angle $\varphi$ locates the material point $N'$ of the pulley $\P'$.} 
\label{figsam}
\end{figure}

\emph{SAM} is represented by the system $\calcS$ sketched in Figure~\ref{figsam} and consisting of:
\begin{itemize}
   \item a pendulum, considered as a material point $A$ of mass $m$,
   \item a counterweight, considered as a material point $B$ of mass $M$,
   \item a thread of length $L$ linking $A$ and $B$,
   \item two pulleys $\P$ and $\P'$ of radius $R$, distant from one another by a predetermined distance $D$.
\end{itemize}

$\calcS$ is studied relative to the Galilean laboratory frame
$\R=\p{O\, \bm{e}_x \,\bm{e}_y \, \bm{e}_z}$ whose
origin $O$ is chosen to correspond to the centre of the pendulum
pulley. Axis $Ox$ corresponds to the pulley revolution axis;
axis $Oy$ is the horizontal direction defined by the pulley
centres ($O$ and $O'$), and oriented from $O$ toward $O'$;
finally, $Oz$ is chosen to correspond, for the sake of convenience, to the
downward direction of the local earth gravity field $\bm{g}$ (vertical).

\medskip

Pendulum $A$ is characterized by its variable length $r = QA$, $Q$ being the
geometrical point where the thread departs from the pulley, and by the angle
$\theta$ formed by $\bm{QA}$ and the downward vertical. Note that $\theta$ as
represented in Figure~\ref{figsam} is a positive angle.

\medskip

Vertical motion of the counterweight $B$ is described by its coordinate $z_B$,
which can be related to the angular position $\varphi$ of any point $N$ on the
pulleys, provided the thread does not slip on the pulley
(in Figure~\ref{figsam}, for the sake of clarity, $N$ is drawn on the pulley
$\P'$ associated to $B$ and thus labelled $N'$). Indeed, under this
assumption, when $B$ is falling down, pulleys are able to rotate in such a way
that the velocity of any point of the pulleys (for instance $N'$) is equal to
the velocity of $B$. Hence:
\begin{equation}\label{firsteq}
\dot{z}_B = R \dot{\varphi}
\end{equation}
Note the difference between the rotation angle $\varphi$ of the pulleys and
$\theta$: the former defines the location of any material point on a pulley,
whereas the latter defines the angular position of $A$, as well as that of the
geometrical point of contact $Q$. This subtlety is due to the necessary
mechanical description of contact in terms of three points \cite{PerezM-2001}:
\begin{itemize}
   \item the geometrical point of contact $Q$,
   \item the point $N$ of the pulley $\P$ and corresponding to $Q$ at time $t$,
   \item the point $K$ of the thread corresponding to $Q$ at time $t$.
\end{itemize}

\begin{figure}[!ht]
\begin{center}
\includegraphics{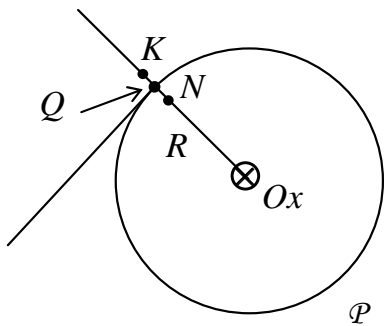}
\end{center}
\caption{Description of the contact between a pulley and the thread with three
points: $K$, $Q$, and $N$. At any time $t$, both $K$ and $N$ touch each other at
$Q$. In the figure, these points are separated for the sake of clarity.}
\label{figcont}
\end{figure}

Figure~\ref{figcont} displays the configuration.
A similar problem can be found in \cite{Pujol-2007}.

\begin{figure}[!ht]
\includegraphics{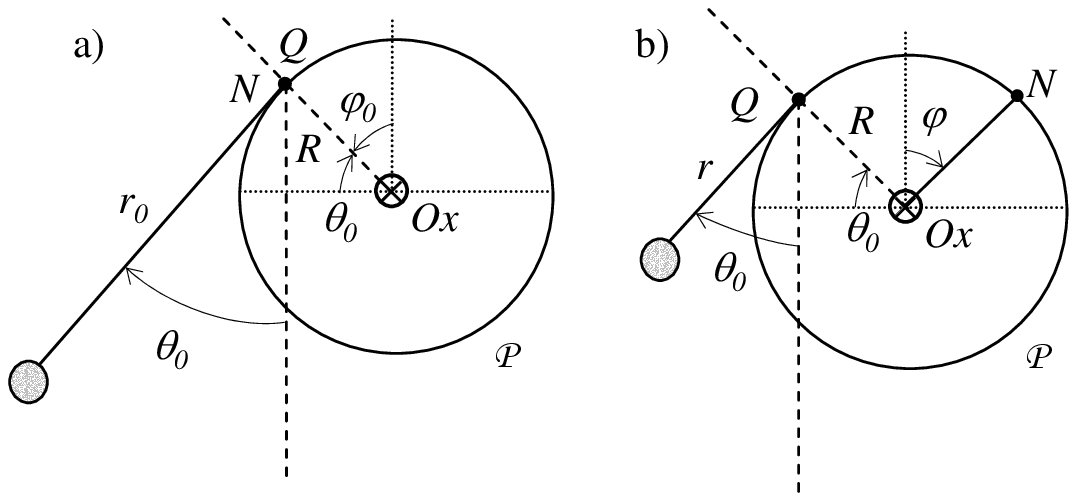}
\caption{Evolution of point $N$ of the pulley $\P$ when $\theta$ is kept
constant. The geometrical point of contact $Q$ is immobile and only $r$ evolves.}
\label{figcontbis}
\end{figure}

A physical way of understanding the difference between $\theta$ and $\varphi$ is to
imagine the following situation. At initial time, assume that $Q$ and $N$ are
superposed: $\theta=\theta_0$ and $\varphi=\varphi_0$
(Figure~\ref{figcontbis}a). If $\theta$ is fixed and $B$ heads downwards with
velocity $\dot{z}_B$, the absence of slippage of the thread on the pulleys
implies that they rotate with angular velocity $\dot\varphi$ given
by \eqref{firsteq}, meaning $N$ is moving while $Q$ remains fixed, and $r$ evolves from
$r_0$ to $r$; at final time, $\varphi \neq \varphi_0$ (Figure~\ref{figcontbis}b).

Since $L$ is supposed to be constant, $z_B$ is directly related to $r$, $D$,
and the lengths $QP$ and $P'Q'$ corresponding to regions where the thread and
the pulleys keep contact. Because
\[
QP=R\left(\frac{\pi}{2}-\theta\right), 
\qquad 
P'Q'=\frac{\pi R}{2},
\]
one has, precisely,
\begin{equation}\label{thirdeq}
L = r + D + \pi R - R \theta + z_B.
\end{equation}
Since $0 = \dot r - R\dot\theta + \dot z_B$, it follows that
\begin{equation}\label{fourtheq}
\dot{\varphi} = \frac{\dot{z_B}}{R} = \frac{R \dot{\theta} - \dot{r}}{R}.
\end{equation}

Hence, $\calcS$ is a system with two degrees of freedom, for
instance $\theta$ and $r$.

\section{Equations of motion of \emph{SAM} with pulleys}
\label{eqmotion}

\subsection{Equations of motion}

Let us determine the equations of motion for \emph{SAM} by taking into account
the pulleys, as opposed to what has been assumed in previous theoretical
studies \cite{Tufillaro-1984,Tufillaro-1985,Tufillaro-1986,Tufillaro-1988, 
AlmeidaMoreira,Tufillaro-1994,Nunes-1995,Yehia}. Indeed, their non-zero radii imply a likely change
in position for $Q$, and the equally likely rotation of $\P$ and $\P'$
around their respective revolution axes apparently deems their
\emph{inertial momentum} $I_p$ a significant dynamic parameter.
Observations will confirm this -- see Section~\ref{resexp}.

\medskip

Lagrange's formalism is used to derive these equations.
The kinetic energy of the system is expressed by:
\[
\E_k = \frac{1}{2} mv_A^2 + \frac{1}{2} Mv_B^2 + 2 \left(\frac{1}{2} I_p \dot{\varphi}^2\right)
\]
The first term sums up the contribution by pendulum $A$, the second is relative to
the counterweight $B$ and the third one corresponds to the rotation of the two
pulleys. We have:
\[
\bm{v}_A = \frac{\rmd \bm{OA}}{\rmd t} = \frac{\rmd \bm{OQ}}{\rmd t}
 + \frac{\rmd \bm{QA}}{\rmd t} ,
\]
where $\bm{OQ}=-R\cos\theta\,\bm{e}_y-R\sin \theta \, \bm{e}_z$ and
$\bm{QA}=-r\sin\theta\,\bm{e}_y+r\cos\theta\,\bm{e}_z$, hence in
the Cartesian base $\p{O\,\bm{e}_x\,\bm{e}_y\,\bm{e}_z}$ we can write
\[
\bm{v} _A =
\left(\begin{array}{c}{0}\\[5pt]{R\dot{\theta}\sin\theta-r\dot{\theta}\cos\theta
-\dot{r}\sin\theta}\\[5pt]{-R\dot{\theta}\cos\theta - r\dot{\theta}\sin\theta +
\dot{r}\cos\theta}
\end{array}\right)
\]
Similarly, $\bm{v}_B = \rmd\bm{OB}/\rmd t$ with $\bm{OB} = z_B \,
\bm{e}_z$. Using \eqref{thirdeq}, one gets
$$
\bm{v}_B = \dot{z}_B \, \bm{e}_z = 
\p{R\dot{\theta} - \dot{r}} \, \bm{e}_z.
$$ 
Finally, using \eqref{fourtheq} and introducing
the \emph{effective total mass} of the system 
$$
M_t = M + m +\frac{2I_p}{R^2},
$$ 
we get
\[
\E_k=\frac{1}{2}M_t(R\dot{\theta}-\dot{r})^2+\frac{1}{2}mr^2\dot{\theta}^2 .
\]
This expression is similar to that obtained when neglecting the pulleys,
except that:
\begin{itemize}
   \item the total mass is now different from $M + m$ by the term $2I_p/R^2$ conveying the rotation of the pulleys;
   \item the counterweight influences pendulum $A$ through its length $r$ and the winding of the rope on the associated pulley.
   The latter influence is considered in the term $R\dot{\theta}$.
\end{itemize}
Potential energy is only due to the Earth's local gravity field.
Dropping an irrelevant additional constant term, we have:
\[
\E_{p,g}=-m\bm{g}\cdot\bm{OA}-M\bm{g}\cdot\bm{OB}=-mgz_A-Mgz_B,
\]
so that
\[
\E_{p,g} = mg(R\sin\theta - r\cos\theta) + Mg(r - R\theta) .
\]
The Lagrangian $\L(r,\theta,\dot{r},\dot{\theta})=\E_k-\E_{p,g}$ of the system is thus:
\[
\L(r,\theta,\dot{r},\dot{\theta})=\frac{1}{2}M_t (R\dot{\theta}-\dot
{r})^2+\frac{1}{2}mr^2\dot{\theta}^2-gr(M-m\cos\theta)-gR(m\sin\theta-M\theta)
,
\]
from which one deduces the conjugate momenta $p_r$ and $p_\theta$ associated to
$r$ and $\theta$ respectively:
\begin{eqnarray*}
p_r&=&\frac{\partial\L}{\partial\dot{r}}=-M_t (R\dot{\theta} - \dot{r})\\
p_\theta &=& \frac{\partial\L}{\partial \dot{\theta}}= M_t R (R\dot{\theta}
 - \dot{r}) + mr^2\dot{\theta} = -Rp_r + mr^2\dot{\theta}
\end{eqnarray*}

The \emph{SAM} Hamiltonian is, in this case, by definition:
\[
\H = \E_k + \E_{p,g} ,
\]
or else, expressed in terms of $p_r$ and $p_\theta$:
\begin{equation}\label{H}
\H(r,\theta,p_r,p_\theta)=\frac{1}{2}\left[\frac{p_r^2}{M_t}+\frac{(p
_\theta+Rp_r)^2}{mr^2}\right]+gr(M-m\cos\theta)+gR(m\sin\theta-M\theta).
\end{equation}
Equations of motion follow from the Hamilton's equations:
\[
{\dot{p_r} = - \frac{\partial \H}{\partial r}
\qquad\hbox{and}\qquad
\dot{p_\theta} = - \frac{\partial \H}{\partial \theta}},
\]
yielding
\begin{equation} \label{eq:motion}
\left\{
\begin{array}{rcl}
\mu _t (\ddot{r} - R\ddot{\theta}) & = &  r\dot{\theta}^2 + g(\cos\theta - \mu)\\
r\ddot{\theta} & = & -2\dot{r}\dot{\theta} + R \dot{\theta}^2 - g\sin\theta
\end{array}
\right.
\end{equation}
with $\mu = M/m$ and $\mu _t = M_t/m = 1 + \mu + \left(2I_p/mR^2\right)$.

\subsection{A more physical way to obtain equations of motion}

An alternative method can be used in order to derive equations of motion
\eqref{eq:motion}. The second of these is obtained by applying
the angular momentum theorem at the mobile point $Q$ in order to cancel
reaction force at this contact point \cite{PerezM-2001}:
\[
\frac{\rmd \bm{L}_Q}{\rmd t}+\bm{v}_Q \times m\bm{v}_A = \bm{QA}
\times m\bm{g} ,
\]
where $\bm{L}_Q = \bm{QA}\times m\bm{v}_A=mr^2\dot{\theta}\,\bm{e}
_x$ and
\[
\bm{v}_Q= \frac{R\dot{\theta}}{AQ}\bm{AQ}, 
\qquad 
\bm{v}_A=\frac{r\dot\theta}{OQ}\bm{OQ}+\frac{\dot{r}}{QA}\bm{QA}.
\] 
We obtain $r\ddot{\theta}=-2\dot{r}\dot{\theta}+R \dot{\theta}^2-g\sin\theta$. 
Taking this into account, the first equation in \eqref{eq:motion} comes from the conservation of mechanical energy in
\emph{SAM}:
\[
\E_m=\frac{1}{2} M_t (R\dot{\theta}-\dot{r})^2+\frac{1}{2} m r^2
\dot{\theta}^2 + mg(R\sin\theta-r\cos\theta) - Mg(R\theta - r) = C
\]
$C$ being a real constant, after derivation with respect to $t$.

\subsection{Comparison with previous studies}

Without any pulley influence, i.e. pulley inertial momentum $I_p = 0$ and
pulley radius $R = 0$, we recover the equations obtained by \cite{Tufillaro-1984}:
\begin{eqnarray}
\left\{
\begin{array}{rcl}
(1 + \mu)\ddot{r} & = & r\dot{\theta}^2 + g(\cos\theta - \mu)\\
r\ddot{\theta} & =& -2\dot{r}\dot{\theta} - g\sin\theta
\end{array}
\right.
\end{eqnarray}
Obviously, if there is no oscillatory motion ($\theta = 0$),
the well-known simple Atwood machine \cite{Atwood-1784} is recovered:
\begin{equation}\label{eq20}
M_t\ddot{r} = g(m-M)
\end{equation}

\section{Description of the experimental apparatus}
\label{expappar}

\begin{figure}[ht]
\begin{center}
\includegraphics{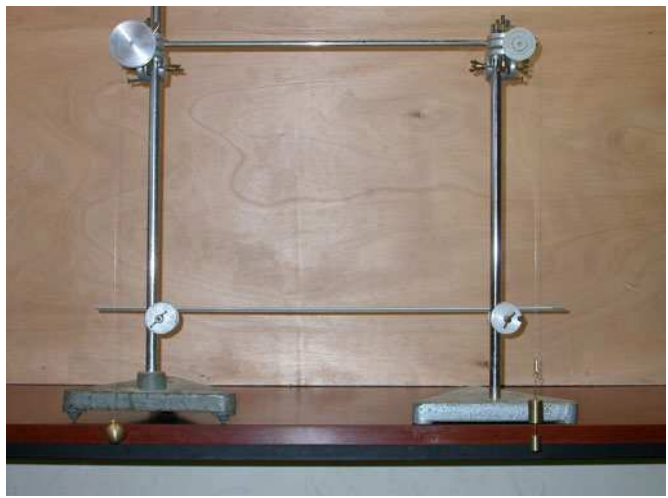}
\end{center}
\caption{Photo of the Swinging Atwood Machine (\emph{SAM}): a pendulum (on the
left) and a counterweight (on the right), linked together by a nylon thread.}
\label{samphoto}
\end{figure}

A physical prototype for \emph{SAM} has been built using two identical pulleys, a nylon thread, a
brass ball as a pendulum and a set of different hook masses acting as
counterweights. The pendulum and the chosen counterweight are linked together by
the nylon thread placed around the pulleys. A photo of \emph{SAM} is displayed
in Figure~\ref{samphoto}.

\subsection{About the pendulum and the counterweight}

\begin{figure}[ht]
\begin{tabular}{cc}
{\includegraphics[width=4.5cm]{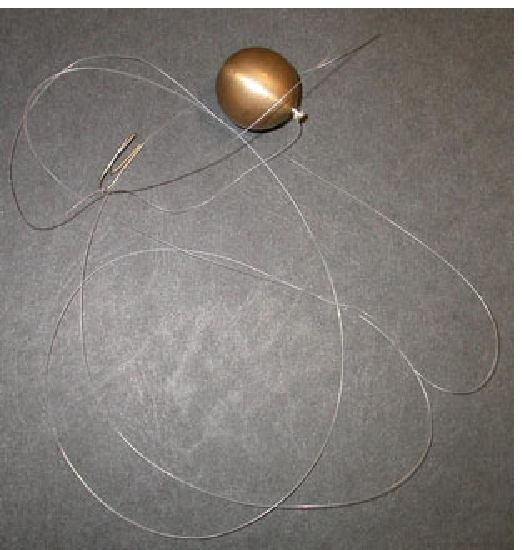}} &
{\includegraphics[width=7.7cm]{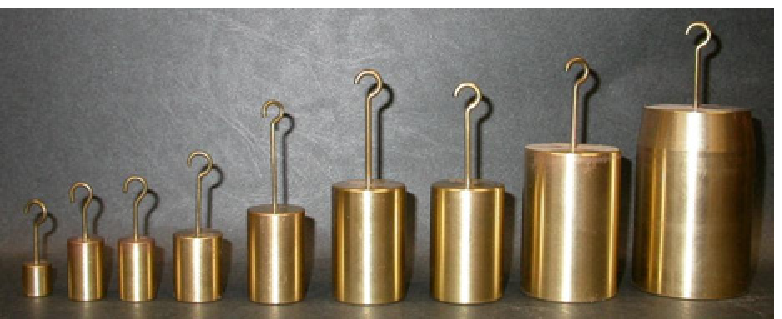}}\\ {(a)} & {(b)}\\
\end{tabular}
\caption{(a) Brass ball (pendulum), nylon thread and paper clip (b) Set of hook
masses used in the experiment (counterweight). From left to right: $10\ut{g}$, $20\ut{g}$, $20\ut{g}$,
$50\ut{g}$, $100\ut{g}$, $200\ut{g}$, $200\ut{g}$, $500\ut{g}$, $1000 \ut{g}$.}
\label{pendphoto}
\end{figure}

Each mass in the experimental device has been measured with a precision scale of
$0.01\ut{g}$ of accuracy. The pendulum is a brass ball with a $30 \ut{mm}$
diameter and a mass $m = 118.36 \ut{g}$.  The picture of the pendulum in
Figure~\ref{pendphoto}a also exhibits a paper clip and the nylon thread,
the latter being solidly tied to the brass ball and the paper clip.

\medskip

The paper clip is secured to the hook of the chosen counterweight, in turn
picked out from nine hook masses whose measured values are $M = 10.01\ut{g}$,
$20.02\ut{g}\, (\times 2)$, $50.05\ut{g}$, $100.10\ut{g}$, $200.22\ut{g} \,
(\times 2)$, $500.51\ut{g}$, and $1000.10\ut{g}$ (Figure~\ref{pendphoto}b).
The relative difference between these values and those engraved in each hook
mass is $0.1 \ut{\%}$; thus, with respect to the orders of magnitude of the
different masses involved in the experimental device, this difference can be
neglected. The values considered are therefore presumed to be those indicated on
the hook mass themselves, namely $M=10 \ut{g}$, $20 \ut{g}\, (\times 2)$, $50
\ut{g}$, $100 \ut{g}$, $200\ut{g}\,(\times 2)$, $500\ut{g}$, and $1000\ut{g}$.
Henceforth, and for the sake of linguistic
simplicity, these hook masses will be called ``counterweights", although \emph{weight}
and \emph{mass} are different notions, however related. This set enables varying the counterweight mass from $10\ut{g}$ (one
mass) to $2\,100\ut{g}$ (addition of all the masses) with a step of $10\ut{g}$
by hooking several masses together. Among these hook masses, one is hung on the
nylon thread by means of the paper clip, whose measured mass is $0.37\ut{g}$. It
is interesting to note that, by a fortunate coincidence, the mass of the paper
clip is equal, with a $0.01 \ut{g}$ difference, to $0.36 \ut{g}$, i.e. the
mass of the brass ball minus $118 \ut{g}$. Therefore, the mass of the brass ball
can be taken as equal to $118 \ut{g}$ and the mass of the paper clip can be
ignored.  Finally, we get, for the pendulum and the counterweight, respectively:
$m = 118 \ut{g}$ and $10 \ut{g} \leq M \leq 2100 \ut{g}$.

\subsection{About the nylon thread and the pulleys}

The thread (Figure~\ref{pendphoto}a) ensures a mechanical coupling between the
pendulum and the counterweight through the two pulleys. The length of the thread
is about one meter and its measured mass of $0.10 \ut{g}$ is negligible compared
to the other masses involved. In addition, the nylon thread is assumed
inextensible. During experimentation, no thread breaking has been reported.

\medskip

\begin{figure}[ht]
\begin{center}
\begin{tabular}{cc}
{\includegraphics{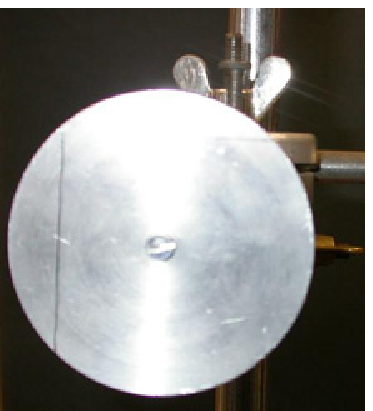}} & {\includegraphics{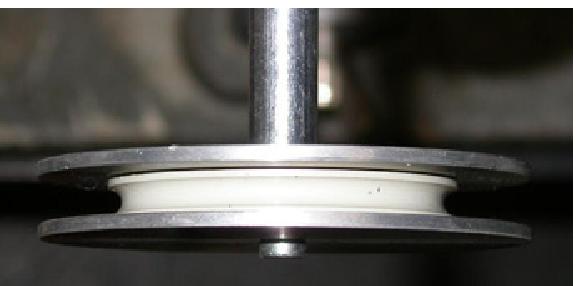}}\\
{(a)} & {(b)}\\
{\includegraphics{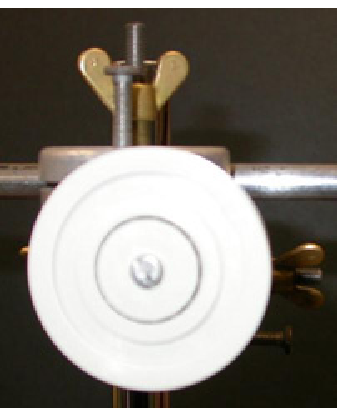}} & {\includegraphics{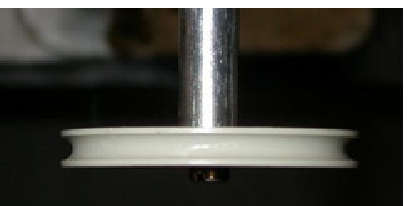}}\\
{(c)} & {(d)}\\
\end{tabular}
\end{center}
\caption{Photos of the two pulleys. (a) Pulley $\P$, front view: a metallic
plate is added which hides the internal part and the ball bearing of the original
pulley. (b) Pulley $\P$, top view: the groove is made deeper by $1\ut{cm}$
with the additional plate; the original pulley is easily recognizable
between the two plates. (c) Pulley $\P'$, front view: the ball bearing
uncouples the immobile internal part and the mobile external one (d) Pulley
$\P'$, top view: photograph of the axis of revolution and the groove of the
pulley. Moreover, the two pulleys are distant enough from one another to avoid a
pendulum-counterweight collision during the motion.}
\label{pulleysphoto}
\end{figure}

Pulleys used are shown in photos of Figure~\ref{pulleysphoto}. They are made up of
two parts: an internal, immobile one bound to the revolution axis, and a mobile,
external one liable to rotate around this axis. These two pulley
components are uncoupled through a ball bearing which, moreover, reduces
mechanical energy dissipation by friction. Pulley radius is $R=2.5 \ut{cm}$ and
that of the motionless part is $1 \ut{cm}$. Pulley $\P$, associated to the
pendulum, has been modified in order to make its groove deeper. Indeed, during
the first experimentations we observed that the thread could rapidly exit the
groove because of the pendulum motion. To avoid this, which could by the way be
dangerous, two metallic plates were added and fixed to the pulley in order to
increase by $1 \ut{cm}$ the depth of the groove (Figures~\ref{pulleysphoto}a
and ~\ref{pulleysphoto}b). It is worth noting that the plates are fixed to the
immobile part of the pulley and are in no way in contact with the mobile one.
The motion of the latter one is thus not affected by such a modification: hence, from a
mechanical point of view, the resulting pulley is identical to the original
one.

\medskip

\subsection{Strengthening of the apparatus}

Figure~\ref{samphoto} also features two horizontal metallic rods binding the two
feet of \emph{SAM}. Their role is to reinforce the machine. Indeed, due to
considerable stress involved in the pendulum and counterweight motions, the
orientation of the two pulleys, as well as the arbitrary distance $D = 57.5 \ut{cm}$ between them,
can change; thus, it could be dangerous not to strengthen the whole device. The lower metallic rod is solidly fixed to the feet
while the upper one is solidly fixed to the revolution axes of the pulleys.
Consequently, we ensured a constant distance between the pulleys whose axes keep
a constant direction; in virtue of such a construction, \emph{SAM} is solid and
operational.

\section{Motion of the pendulum: experimental results}
\label{resexp}

\subsection{Experimental measure of $I_p$}

The presence of the inertial momentum $I_p$ of a pulley in \eqref{eq:motion} renders an
experimental determination thereof necessary. This measure was made using a
simple Atwood machine where the heavier mass ($M=130 \ut{g}$) fell down from a
convenient height $h=1.66 \ut{m}$; the lighter mass (the brass ball) being $m =
118 \ut{g}$. Using equation \eqref{eq20}, one gets:
\[
I_p=R^2\left[\frac{(M-m)g(\Delta t)^2}{2h}-(M+m)\right].
\]
where $\Delta t$ is the fall duration. 
Through a set of ten measures with a chronometer of $0.01 \ut{s}$ of accuracy,
the mean fall duration found is $\left\langle\Delta t\right\rangle=2.70\pm 0.01\ut{s}$. 
$\left\langle I_p\right\rangle=6.85\times 10^{-6} \ut{kg\cdot m^2}$ ensues. 
Concerning the value of $I_p$, errors are due to the measure of
both $\Delta t$ and the positions of $M$ at the initial and
final times -- that is, the determination of $h$. Uncertainties are mainly due to the
determination of the final time, which must correspond to the falling distance
$h$ as precisely as possible; initial and final positions are determined with an
error of $0.1 \ut{cm}$, which compared to the value of $h$ can be neglected.
Consequently, uncertainties in position determinations are disregarded and the
error in the measure of $I_p$ can be reasonably associated to the uncertainty in
$\Delta t$ ($0.4 \hbox{ \%}$). Thus, precision on $I_p$ is twice that of
$\Delta t$, hence about $1 \hbox{ \%}$; absolute uncertainty is thus $0.07\times
10^{-6} \ut{kg\cdot m^2}$. Therefore, we can write:
\[
I_p = (6.85 \pm 0.07) \times 10^{-6} \ut{kg\cdot m^2}
\qquad\hbox{or}\qquad
I_p = 6.85 \times 10^{-6} \ut{kg\cdot m^2} \pm 1 \hbox{ \%}.
\]

\subsection{Experimental results}

The motion of the pendulum has been filmed for various $\mu$-values and initial
conditions $(r_0\,;\,\theta_0)$. Then, using the ``Synchronie" software and
focusing on each film image by image, a pointer enabled us to pick up pendulum
positions and record them. Such a process is necessarily a source of errors, as
it is sometimes difficult to locate the pendulum exactly, especially if
velocity is high. The errors introduced by such a procedure are not simple to estimate. However,
trajectories have been correctly recorded, as comparisons with numerically-simulated theoretical
results will show (see Section~\ref{comptheo}).

\subsubsection{Case $\mu_{\mathrm{theo}} = 3$}\label{5.2.1}

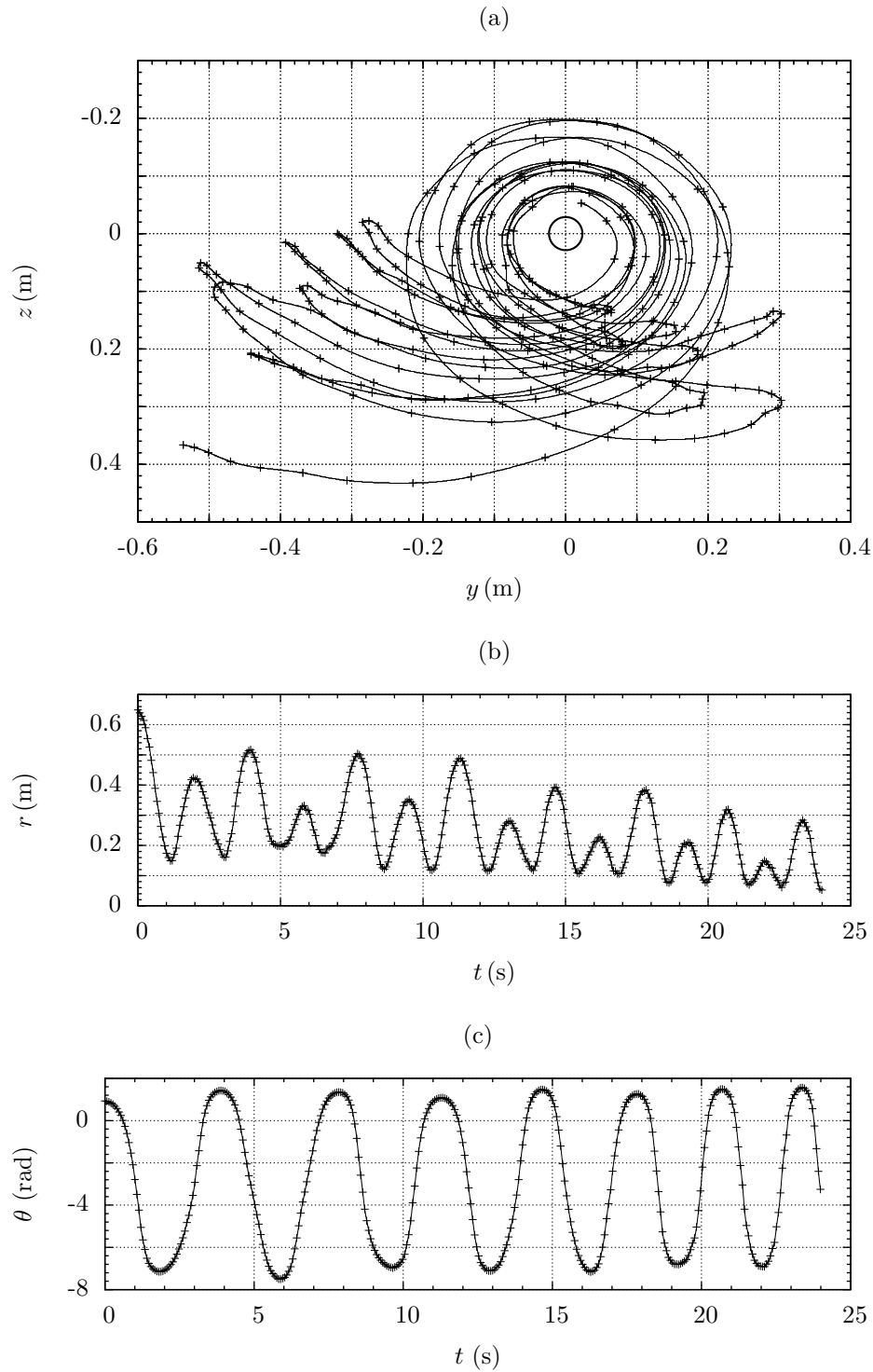
\begin{figure}[ht]
\begin{tabular}[t]{c}
\hspace{-1mm}{\input{Figure7a}}\\
\hspace{-1mm}{\input{Figure7b}}\\
\hspace{-1mm}{\input{Figure7c}}\\
\end{tabular}
\caption{Case $\mu_{\mathrm{exp}}=2.966$ with initial conditions $r_0=0.649\ut{m}$ and
$\theta _0=53.5^{\circ}$. (a) Experimental positions (black crosses) and
interpolated pendulum trajectory (solid line). The Cartesian coordinates of the
initial position are $y=-0.54\ut{m}$ and $z=0.37\ut{m}$ and those of the final
position are $y=0.02\ut{m}$ and $z=-0.05\ut{m}$. The pulley is represented by
the circle whose centre is at the origin of coordinates $(0\,;\,0)$.
(b) Experimental positions (black crosses) and interpolated curve (solid line)
for the evolution of $r$. (c) Same as (b) but for the angle $\theta$. }
\label{traj}
\end{figure}

Since previous studies focused mainly on the particular and theoretical case
$\mu_{\mathrm{theo}}=3$, this was the first one we experimentally addressed. In fact, the
masses available only allowed us to \emph{approach} $\mu_{\mathrm{theo}}$: with $m=118 \ut{g}$
and $M=350 \ut{g}$ one obtains $\mu_{\mathrm{exp}}=2.966$, which is the closest value to
$\mu_{\mathrm{theo}}$. The sampling time step has been $67 \ut{ms}$. Motion
has been researched for four different initial conditions
$(r_0\,;\,\theta _0)$=$(0.649\,;\,53.5)$, $(0.710\,;\,66.5)$,
$(0.854\,;\,68.3)$
and $(0.867\,;\,51.1)$; $r_0$ is in meters, $\theta_0$ in degrees. The
motion of the pendulum presents the same pattern and characteristics for all
these conditions, so only the trajectory for the first initial conditions is
shown in Figure~\ref{traj}a. All in all, $359$ sampling times have been recorded.
The pendulum has a planar revolving trajectory around the pulley and presents an
asymmetry with respect to the vertical direction. Note that the pendulum becomes
closer and closer to the pulley as a consequence of dissipative phenomena and is
bound to end up knocking against it. Phenomena qualifying as dissipative are, to our knowledge:
the friction between the thread and the pulley, the air friction on the pendulum
and the counterweight as well as friction inside the ball bearing. Evolutions of
the length of the pendulum $r$ and angle $\theta$ are displayed in
Figure~\ref{traj}b and Figure~\ref{traj}c respectively. The asymmetry of the
trajectory and dissipation are observable in the evolution of $r$ since this
variable exhibits different minimal and maximal amplitudes which decrease in function of time $t$.

\medskip

The Fourier analysis (non-displayed) of the data shows that, 
for the behaviour of $\theta$, the most relevant harmonic is the constant 
term, followed by harmonics $7$, $6$ and $8$. They account, respectively, for $0.401$, $0.311$,
$0.117$, and $0.091$ of the total variation. Of course, the contributions of
harmonics $6$ and $8$ are due to leakage of the $7^{\mathrm{th}}$ one. The amplitude of the
$6^{\mathrm{th}}$ harmonic is larger than that of the $8^{\mathrm{th}}$, showing that the true
dominant average frequency is slightly less than $7$ times the basic frequency.
From the number of data and the sampling step time, a basic frequency
of $0.042 \ut{Hz}$ follows. Hence, an average value of the dominant frequency can be
estimated to be equal to $0.294 \ut{Hz}$. However, as is clear by looking at
the maxima of the angle, the frequency changes with time. The spacing between
successive maxima takes the approximate values $3.820\ut{s}, 3.909\ut{s}, 3.379
\ut{s}, 3.311\ut{s}, 3.173\ut{s}, 2.780\ut{s}$, and $2.643\ut{s}$. Hence, the
instantaneous frequency changes from about $0.262\ut{Hz}$ to about $0.378
\ut{Hz}$. The explanation is simple: the dissipation reduces the energy and the
length of the pendulum becomes shorter, increasing the frequency.

\medskip

For the radius, the major contribution comes from the constant term, followed by
harmonics $13$ and $7$. They account for $0.810, 0.037$, and $0.026$ of the
total variation. Comparing the plot of $r$ as a function of time with that of
$\theta$, one can see a doubling in the number of maxima. The explanation is simple:
largest maxima occur at the left part of the plot of the orbit. Then, a minimum
is reached when $\theta=-\pi$ (i.e., upwards), followed by a maximum to the
right, and a new minimum at $\theta=-\pi$ to reach a larger maximum on the left.
The spacing between successive larger maxima of $r$ is very close to the one
observed for $\theta$. As mentioned, the largest non-constant harmonics are the
$13^{\mathrm{th}}$ and the $7^{\mathrm{th}}$, not in a 2-to-1 ratio. This is related to the fact that
the ``best'' estimate of the main frequency for $\theta$ is slightly less than $7$
times the basic frequency. The closest integer to the double would be $13$ rather
than 14.

\medskip

The decrease in energy can be calculated as follows. From the experimental data,
the values of $\dot r$ and $\dot \theta$ can be computed. For that, we have used
two independent methods. The first one is simply numerical differentiation with a
central formula. The second one aims at filtering errors in the data as well.
A Discrete Fourier Transform has been computed and harmonics up to order $32$
have been retained. Then, it is possible to check that the reconstruction agrees
quite well with the initial data and one can compute the values of $\dot r$ and
$\dot\theta$ using these Fourier expansions. To prevent leakage due to
the fact that the data at the ends of the interval are quite different
(Figs.~\ref{traj}b and c), which originates the well-known $O(1/n)$ decrease of
the order of magnitude of the $n^{\mathrm{th}}$ harmonic,
different procedures have been used, but the results are essentially the same.
They also show a reasonable agreement using the first and second methods.

\clearpage

\medskip

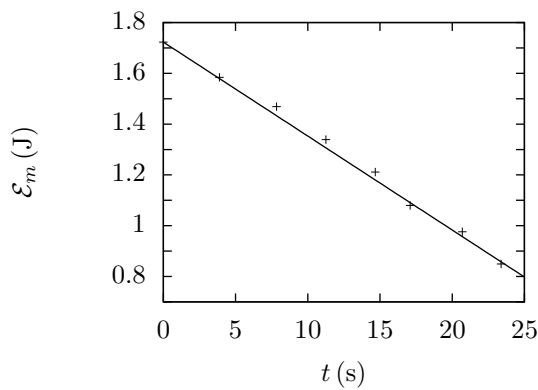
\begin{figure}[ht]
\input{Figure8}
\vspace{-3mm}
\caption{Values of the energy (crosses) occurring when $\theta$ reaches a
maximum at the left part of Figure~\ref{traj}a. We have also displayed the
function $\E_m = 1.723 - 0.037\,t$.}
\label{en3}
\end{figure}

When $\dot r$ and $\dot\theta$ are available, one can compute $p_r$ and $p_
{\theta}$ and subsequently the value of the energy. The values for which $\theta$ reaches a
maximum (i.e., on the left of Figure~\ref{traj}a) are shown in Figure~\ref{en3}.
The rate of decrease of the energy is about $0.037 \ut{J\cdot s^{-1}}$.

\subsubsection{Case $\mu_{\mathrm{theo}}=1.5$}\label{5.2.2}

\begin{figure}[ht]
\begin{tabular}[t]{c}
\hspace{-1mm}{\input{Figure9a}}\\
\hspace{-1mm}{\input{Figure9b}}\\
\hspace{-1mm}{\input{Figure9c}}\\
\end{tabular}
\caption{Same as Figure~\ref{traj} but for $\mu_{\mathrm{exp}}=1.525$ and initial
conditions $r_0=0.484\ut{m}$ and $\theta_0=87.0\ut{^\circ}$. In (a), the
Cartesian coordinates of the initial position are $y=-0.485\ut{m}$ and $z=0$~;
for the final position, one has $y=-0.330\ut{m}$ and $z=0.195\ut{m}$. The origin
$(0\,;\,0)$ corresponds to the centre of the pulley (not represented).}
\label{otraj}
\end{figure}
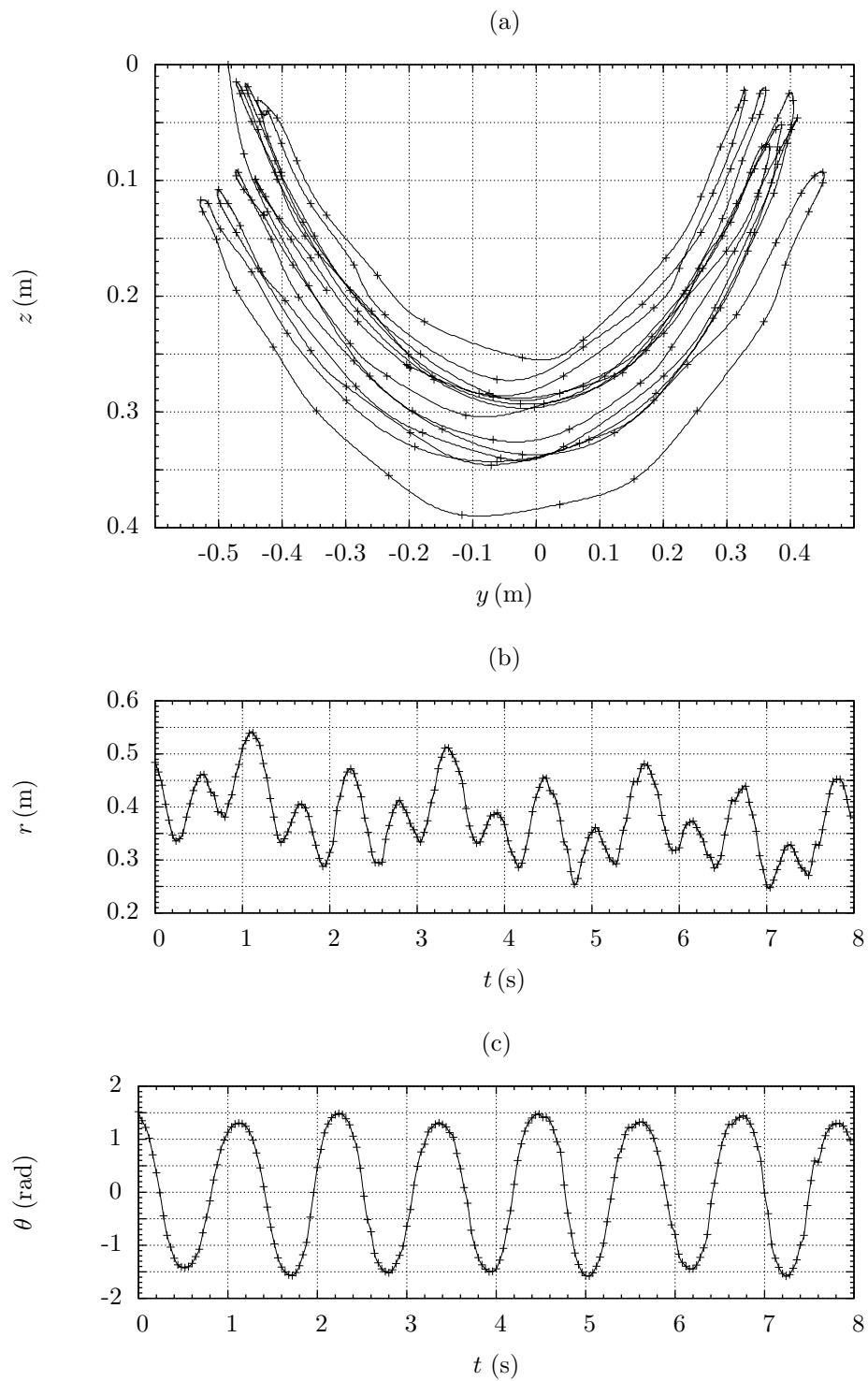

The experimental value of $\mu$ closest to $\mu_{\mathrm{theo}}=1.5$ is $\mu_{\mathrm{exp}}=1.525$,
given by $M=180 \ut{g}$. Proceeding as in the above case allows us to retrieve the
experimental trajectory and the evolution of the degrees of freedom. In this
case, time step is $40\ut{ms}$. Two initial conditions have been considered:
$(r_0\,;\,\theta_0)$=$(0.484\ut{m}\,;\,87.0^{\circ})$ and $(0.621\ut{m}\,;\,87.7^{\circ})$.
Since they produce the same dynamic behaviour, only the first one is displayed
(Figures~\ref{otraj}a,~\ref{otraj}b and \ref{otraj}c respectively).
A slight asymmetric trajectory with respect to the vertical direction and an
evolution of $\theta$ close to periodic with a period around $1.1-1.2\ut{s}$
can be observed. For the evolution of $r$, asymmetry and slight dissipation are
also observed.

\medskip

A study similar to that of $\mu_{\mathrm{exp}}=2.966$ is performed. The main contribution~to
the Fourier analysis of $\theta$ comes from harmonic number $7$ which accounts
for $0.912$ of the total variation with a frequency of approximately
$0.875\ut{Hz}$, although a better value for the average frequency seems to be
$0.89\ut{Hz}$. Harmonics $14$ and $21$ also play a relevant role. As we did for
$\mu_{\mathrm{exp}}=2.966$, we can consider the spacing between successive maxima of
$\theta$ which takes the values $1.125\ut{s}$, $1.119\ut{s}$, $1.114\ut{s}$,
$1.110\ut{s}$, $1.156\ut{s}$, $1.133\ut{s}$ and $1.063 \ut{s}$,
showing a decreasing trend with irregularities.

\medskip

For $r$, the largest harmonic is the constant term which accounts for $0.971$ of
the signal. If we skip this term, harmonics $7$, $14$ and $3$ are
clearly seen. They contribute to $0.30$, $0.29$, and $0.14$ of the signal minus
the constant part.

\clearpage

\begin{figure}[ht]
\input{Figure10}
\caption{Values of the energy (crosses) occurring when $\theta$ reaches a
minimum at the right part of Figure~\ref{otraj}a. We have also displayed the
function $\E_m=0.751-0.024\,t$.}
\label{en15}
\end{figure}
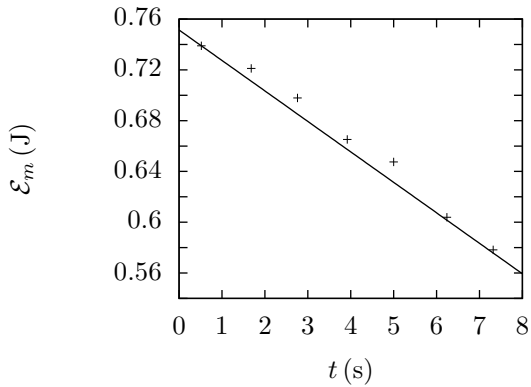

The decrease of the energy as a function of time has been displayed in
Figure~\ref{en15}, this time using the values of the energy computed at the
minima of $\theta$, on the right part of Figure~\ref{otraj}a. Now the rate of
decrease is about $0.024\ut{J\,s^{-1}}$. In this case, using the filtered
Fourier methods gives better results, because of the large changes in position
with a time step of $40\ut{ms}$.

\subsubsection{An unbounded case: $\mu_{\mathrm{theo}} = 1$}

\begin{figure}[ht]
\begin{tabular}[t]{c}
\hspace{-1mm}{\input{Figure11a}}\\
\hspace{-1mm}{\input{Figure11b}}\\
\hspace{-1mm}{\input{Figure11c}}
\end{tabular}
\caption{Same as Figure~\ref{traj} but for $\mu_{\mathrm{exp}}=1.017$ and initial
conditions $r_0=0.120\ut{m}$ and $\theta_0=77.9\ut{^\circ}$. In (a), the
Cartesian coordinates of the initial position are $y=-0.122\ut{m}$ and $z=0$~;
for the final position, one has $y=-0.116\ut{m}$ and $z=1.167\ut{m}$. The
origin $(0\,;\,0)$ corresponds to the centre of the pulley (not represented).}
\label{ootraj}
\end{figure}
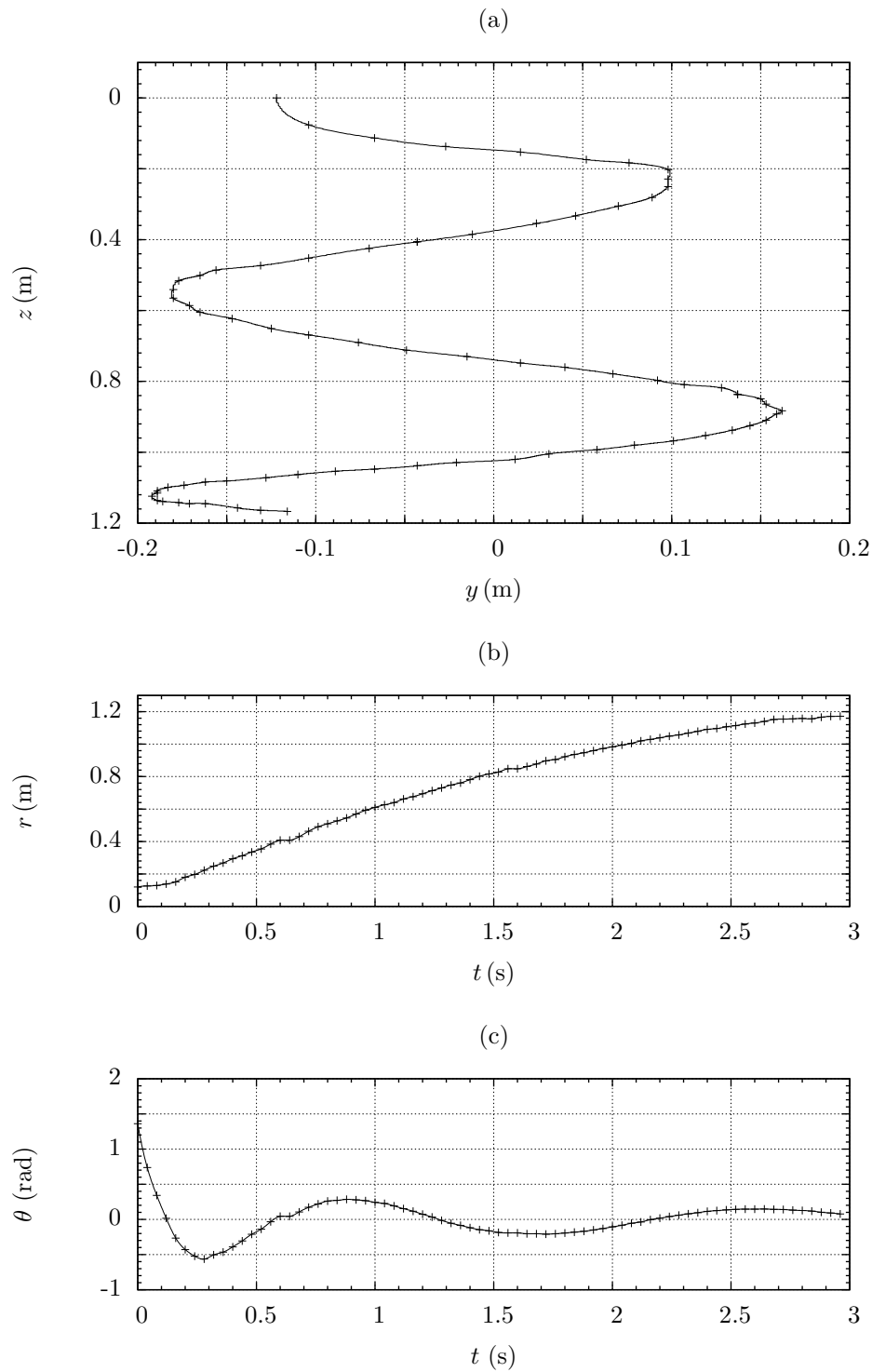

In this situation, the experimental value of $\mu$ is $\mu_{\mathrm{exp}}=1.017$ and
three initial conditions have been considered: $r_0=0.120\ut{m}$, $0.263\ut{m}$
and $0.477\ut{m}$ for $\theta_0=77.9^{\circ}$. Again, trajectories present the
same pattern, so only one is shown (Figure~\ref{ootraj}a). They are
characterized by an increase in $r$ and $\theta$ oscillations around the
vertical $(\theta=0)$ with a decreasing amplitude of $\theta$
(Figures~\ref{ootraj}b and c respectively). In Figure~\ref{ootraj}b, $r$ appears
to approach a linear increase in time: $r=v_z t$ where $v_z$ is the velocity
along the vertical.

\clearpage

\section{Numerical solution of \emph{SAM} equations of motion}
\label{comptheo}

\subsection{Theoretical trajectories}

Equations of motion \eqref{eq:motion} have been numerically integrated for the same initial
conditions and values of the parameter $\mu$ as above in order to compare the
theoretical trajectories, displayed in Figure~\ref{trajtheo}, with the
experimental ones. Computation of the theoretical trajectories has been
performed by using both a Taylor integration method and a variety of Runge-Kutta methods of different orders with
step-size control to integrate the equations of motion.

\medskip

From a general point of view, the theoretical trajectories seem quite
similar to the experimental ones. However, some slight
differences can be detected. Firstly, it is obvious that dissipative phenomena,
though experimentally reduced, play a non-negligible role since convergence of
the pendulum towards the pulley for the first case (Fig.~\ref{traj}b),
decrease of $r$ for the second one (Fig.~\ref{otraj}b) and
relatively slow increasing of $r$ for the third one (Fig.~\ref{ootraj}b)
are clearly associated to energy dissipation. Friction will be studied a bit
further in Subsection~\ref{comptheo}.3 and much more in future works. Secondly,
it must be said that for the case $\mu_{\mathrm{exp}}=2.966$ the pendulum touches the
nylon thread at each revolution, an effect which could be included into the
equations of motion through a dissipative term; however, it seems quite difficult to
introduce such an effect in a realistic manner.

\medskip

An important point concerns the comparison between these trajectories and those
of \cite{Tufillaro-1984}. For the first case for instance, Tufillaro's
trajectories are symmetrical with respect to the vertical direction,
as opposed to the above ones (Figures~\ref{traj}a and ~\ref{trajtheo}a).
Clearly, this asymmetry is due to the influence of the pulleys.

\subsection{Influence of the pulleys on the motion}

Pulleys can influence the motion of \emph{SAM} through their dimension (since
radius $R\neq 0$) and their rotation (since $I_p\neq 0$).

\subsubsection{Influence of the radius: $\mu_{\mathrm{exp}}=2.966$} 

Figure~\ref{trajinmom}
sketches the trajectory for $\mu_{\mathrm{exp}}=2.966$ when $I_p=0$ for different values
of increasing $R$; i.e. the pendulum pulley $\cal{P}$ has a non-negligible
radius but pulleys are not allowed to rotate. The first figure corresponds to
the symmetrical Tufillaro trajectory $(R=0,I_p=0)$. Ostensibly,
the larger $R$ is, the more significant the asymmetry becomes; for $R=5\ut{cm}$,
the pendulum hits the pulley before completing one revolution.

\subsubsection{Influence of the inertial momentum: $\mu_{\mathrm{exp}} = 2.966$}

Figure~\ref{trajrpim} sketches the trajectory for $\mu_{\mathrm{exp}}=2.966$ for
increasing values of $I_p$ with a value of pulleys radius taken to the real
value $R=2.5\ut{cm}$. The trajectory of the pendulum is visibly modified: it describes more irregular trajectories and fills more space
as $I_p$ increases.
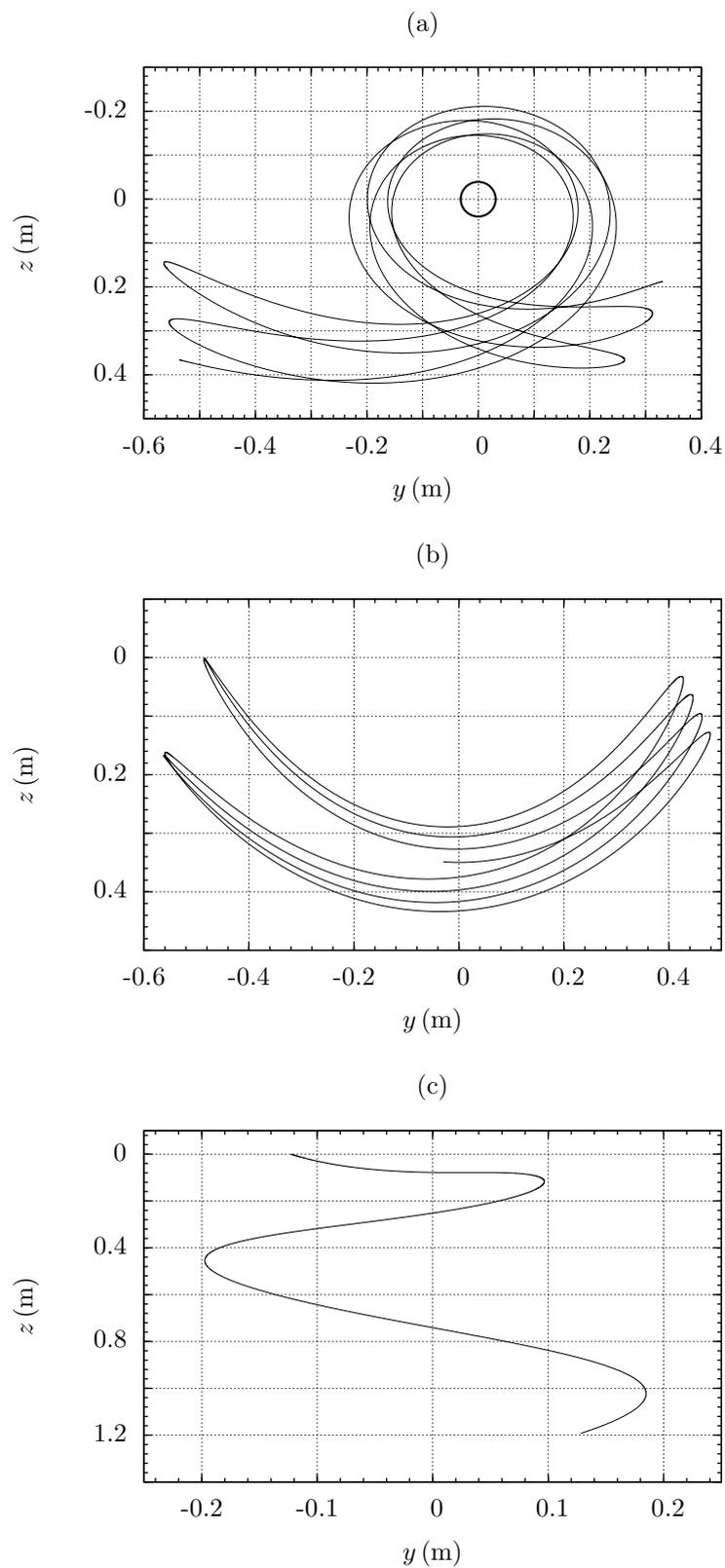
\begin{figure}[ht]
\begin{tabular}[t]{c}
{\input{Figure12a}}\\
{\input{Figure12b}}\\
{\input{Figure12c}}
\end{tabular}
\caption{Numerical trajectories obtained by solving equations of motion \eqref{eq:motion}.
(a) $\mu_{\mathrm{exp}}=2.966$ and for the same initial conditions as Figure~\ref{traj}a.
(b) $\mu_{\mathrm{exp}}=1.525$ and for the same initial conditions as
Figure~\ref{otraj}a. (c) $\mu_{\mathrm{exp}}=1.017$ and for the same initial conditions
as Figure~\ref{ootraj}a.
}
\label{trajtheo}
\end{figure}

\clearpage

\begin{figure}[ht]
\begin{tabular}[t]{cc}
\hspace{-3mm}{\input{Figure13a}}&\hspace{-2mm}{\input{Figure13b}}\\
\hspace{-3mm}{\input{Figure13c}}&\hspace{-2mm}{\input{Figure13d}}
\end{tabular}
\caption{Theoretical trajectories of the pendulum with fixed pulleys of different
radii $R$ for $\mu_{\mathrm{exp}}=2.966$ and initial condition
$(r_0\,;\,\theta_ 0)$=$(0.649\ut{m}\,;\,53.5 ^{\circ})$.
(a) $R=0$, (b) $R=2.5\ut{cm}$, (c) $R=4\ut{cm}$ and (d) $R=5\ut{cm}$.
In the latter subplot, the pulley has been represented by the
circle whose centre is at the origin of coordinates in order to see the
collision between the pendulum and the pulley.}
\label{trajinmom}
%
\begin{tabular}[t]{cc}
\hspace{-3mm}{\input{Figure14a}}&\hspace{-2mm}{\input{Figure14b}}\\
\hspace{-3mm}{\input{Figure14c}}&\hspace{-2mm}{\input{Figure14d}}
\end{tabular}
\caption{Theoretical trajectories of the pendulum with pulleys of radius $R=2.5
\ut{cm}$ and different inertial momenta $I_p$ for $\mu_{\mathrm{exp}}=2.966$ and initial
condition $(r_0\,;\,\theta_0)$=$(0.649\ut{m}\,;\,55.7^{\circ})$.
(a) $I_p=0$; this subplot which is the same as Figure~\ref{trajinmom}b has been
repeated for the sake of clarity, (b) $I_p=6.85\times 10^{-6}\ut{kg\cdot m^2}$,
(c) $I_p=13.70\times
10^{-6}\ut{kg\cdot m^2}$ and (d) $I_p=27.40\times 10^{-6}\ut{kg\cdot m^2}$.}
\label{trajrpim}
\end{figure}

\clearpage

\begin{figure}[ht]
\begin{tabular}[t]{cc}
\hspace{-3mm}{\input{Figure15a}}&\hspace{-2mm}{\input{Figure15b}}\\
\hspace{-3mm}{\input{Figure15c}}&\hspace{-2mm}{\input{Figure15d}}
\end{tabular}
\caption{Theoretical trajectories of the pendulum with fixed pulleys of different
radii $R$ for $\mu_{\mathrm{exp}}=1.525$ and initial condition
$(r_0\,;\,\theta_0)$=$(0.484\ut{m}\,;\,87.0^{\circ})$. (a) $R=0$,
(b) $R=2.5\ut{cm}$, (c) $R=5\ut{cm}$ and (d) $R=10\ut{cm}$.}
\label{otrajtheo}
%
\begin{tabular}[t]{cc}
\hspace{-3mm}{\input{Figure16a}}&\hspace{-2mm}{\input{Figure16b}}\\
\hspace{-3mm}{\input{Figure16c}}&\hspace{-2mm}{\input{Figure16d}}
\end{tabular}
\caption{Theoretical trajectories of the pendulum with pulleys of radius $R=2.5
\ut{cm}$ and different inertial momenta $I_p$ for $\mu_{\mathrm{exp}}=1.525$ and initial
condition $(r_0\,;\,\theta_0)$=$(0.484 \ut{m}\,;\,87.0^{\circ})$.
(a) $I_p=0$; this subplot which is the same as Figure~\ref{otrajtheo}b has been
repeated for the sake of clarity, (b) $I_p=6.85\times 10^{-6}\ut{kg\cdot m^2}$,
(c) $I_p=13.70\times 10^{-6}\ut{kg\cdot m^2}$ and (d) $I_p=54.80\times
10^{-6}\ut{kg\cdot m^2}$.}
\label{ootrajtheo}
\end{figure}

\clearpage

\subsubsection{Influence of the pulleys: $\mu_{\mathrm{exp}} = 1.525$}

In this case, results of variations in $R$ with $I_p=0$ (Figure~\ref{otrajtheo})
and $I_p$ variations for $R=2.5\ut{cm}$ (Figure~\ref{ootrajtheo}) are similar
in that trajectories are modified more visibly as $R$ and $I_p$ increase.
However, the influence of $I_p$ and $R$ on \emph{SAM} motion depends on the value
of $\mu$ considered. If $R$ increases, the pulleys being fixed, the brass ball
ends up hitting the pulley but asymmetry does not becomes more and more
important. If $I_p$ increases, pulley dimensions being fixed, the pendulum
evolves in a much more limited space.

\subsection{Poincar\'e maps and rotation number}
To have a global view of the dynamics of \emph{SAM},
we have computed Poincar\'e maps $\P_m$ 
on suitable Poincar\'e sections. We note that the Hamiltonian
\eqref{H} is not $2\pi$-periodic because of the linear term in $\theta$.
Be that as it may, our Poincar\'e section $\Sigma$ defining $\P_m$ is
given by the coiling of $\theta$ through multiples of $2\pi$, with $\dot{\theta}>0.$ But,
contrarily to \emph{SAM} without pulleys (\cite{Tufillaro-1985}), one has to distinguish
between cuts through different multiples of $2\pi$. Plus, one can not superimpose the
different sheets.

\medskip

Some types of ``escape'' can occur. The main source thereof is $r$ going to
zero. Other relevant sources of escape are $r$ increasing too much or $\left|\theta\right|$
becoming too large. All orbits leading to some of these escapes are deleted.

\medskip

\begin{figure}[!ht]
\begin{tabular}{cc}
\psfrag{pr}{$p_r$} \psfrag{r}{$r$}
\hspace{-5mm}\epsfig{file=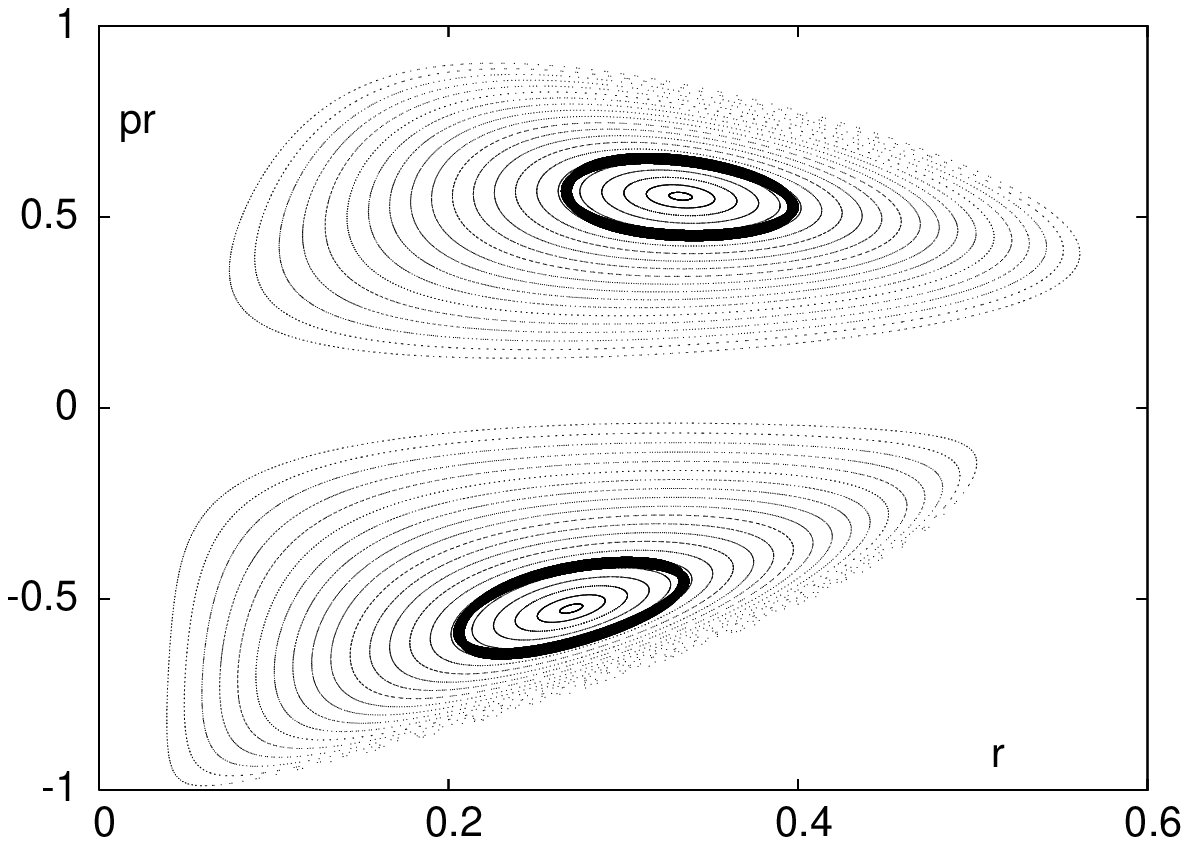,width=6.5cm}&
\psfrag{pr}{$p_r$} \psfrag{r}{$r$}
\hspace{-5mm}\epsfig{file=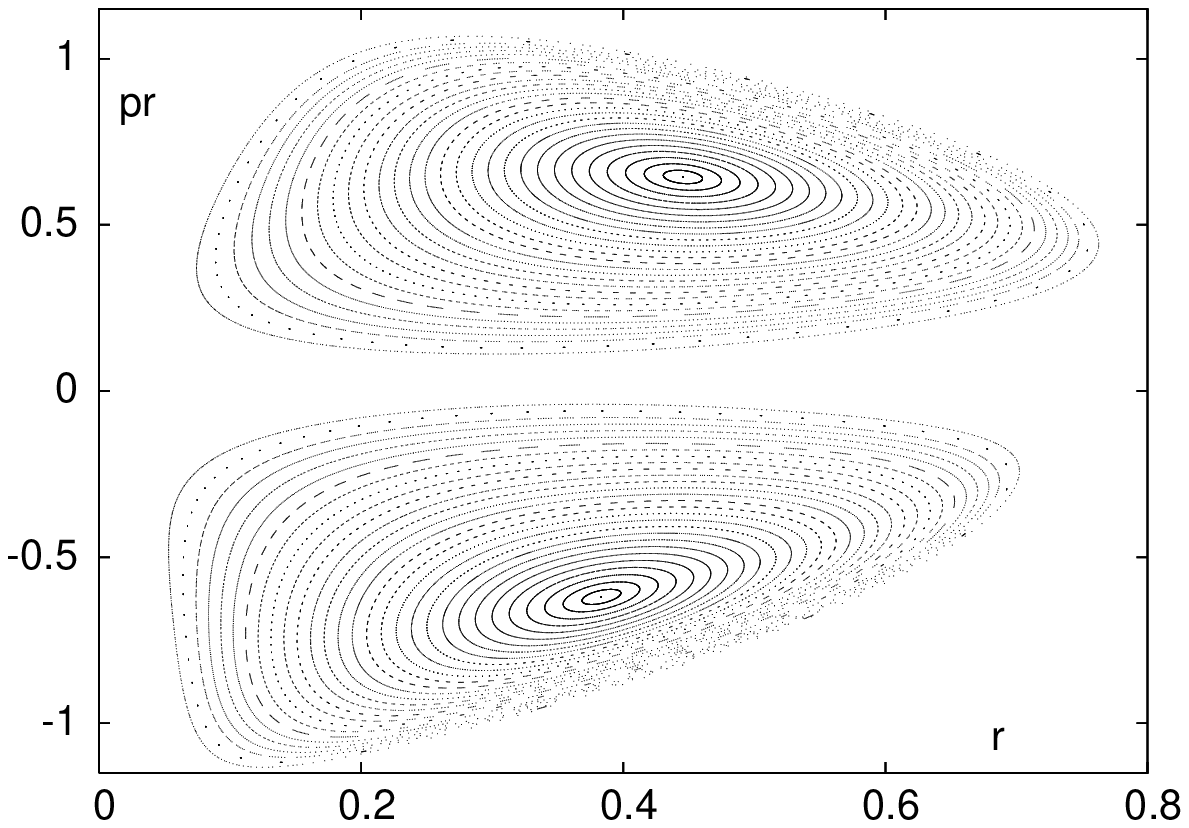,width=6.5cm} \\
\psfrag{pr}{$p_r$} \psfrag{r}{$r$}
\hspace{-5mm}\epsfig{file=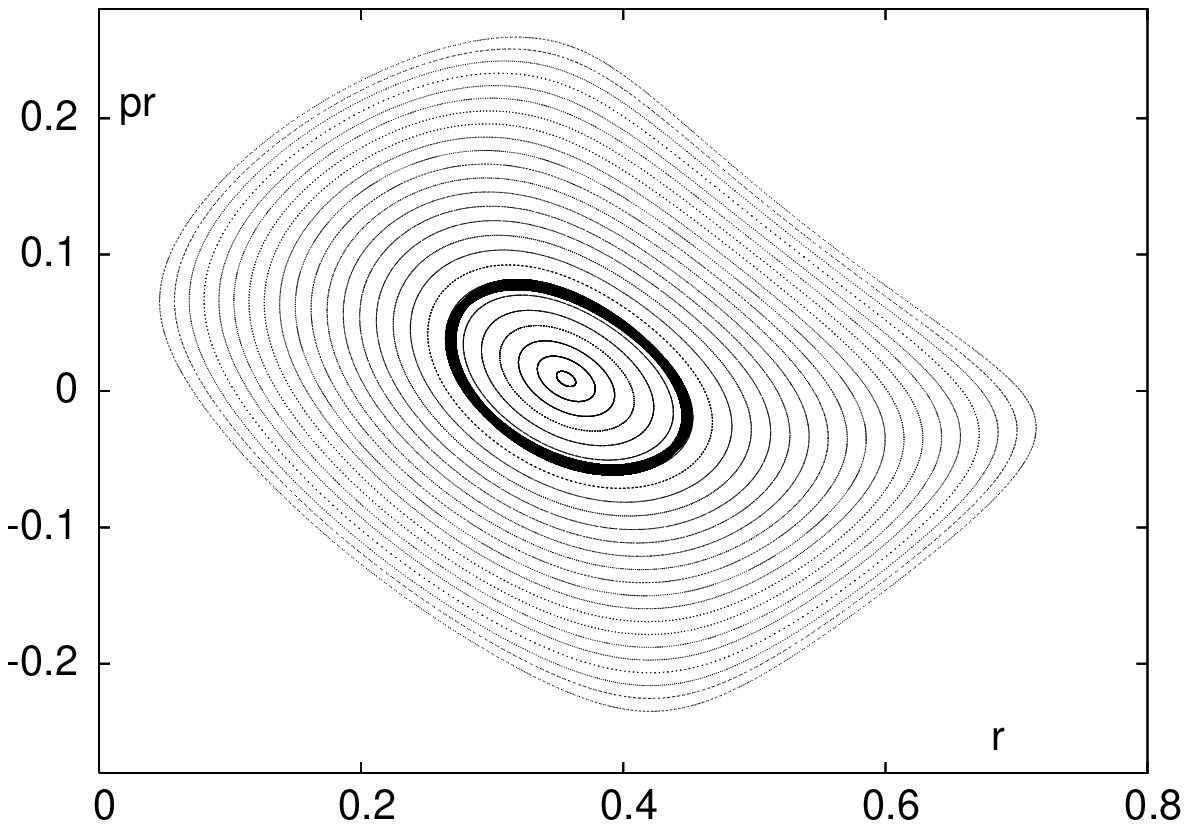,width=6.5cm} \\
\end{tabular}
\caption{Iterates of Poincar\'e maps for $M=350 \ut{g}$ (top row) and $M=180
\ut{g}$ (bottom plot) on the levels of energy of the physical experiments
described in the text. The variables $(r\,;\,p_r)$, as usual in $\ut{m}$ and
$\ut{kg\cdot m\cdot s^{-1}}$, respectively, are shown for $\theta=0$ (mod
$2\pi$) and $\dot{\theta}>0$. On the top left plot, the points leave the section
with $p_r>0$ and $\theta=0$, and those with $p_r<0$ are in
$\theta=-2\pi$. On the right top row, top initial points are in $\theta=0$ with
$p_r<0$, and those showing in $p_r>0$ have $\theta=2\pi$. On the bottom
plot, all intersections occur at $\theta=0$. The thick curves are the
intersections of the theoretical $2\mathrm{D}$ tori corresponding to the physical
experiments with the section $\Sigma$.}
\label{fig:poinc}
\end{figure}

To compare with the experiments, we present some examples for
$M=350 \ut{g}$ and $M=180 \ut{g}$. As levels of energy, one has taken the values
corresponding to the experiments described in \ref{5.2.1} and \ref{5.2.2}, that is,
$(r_0\,;\,\theta _0)$=$(0.649\ut{m}\,;\,53.5^{\circ})$
and $(r_0\,;\,\theta_0)$=$(0.484\ut{m}\,;\,87.0 ^{\circ})$,
respectively, with zero initial velocity. Given $r$, $p_r$ in $\Sigma$ and
$\theta=0$, the value of $p_\theta$ is recovered from the energy level. Figure
\ref{fig:poinc} shows some results. All these massive computations use Taylor
integration methods, in order to ensure a very good conservation of the energy (see,
e.g., \cite{S2007}).

\medskip

The top left plot corresponds to $M=350 \ut{g}$, leaving $\Sigma$ with $p_r>0$.
The points on $\Sigma$ with $p_r<0$ 
correspond to cuts through
$\theta=-2\pi$, in agreement with the description of motion in \ref{5.2.1}. 
To produce the Poincar\'e map, we first computed the periodic orbit
as a fixed point in $\Sigma\cap\left\{p_r>0\right\}$. The approximate values
$(r^*\,;\,p_r^*)$ of the fixed point are
$(0.332814\ut{m}\,;\,0.554330\ut{kg\cdot m\cdot s^{-1}})$. 
This periodic orbit can also be
obtained by starting the motion from rest at $(r_0\,;\,\theta _0) \approx
(0.61316\ut{m}\,;\,64.032^{\circ})$, not too far from the values used in the
experiment. The curve drawn with large dots around the periodic orbit shows the iterates
of $\P_m$ corresponding to the data used in the experiment. It is clear that
the theoretical, non-dissipative, motion seems to be in a $2\mathrm{D}$ torus, but
numerical computations can never exclude the possibility of having a
periodic orbit with very long period or a tiny chaotic zone. The intersection of
this torus with $\theta=-2\pi$ is also shown in the region $p_r<0$. The
asymmetry is ostensible in the plot, and is due to the effect of the pulleys. To produce
the full plot, we have taken initial conditions on $\Sigma$ with $r=r^*$ and
different values of $p_r$ starting at $p_r^*$. From some value of $p_r$ onwards,
iterates escape. We do not exclude the presence of tiny islands outside the last
invariant curve shown.

\medskip

For the sake of completeness, we show in the top right plot the Poincar\'e iterates
leaving $\Sigma$ in the region $\left\{p_r<0\right\}$. 
Points appearing in $\brr{p_r>0}$ are on the sheet $\brr{\theta=2\pi}$.
The fixed point in $\brr{p_r<0}$ is approximately
$(0.383367\ut{m}\,;\,-0.620020\ut{kg\cdot m\cdot s^{-1}})$.

\clearpage

The bottom plot corresponds to $M=180 \ut{g}$. In that case, only intersections
having $\theta=0$ are found. The fixed point is $(r^*\,;\,p_r^*) \approx
(0.356386\ut{m}\,;\,0.008848\ut{kg\cdot m\cdot s^{-1}})$,
which can also be obtained leaving from
$(r_0\,;\,\theta_0) = (0.53313\ut{m}\,;\,78.596^{\circ})$,
again not too far from the values used in the experiment.
As before, the curve drawn with large points around the periodic
orbit would be the one obtained for the physical experiment without dissipation
and, as expected, denotes motion on a $2\mathrm{D}$ torus.

\medskip

A useful tool to understand the dynamics of Area-Preserving Maps and, in
particular, Poincar\'e maps such as the ones displayed, is the \emph{rotation number} $\rho$
for the map restricted to invariant curves. Despite the fact that the rotation number still
exists for periodic orbits of $\P_m$ and for the eventual islands around them
(thereupon being rational), it is not defined, in general, for orbits with
chaotic dynamics. The method used for the computation is topological and based
on the order of the arguments of the iterates with respect to the central fixed
point $(r^*\,;\,p_r^*)$ of $\P_m$. The procedure computes two estimates
$\rho_{\mathrm{inf}}$ and $\rho_{\mathrm{sup}}$ such that
$\rho_{\mathrm{inf}} \leq \rho\leq \rho_{\mathrm{sup}}$. If for some
orbit one has $\rho_{\mathrm{inf}}>\rho_{\mathrm{sup}}$ this proves
$\rho$ not defined. If the number of $\P_m$ iterates is $N$, the
typical errors, when $\rho$ exists, are $O\p{N^{-2}}$ for constant type rotation
numbers. See the Appendix in \cite{SNS2009} for details and a complete analysis
of the error depending on the Diophantine properties of $\rho$.

\medskip

\begin{figure}[!ht]
\begin{tabular}{cc}
\psfrag{rho}{$\rho$} \psfrag{pr}{$p_r$}
\hspace{-4mm}\epsfig{file=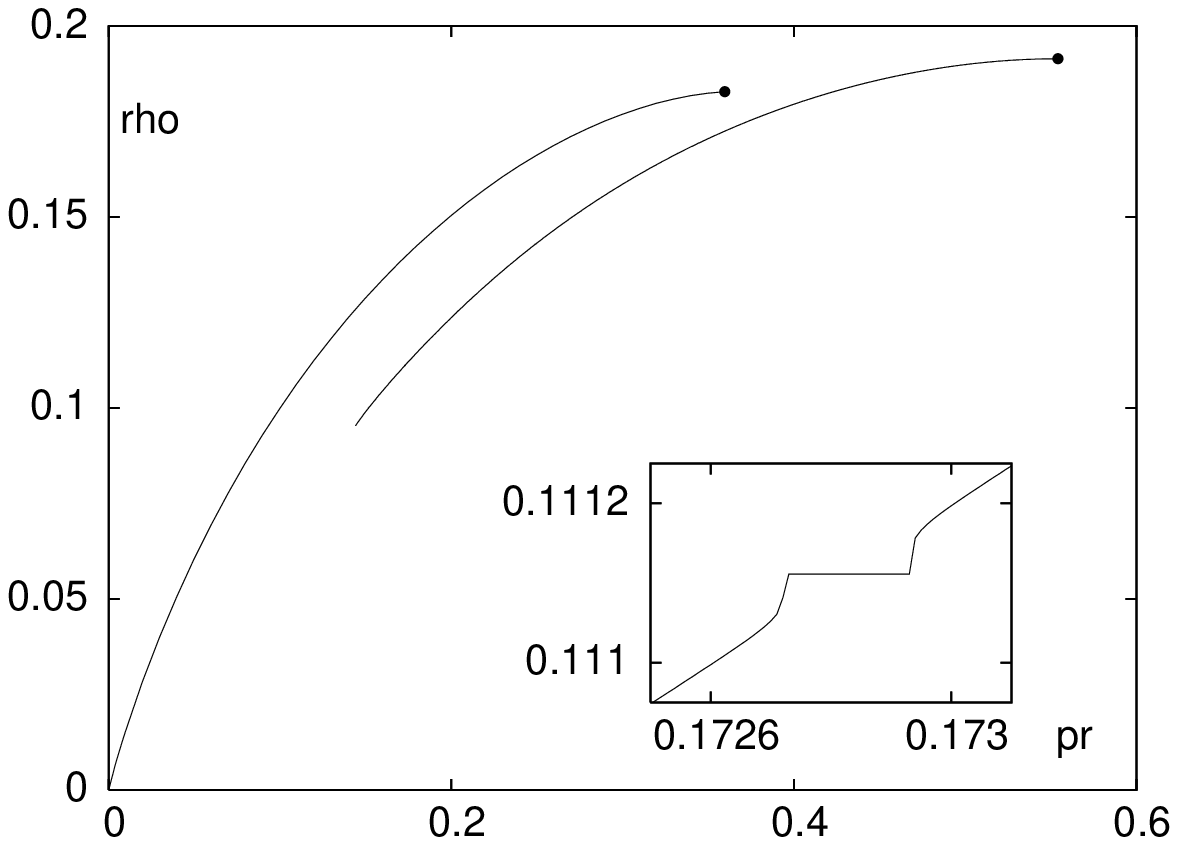,width=6.5cm}&
\psfrag{rho}{$\rho$} \psfrag{pr}{$p_r$}
\hspace{-4mm}\epsfig{file=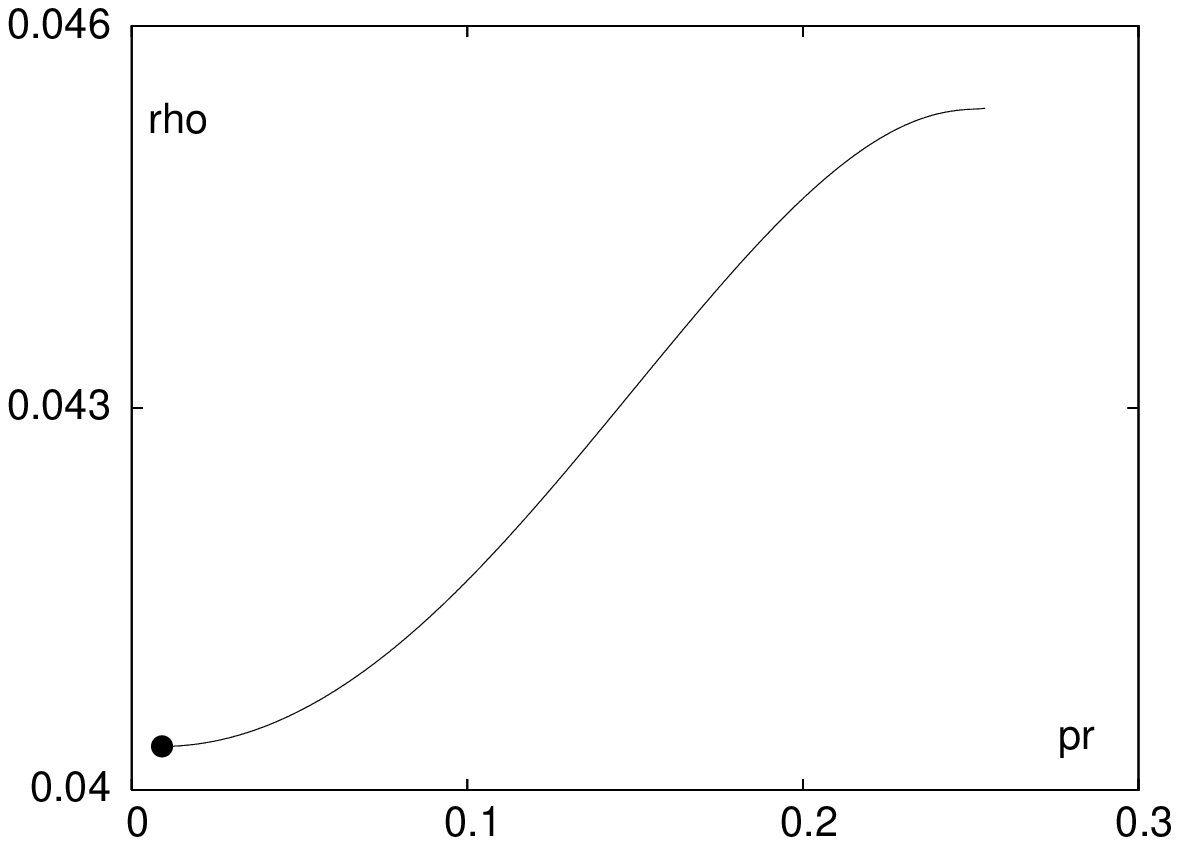,width=6.5cm} \\
\end{tabular}
\caption{Plots of rotation number for different cases. Left plot: the rotation
number $\rho$ as a function of the initial value of $p_r$ with $r=r^*$ for
$M=350\ut{g}$ and also for the pulley-less case with $\mu=3$. A zoom-in
around a resonance is shown in the inset. Right plot: $\rho$ as a function of
$p_r$ for $M=180\ut{g}$. See the text for additional details.}
\label{fig:numrot}
\end{figure}

In Figure \ref{fig:numrot}, we show results corresponding to the Poincar\'e maps
displayed in Figure \ref{fig:poinc} top and bottom. The computations are done
starting at initial points of the form $(r^*\,;\,p_r)$ with $p_r=p_r^*-j\Delta$,
$j=1,2,\ldots$ with a small step $\Delta$. As successive iterates fall in
$\Sigma$ for values of $\theta$ alternating between $0$ and $-2\pi$, the map
${\cal{P}}_m^2$ has been used instead of ${\cal{P}}_m$. On the left plot, we
display $\rho$ as a function of $p_r$ for $M=350 \ut{g}$. This is the curve
which has a large dot near the upper right corner. The point corresponds to
$p_r=p_r^*$ and the limit rotation number. We see a decreasing rotation number
when $p_r$ decreases up to a value $p_r\approx 0.143982\ut{kg\cdot
m\cdot s^{-1}}$. Beyond that point, the $\P_m$ iterates escape. The other curve, also shown here for comparison, corresponds to
the pulley-less case with $\mu=3$ and will be described later.

\medskip

In fact, what seems a nice curve for $M=350 \ut{g}$ should have, generically,
a ``devil's staircase'' structure. That is, there are infinitely many intervals
in which $\rho\in\mathbb{Q}$ and it is constant. They correspond to islands
around elliptic fixed points. Some of these rational values, such as $1/6,\,1/7,\,
1/9,\,3/20,\,\ldots$ (or resonances) have been detected. But they are very
narrow. As an example, the inset in the left plot of Figure \ref{fig:numrot}
shows the behaviour of $\rho$ in an interval of $p_r$ whose width is lower than 
$10^{-3}\ut{kg\cdot m\cdot s^{-1}}$ and $\rho$ around $1/9$, which illustrates
a typical pattern when crossing a resonance through an island.

On the right hand side of Figure \ref{fig:numrot}, we show the results for
$M=180 \ut{g}$. As before, we use initial points $(r^*\,;\,p_r)$ with $p_r$
going away from the fixed point of $\P_m$. $\P_m^2$ has also been used instead
of $\P_m$ because the latter is close to a symmetry with respect to
$(r^*\,;\,p_r^*)$, hence the displayed rotation number is small. It
increases monotonously as we move away from the fixed point, also marked as a
large dot. Now, however, the intervals with $\rho\in\mathbb{Q}$ constant are
extremely narrow. We observed a few resonances, such as $2/45$ and $5/116$,
checking that the width of $p_r$-intervals is below $10^{-6}\ut{kg\cdot m\cdot s^{-1}}$.

\medskip

We want to mention now that the pattern of $\rho$ as a function of $p_r$ is a
clear indication of non-integrability. Indeed, let us first look at the left
plot in Figure \ref{fig:numrot}. We have also shown the results for \emph{SAM}
without pulleys and $\mu=3$, the integrable case. The level of energy is the one
corresponding to $(r_0\,;\,\theta_0)=(0.25\ut{m}\,;\,0.0^{\circ})$, similar to
the kind of reference orbits used in \cite{MS2009}. A periodic orbit is found
near $(r^*\,;\,p_r^*)=(0.125\ut{m},0.369587\ut{kg\cdot m\cdot s^{-1}})$.
Using $r=r^*$ and $p_r$ between $p_r^*$ and
zero, we check that $\rho$ decreases to zero. Under ${\cal{P}}_m^2$ there is a
``separatrix'' bounded by $r=0,\,p_r=0$ and a curve of the form
$r=0.25-\alpha p_r^2$ for a suitable $\alpha$. The system being integrable, the
map $\P_m^2$ has a first integral $I$ and the iterates are on level curves
of $I$. When approaching the separatrix, the dynamics slows down near $r=0$ and
$p_r$, hence $\rho$ is very small.

\medskip

But in the pulley case, the rotation number ceases to exist at a value which is
unrelated to any separatrix. This is against the typical behaviour of integrable
maps.

\medskip

One should expect chaotic dynamic regions in a resonance zone in \emph{SAM} with
pulleys. Beyond the islands around periodic
elliptic points, there are homoclinic tangles associated to the hyperbolic
zones. Attempts to visualize them lead to the conclusion that the size of
the ``chaotic zones'' is, at most, of the order of magnitude of the round-off
errors with standard double precision computation. Hence, the escape remains a
main evidence of non-integrability.

\subsection{Remarks on dissipation}

As we have already noted, experiments show non-negligible dissipative phenomena
which decrease mechanical energy during the motion. Hence, equations of motion
such as \eqref{eq:motion} which do not contain any dissipation term yield by no means a
complete description of \emph{SAM} dynamics.

\medskip

In order to show that the observed convergence of the pendulum towards the
pulley (Figure~\ref{traj}a) is due to dissipation and not to a too short integration time interval, the
corresponding theoretical trajectory (Figure~\ref{trajtheo}a)
has been integrated for much larger intervals. Trajectories are not shown
but we can say that the limited region of space occupied by the pendulum in
Figure~\ref{trajtheo}a is progressively filled as time is running. The
corresponding Tufillaro trajectory integrated for larger time is in addition
symmetrical.

\medskip

Four main different sources of dissipation can be noted: air friction on the
pendulum and the counterweight, thread friction on the pulley grooves due to an
imperfect absence of slippage condition, and dissipation due to the ball bearing
of the pulleys. This last source is implicitly included in the equation of
motion through the measure of $I_p$ since it has been determined using a pulley
(see Section~\ref{resexp}.1).

\medskip

An estimation of air friction can be performed by comparing the weight of the
pendulum, $mg$, to the drag force exerted by air on the pendulum, whose
expression is:
\[
F_d = C_d \frac{\rho_a v^2}{2} S
\]
where $C_d$ is the drag factor, $\rho_a=1.29\ut{kg\cdot m^{-3}}$ the air density
at ambient temperature $T = 298\ut{K}$, $v$ the velocity of the ball, and $S$
the effective surface of the pendulum. For a spherical ball, $C_d \approx 0.4$.
Since $v\approx 1\ut{m\cdot s^{-1}}$, $S=\pi D_b^2/4$ with $D_b=30\ut{mm}$, the
drag-weight ratio is of the order of $2\times 10^{-4}$. One gets the same order
of magnitude for the cylindrical counterweight for which $C_d \approx 1$.
Dissipation will be the detailed topic of a further paper.

\section{Non-integrability of \emph{SAM} with pulleys}\label{integ}

In this Section, a rigorous and original analytical proof of non-integrability of SAM is performed 
in order to complete the above rotation number and Poincar\'e section analysis. 
Needless to say, this non-integrability result is fundamentally different, both in approach and scope, 
from the numerical and graphical inference.

Since \emph{SAM} may be expressed in terms of symplectic formalism, in order to
detect or predict chaotic behaviour, it is pertinent to recall some concepts
related to the integrability of Hamiltonian systems in the sense of
Liouville-Arnold.

\subsection{Algebraic background for studying integrability}

\subsubsection{Linear and Hamiltonian integrability}
\label{linham}

\paragraph{Differential Galois Theory.} See \cite{Morales},
\cite{SingerVanderPut} and \cite{Humphreys} for more information. Given a
\emph{linear} differential system, with coefficients in a differential field
$\left(K,\partial\right)$ whose field of constants $\mathcal{C}$ is
algebraically closed (e.g. $\left[\mathbb{C}\!\left(t\right),\,
\dd/\dd t\right])$,
\begin{equation} \label{HODE}
\partial\bm{y}=A\left( t\right) \bm{y},
\end{equation}
an algebraic group $G$ exists, called the \emph{differential Galois group} of
\eqref{HODE}, acting over the $\mathbb{C}$-vector space $\left\langle
\bm{\psi}_1,\dots,\bm{\psi}_n \right\rangle$ of solutions of \eqref{HODE} as a
linear transformation group over $\mathbb{C}$. Furthermore, $G$ contains the
monodromy group of \eqref{HODE}. The Galoisian formalism proves useful here due
to the following: \emph{\eqref{HODE} is integrable (i.e. its general solution
can be written as a finite sequence of quadratures, exponentials, and algebraic
functions) if, and only if, the identity component $G^0$ of the differential
Galois group $G$ of (\ref{HODE}) is solvable.}

\medskip

Everything said in the previous paragraph may be obtained,
\emph{mutatis mutandis}, for linear homogeneous differential equations
\[ a_n\p{t}\frac{\dd ^n}{\dd t^n}y+a_{n-1}\p{t}\frac{\dd^{n-1}}{\dd t^{n-1}}y+
\dots + a_1 \p{t}\frac{\dd}{\dd t}y + a_0\p{t} y = 0. \]
In Subsection \ref{boucherweilsection}, we will denote Galois groups in this
$\mathrm{Gal}\p{L}$ accordingly, $L\in\nc\p{t}\left[\dd/\dd t\right]$ being
the corresponding differential operator. See \cite{Magid} for more details.

\paragraph{Liouville-Arnold integrability.} On the other hand, we call a
\emph{Hamiltonian} system $\bm{\dot{q}}=\partial H/\partial\bm{p},\;\bm{\dot{p}}
=-\partial H/\partial\bm{q}$, whether or not linear, \emph{meromorphically
integrable (in the sense of Liouville-Arnold)}, if it has as many independent
meromorphic first integrals in pairwise involution as degrees of freedom.
Same applies in the above definition, \emph{mutatis mutandis}, substituting
\emph{algebraic}, \emph{rational} or any other function class in for
\emph{meromorphic}. For the sake of simplicity, conjugate canonical variables
will be henceforth assembled in a single
vector $\bm{z}=\p{\bm{q},\bm{p}}$ and the Hamiltonian system will be written in
compact form $\bm{\dot{z}}=X_H\p{\bm{z}}$. 
\emph{Everything is considered in the complex analytical setting from this point on.}

\subsubsection{Morales-Ramis-Ziglin Theory}\label{MRZtheory}

For each integral curve $\Gamma=\left\{\tilde{\bm{\phi}}(t):t\in I\right\}$
of a given autonomous dynamic system in dimension $m$
\begin{equation} \label{DS}
\bm{\dot{z}} = X\left( \bm{z}\right) ,
\end{equation}
the \emph{variational equations} of order $k$ for \eqref{DS} along $\Gamma$,
$\mathrm{VE}^k_{\Gamma}$, are satisfied by $\Xi_k:=
\partial^k\tilde{\bm{\phi}}/\partial\bm{z}^k$ -- see, e.g., \cite{MRS2007} for explicit expressions of $\mathrm{VE}^k_{\Gamma}$
for general $k$ in terms of vectors $\bm{k}=(k_1,\ldots,k_m)\in\nz_+^m$ such that $k=k_1+\ldots +k_m$. 
Note that $\partial^k \tilde{\bm{\phi}} / \partial \bm{z}^k$ is an
abridged notation for the $(k+1)$-dimensional matrix of all partial
derivatives of $\tilde{\bm{\phi}}$. In other words, it's a vector for $k=0$, a
matrix for $k=1$, etc.

In particular, for $k=1$ and denoting the matrix of the first-order variational equations simply by $\Xi$, we obtain
\begin{equation}
\label{VE}
\tag{$\mathrm{VE}_{\Gamma}$}\dot{\Xi}=X'\left(\tilde{\bm{\phi}}\right)
\Xi.
\end{equation}
We thus have:

\begin{itemize}
\item a \emph{(generally nonlinear)} system
\eqref{DS} and
\item a \emph{linear system} \eqref{VE} linked to (\ref{DS}).
\end{itemize}
The hallmark theorem in this approach connects the two notions of solvability
listed in \ref{linham}, namely as applied to a Hamiltonian $X_H$ and its
variational equations $\mathrm{VE}_{\Gamma}$, along an integral curve $\Gamma$
of $X_H$. The whole theory is actually the \textit{ad hoc} implementation of the
following heuristic principle: if a Hamiltonian is integrable, then its
variational equations must also be integrable.

\medskip

We assume $\Gamma$, a Riemann surface, may be locally parameterized in a disc
$I\subset\mathbb{C}$ of the complex plane; we may now complete $\Gamma$ to a new
Riemann surface $\overline{\Gamma}$, as detailed in \cite[\S 2.1]{MoralesRamis}
(see also \cite[\S 2.3]{Morales}), by adding equilibrium points, singularities
of the vector field and possibly $t=\infty$.
\begin{thm}[J Morales-Ruiz \& J-P Ramis, 2001] \label{moralesramis}
Let $H$ be an $n$-degree-of-freedom Hamiltonian having $n$ independent first
integrals in pairwise involution, defined on a neighbourhood of an integral
curve $\overline{\Gamma}$. Then, the identity component $\mathrm{Gal}
\left(\mathrm{VE}_{\overline{\Gamma}}\right)^0$ is an abelian group (i.e.
$\mathrm{Gal}\left(\mathrm{VE}_{\overline{\Gamma}}\right)$ is
\emph{virtually abelian}).
\end{thm}
See \cite[Corollary 8]{MoralesRamis} or \cite[Theorem 4.1]{Morales} for a
precise statement and a proof.

\subsubsection{Differential operators. A primer in the Boucher-Weil Theorem}
\label{boucherweilsection}

\paragraph{Linear differential equations.}
See \cite[\S 2]{SingerUlmer} for more details.
Let
\[
L=a_n\p{\frac{\dd}{\dd t}}^n+a_{n-1}\p{\frac{\dd}{\dd t}}^{n-1}+\dots+a_0\,;
\quad a_n,\ldots,a_0\in\nc\p{t}
\]
be a differential operator with coefficients in the field of formal Laurent
series. If $L\p{y}=0$ has a solution of the form $y=t^{\rho}\sum_{k\ge 0}c_kt^k$
and $c_0\ne 0$, the formal substitution of $y$ into the differential equation
yields the vanishing of all powers of $t$, the smallest one among them -- we
call the equation $P\p{\rho}=0$ derived from the latter vanishing the
\emph{indicial equation} (at $0$), the roots of which are usually called
\emph{exponents} of $L\p{y}$ (at $0$). In particular we
can also define the indicial equation at infinity by means of the
transformation $f=1/t$ and expansion around $f=0$.

\medskip

It is a known fact (\cite[Lemma 2.1]{SingerUlmer}) that the degree of $P\p{\rho}
=0$ is at most $n$. A singular point $c$ of $L$ is called \emph{regular
singular} if $\deg P\p{\rho}=n$. A linear differential equation $L=0$ with only
regular singular points (including $\infty$) is called \emph{Fuchsian}.

\medskip

We call $L$ \emph{reducible} if it factors in a product of operators of
smaller positive order. An operator $L$ admits a first order factor
$\dd/\dd t-f$, $f\in\nc\p{t}$ if and only if $L\p{y}=0$ admits a solution $y$
such that $\dot{y}=fy$; in particular:
\begin{lem}\label{redexp}
If $L$ is of order $2$: $L$ is reducible if, and only if, it has an
\emph{exponential} solution, i.e. a solution whose logarithmic derivative is
rational. $\square$
\end{lem}

\begin{lem}[{\cite[\S 3.1.2]{SingerUlmer}}]\label{singul} If $L$ is Fuchsian,
every exponential solution must be of the form $\tilde{y}=\prod_{i=1}^s(t-t_i)
^{e_i}P\left(t\right), $ where $P\in\mathcal{C}\left[X\right]$ and $t_1,\dots,
t_s$ are finite singularities of $L$ with exponents $e_1,\dots,e_s$, whether or
not integers.
\end{lem}

\paragraph{Normal variational equations}
Let
$$
J=\left(\begin{array}{cc}0&\mathrm{Id}_n\\-\mathrm{Id}_n&0\end{array}\right)
$$
be the canonical symplectic matrix. Given a Hamiltonian system
$\bm{\dot{z}}=X_H\p{\bm{z}}=J\nabla H\p{\bm{z}}$ expressed in Darboux canonical
coordinates $\bm{z}=\p{\bm{q},\bm{p}}$, system \eqref{VE} reads $\dot{\Xi}=
JH''\p{\tilde{\bm{z}}}\Xi$ along $\Gamma=\brr{\tilde{\bm{z}}\p{t}}$.

\paragraph{Gauge transformations.} $A\mapsto P\left[A\right]:= P^{-1}\p{AP-\dot
{P}}$, $P$ being a given symplectic matrix, may be used to reduce \eqref{VE} by
selectively vanishing a number of entries in $J H''\p{\tilde{\bm{z}}}$ (see 
\cite[\S 5.2]{BoucherPhD}, \cite[\S 4.1]{Morales}, \cite{AparicioWeil}). A
typical first choice consists of symplectic matrices of the form
\begin{equation}\label{gauge}
P\p{t} = \p{
\begin{array}{cccc}
\displaystyle{\frac{\dd}{\dd t}\tilde{\bm{z}}} & \bm{c}_2 & \bm{c}_3 & \bm{c}_4
\end{array} },
\end{equation}
in order to induce a row and a column of zeroes in the variational matrix. When
such is the case, \eqref{VE} acquires a ``decoupled'' appearance and a system of
order $n-2$ may be extracted therefrom. Such a system is usually called a
\emph{normal variational system} along $\Gamma$. Let us denote it by $\mathrm{NVE}_{\Gamma}$.

\medskip

As is always the case with all differential systems, a \emph{cyclic vector}
(\cite{BoucherPhD}, \cite{Singer}) may be used to obtain a linear differential
equation $L\p{y} = 0$ of order $n-2$ equivalent to $\mathrm{NVE}_{\Gamma}$.

\medskip

The result central to this paragraph, and a particularly useful consequence of
Theorem \ref{moralesramis}, is the following (see also \cite[Proposition 9 \&
Theorem 8 (\S 5.3)]{BoucherPhD}):
\begin{thm}[\textbf{D Boucher \& J-A Weil, \cite[Criterion 1]{BoucherWeil}}]
\label{boucherweil}
Let $X_H$ be a Hamiltonian system and $L$ its normal variational operator along
a given particular solution. If $L$ is irreducible and displays logarithms in a
formal solution, then $\mathrm{Gal}\left(L\right)^0$ is not abelian, i.e. $X_H$
is not integrable. $\square$
\end{thm}

Typically, the most difficult part in trying to apply Theorem \ref{boucherweil}
is to check the irreducibility condition.

\subsection{Statement of the main result}

We recall $M_t=M+m+2I_p/R^2$ and
\[
\mathcal{H} = \frac12 \left[\frac{p_1^2}{M_t} + \frac{(p_2+Rp_1)^2}{mq_1^2}
\right] + gq_1(M-m\cos q_2)-gR(Mq_2-m\sin q_2),
\]
where $q_1=r$, $q_2=\theta$, $p_1=p_r$, and $p_2=p_{\theta}$.
The main result in this Section is the following:
\begin{thm}[\textbf{Non-integrability of \emph{SAM} with massive pulleys}]
\label{main}
For every physically consistent value of the parameters, regardless of $I_p$ and
$R$, $X_{\mathcal{H}}$ is mero\-morphically non-integrable.
\end{thm}

This is a complement to what has already been proved for \emph{SAM}
\emph{without pulleys}, i.e. the limit case $I_p=0$, $R=0$ and $M_t=M+m$:
\begin{equation}\label{Hw}
\mathcal{H}_w = \frac12\left(\frac{p_1^2}{M_t}+\frac{p_2^2}{mq_1^2}\right) +
gq_1\left(M-m\cos q_2\right);
\end{equation}
in that case, the following held:
\begin{thm}
[\textbf{\emph{SAM} without massive pulleys}]\label{previous}
\begin{itemize}
\item[\phantom{1}]
\item[\textbf{1}.] \emph{(\cite[Theorem 1]{Casasayas-1990}, \cite[Equation (16)]{Yehia}, \cite{
AlmeidaMoreiraSantos}; see also Remark \ref{finalrems}(2))} If $M>m$ and $$\mu = \frac Mm \ne \mu_p:= \frac{p(p+1)}{p(p+1)-4}$$ for
every $p\in\mathbb{Z},p\geq 2$, then Hamiltonian $X_{\mathcal{H}_w}$ is
non-integrable. In particular, if $\mu\in\left(3/2,\,3\right)\cup(3,\,\infty)$,
it is non-integrable.
\item[\textbf{2}.] \emph{(\cite[Equation (16)]{Tufillaro-1986})}
For $p=2$, $\mu=\mu_2=3$, $X_{\mathcal{H}_w}$ is integrable and has the
following first integral:
\[
I = q_1^2\dot{q_2}\left(\dot{q_1}c-\frac{q_1\dot{q_2}}{2}s\right) + gq_1^2sc^2 =
 gq_1^2c^2s + p_2\frac{p_1q_1c - 2p_2s}{4m^2q_1},
\]
where $c=\cos{\p{q_2/2}}, s=\sin{\p{q_2/2}}$.
\item[\textbf{3}.] \emph{(\cite[Theorem 4]{MS2009})} The degenerate cases
$\mu_p$, $p\geq 2$ referred to in item \textbf{1} are non-integrable.  $\square$
\end{itemize}
\end{thm}

The last case is significantly more difficult; it relies on the
higher order variational equations \cite{MRS2007} and uses
techniques introduced in \cite{MS2008}.

\subsection{Proof of Theorem \ref{main}}

We have two particular solutions for \emph{SAM} with massive
pulleys:
{\small \begin{eqnarray*}
\mbox{\boldmath${z}$\unboldmath}_{0,R}(t)& = &\frac{g(M-m)}2\left({\frac{t(1-t)}
{M_t}\,,\,\,0\,,\,\,1-2t\,,\,\,R(2t-1)}\right),\\
\mbox{\boldmath${z}$\unboldmath}_{\pi,R}(t) & = & \frac{g(M+m)}2\left(
{\frac{t(1-t)}{M_t}\,,\,\frac{2\pi}{g\p{M+m}}\,,\,1-2t\,,\,R(2t-1)}\right).
\end{eqnarray*}}
System \eqref{VE} around $\Gamma=\brr{\bm{z}_{\pi,R}(t)}$ takes the form
$\dot{\Xi}=A\Xi$ with
\[ A=\left(\begin{array}{cccc}0&0&1/M_t+{R}^{2}a_1&Ra_1\\
\noalign{\medskip}0&0&Ra_1&a_1 \\\noalign{\medskip}0&0&0&0\\\noalign{\medskip}0
&a_2&0&0\end {array} \right)\,,\]
where
$$
a_1=\frac{4M_t^2}{mg^2(M+m)^2t^2(t-1)^2}
\qquad\hbox{and}\qquad
a_2=-\frac{mg^2(M+m)t(t-1)}{2M_t}.
$$
Notably, \eqref{VE} decouples
without the need for an additional gauge transformation such as \eqref{gauge} as
is usual and customary (\cite{BoucherPhD}, \cite{BoucherWeil}, \cite{Morales}),
and as would be the case if $\Gamma=\brr{\bm{z}_{\pi,0}(t)}$. 
See Remark \ref{finalrems} \textbf{1} below. System
$\mathrm{NVE}_{\Gamma}$ takes the form
$\dot{\Phi}=B\Phi$ with
\[
B = \left(
\begin{array}{cc}
0&a_1\\\noalign{\medskip}
a_2&0
\end {array}
\right).
\]

A necessary condition of integrability is the virtual abelianity (see Theorem
\ref{moralesramis}) of $\mathrm{Gal}\p{\mathrm{NVE}_{\Gamma}}$, $\mathrm{NVE}_
{\Gamma}$, by means of a \emph{cyclic vector} $\bm{c}$ and the subsequent
gauge transformation given by $\Phi=Q^{-1}\tilde{\Phi}$, where
$Q=\p{\begin{array} {cc}\bm{c}&\bm{\dot{c}}+B^T\bm{c}\end{array}}^T$
(\cite[\S B.4]{BoucherPhD}), will take the form
\[
\frac{\dd}{\dd t}\tilde{\Phi} =
\left(
\begin{array}{cc}
0&1\\[5pt]
-\displaystyle{\frac{2M_t}{(M+m)(t-1)t}}& -\displaystyle{\frac{2(2t-1)}{(t-1)t}}
\end{array}
\right)
\tilde{\Phi},
\]
in presence of a constant cyclic vector $\bm{c}=\p{a,0}$. The above system is
obviously equivalent to the following \emph{hypergeometric} (\cite[\S 15.5]
{Abramowitz}), hence Fuchsian differential operator:
\[L= \p{\frac{\dd}{\dd t}}^2+\frac{2(2t-1)}{t(t-1)}{\frac{\dd}{\dd t}}+
\frac{2 M_t}{(M+m)t(t-1)}.\]

Let us now check the virtual non-commutativity of $\mathrm{Gal}\p{L}$. $L$ has
three singularities: $0$, $1$, $\infty$. At $t=0$ or $t=1$, we have local
exponents $-1$ and $0$. The formal solution at $t=0$ has logarithms except in
two cases:
$m=0$, $M_t=M+m$, both discarded in our case since they would correspond to no
small mass and no pulley, respectively. Indeed, a particular solution is
\[
\tilde{y}_1\p{t} = {}_2F_1\p{\frac{3}{2}-\frac{[9(M+m)-8M_t]^{1/2}}
{2(M+m)^{1/2}}\,,\,
\frac{3}{2}+\frac{[9(M+m)-8M_t]^{1/2}}{2(M+m)^{1/2}}\,;\,2\,;\,t},
\]
where
$$
{}_2F_1\p{a,b\,;\,c\,;\,t}=\sum_{k=0}^{\infty}\frac{(a)_k(b)_k}{(c)_k}\,
\frac{t^k}{k!}
$$
is the \emph{Gauss hypergeometric function} (\cite[\S 15.1]{Abramowitz}),
$(a)_k=a(a+1)\cdots(a+k-1)$ being the \emph{Pochhammer symbol}. This first
solution has a local expansion around $t=0$ of the form
$$
\tilde{y}_1 = 1+\frac {M_t}{M+m}t+\frac{M_t\left[2\p{M+m}+M_t\right]}{3\p{M+m}^2}t^2 +
O\p{t^3}.
$$
We may then obtain an expansion for a second formal solution
$$
\tilde{y}_2 = -\frac{1}{t} - \frac { M_t }{M+m} +
\left\{3+\frac{2M_t\left[M_t-3(M+m)\right]}{\p{M+m}^2}\right\}t +
 O\p{t^2}+2\frac {M+m-M_t }{M+m}  \tilde{y}_1 \ln t.
$$
Keeping the Boucher-Weil Theorem \ref{boucherweil} in mind, and in presence of
the logarithm in $\tilde{y}_2$, there is obstruction to integrability if $L$ is
irreducible.

\medskip

Let us assume it is reducible. In virtue of Lemma \ref{redexp}, $L$ admits an
exponential solution. We recall that $L$ is Fuchsian. The expansion of an
exponential solution around $t=0$ does not contain logarithms, although we have
shown there are formal solutions with logarithms around the singularities -- as
has been seen explicitly for $t=0$. Thus, those without logarithms correspond to
the maximal exponents (\cite{WhittakerWatson}), hence the admissible exponents
at the finite singularities are all $0$. Hence, in virtue of Lemma \ref{singul},
the only possible form for an exponential solution $\tilde{y}$ would be that of
a polynomial solution; let $N$ be its degree. Expanding $\tilde{y}$ in
increasing powers of $t^{-1}$, $-N$ is the exponent of the leading term, hence
an exponent at infinity: $\sum_{k=0}^Na_kt^k=(1/t)^{-N}\sum_{k=0}^N
a_{N-k}(1/t)^k$.

\medskip

Now, the exponents at infinity are the two roots of the indicial
equation $\rho ^2-3\rho+2M_t/(M+m)$. Since $-N$ is such a
root, this means $M_t=-N(N+3)(M+m)/2$ with $N$ positive.
Therefore $M_t$ would be negative, which is physically irrelevant.

\medskip

Hence follows that the Swinging Atwood Machine  system with massive pulleys is
{\em always} non-integrable with meromorphic first integrals. $\square$

\begin{rems} \label{finalrems}
\begin{enumerate}
\item[\phantom{0}]
\item[\textbf{1.}] Intriguingly, the solution used for our proof was
$\bm{z}_{\pi,R}$, which, at least for the
pulley-less case $R\to 0$, $M_t\to M+m$, and although
mathematically plausible, has no actual physical significance. The
solution which would be physically possible for all values of $R$,
$\bm{z}_{0,R}$, posed further problems with regards to system
\eqref{VE} and was finally discarded in our proof. It is worth
mentioning, however, that in the case without pulleys,
\cite{MS2009} used precisely the corresponding form of the latter
solution $\bm{z}_{0,0}$.

\item[\textbf{2.}] The same proof given for Theorem \ref{main} may be obtained,
analogously, for \emph{SAM} without pulleys $\mathcal{H}_w$, $\mu>1$, and the
proof yields precisely item 1 in Theorem \ref{previous}. Indeed, by using the
``classical'' solution $\bm{z}_{0,0}$ (which corresponds to the original Atwood
machine) and an adequate gauge transformation, we obtain the normal variational
equation:
$$
-\frac{2M_t}{(M-m)t(t-1)}\,y(t) - \frac {(2t-1)}{(t-1)t}\,
\frac{\dd}{\dd t}y(t)+ \frac{\dd^2}{\dd t^2}y(t) = 0.
$$
This is a Gauss hypergeometric equation. Local solutions are
\[
t^2-\frac 23\left(\frac{M_t}{M -m}+1\right)\,t^3 + O\left( {t}^4\right),
\]
and
\[
-\frac{(M-m)\left[2t M_t+(M-m)\right]}{2M_t\left[M_t+(M-m)\right]} +
O\left(t^2\right) +
\left[t^2-\frac 23\left(\frac{M_t}{M-m}+1\right) t^3 +
O\left(t^4\right) \right]\,\ln t.
\]
The degenerate cases
$M_{t}=0$ and $M_{t}+M-m=0$ are of course not physically
acceptable. The exponents at zero are $0$ and $2$, the exponents at
$t=1$ are also $0$ and $2$, and the exponents at infinity are the
roots of the polynomial $P_{\infty}(X)$ where $P_{\infty}(X)=
\left( M-m \right) {X}^{2}+3\left( M-m \right) X-2\,M_t$. Same as
in the proof of Theorem \ref{main}, reducibility would occur only
for a polynomial solution, i.e. in presence of an integer $N$ such
that $P_{\infty}(N)=0$, implying
$$
M_{t}=\frac{1}2\left( N+4\right) \left( N+1 \right)  \left( M-m \right).
$$
Setting $M_t =M+m$ (that is, the pulley-less case) we would have an equation for
$\mu$ whose solution would be
$$
\mu = \frac{N^2+5N+6}{N^2+5N+2} = \frac{(N+3)(N+2)}{(N+3)(N+2)-4},
$$
obviously equivalent to the condition in item 1 in Theorem \ref{previous} for
$N=p-2$.

\item[\textbf{3.}] For $\mu=1$, however, the closest thing to such a proof as
that sketched in item \textbf{2} is discarding the existence of first integrals
with a meromorphic growth at infinity, e.g. rational ones.  This is due to the
fact that the normal variational equation around particular solution
$ \tilde{\bm{z}} \p{t} = \p{-gt/2\,;\,0\,;\,-gm\,;\, 0} $ is a \emph{Hamburger equation},
\[
-\frac 2t\,y(t)-\frac 1t\,\frac{\dd}{\dd t}y(t) + \frac{\dd^2}{\dd t^2}y(t) = 0,
\]
i.e. an equation with exactly two singularities: a regular one at zero and an
irregular one at infinity \cite[\S 17.6]{Ince}; the solutions are
almost Bessel functions: the general
solution is:
\[
y(t) = C_1t\,I_2\left(2\sqrt{2t}\right) + C_2t\,K_2\left(2\sqrt{2t}\right),
\qquad
C_1,C_2\in \mathbb{C},
\]
where, $n\in \mathbb{Z}$ given, $I_n\left( {z}\right) $ and $K_n\left( {z}
\right) $ are the \emph{modified Bessel functions of the first and second
kind}, respectively (\cite[p. 416]{Arfken}, \cite[p. 376]{Abramowitz},
\cite[p. 185]{Watson}):
\[
I_n(z) = \frac 1{2\pi \mathrm{i}}\int_{\mathbb{S}^1}\frac{
\exp\left[ (z/2)\,\left({t+1/t}\right)\right] }{t^{n+1}}\,\dd t,
\qquad
K_n(z) = \frac \pi 2\,\frac{I_{-n}\left({z}\right) - I_n\left({z}\right)}
{\sin (n\pi)},
\]
both having a branch cut discontinuity in the complex $z$ plane running from
$-\infty $ to $0$ (although $I_n$ is regular at $0$, whereas $K_n$ has a
logarithmic divergence at $0$). This assures the presence of a non-trivial
Stokes multiplier at infinity for the variational equation when $\mu=1$. This
implies the following important conclusion: \emph{the system is not integrable
with first integrals which are rational functions of $r$, $\theta$,
$\cos\theta$, $\sin\theta$, $p_r$, $p_{\theta}$, where $\theta$ belongs to a
neighbourhood of $0$}.

Indeed, any first integral is a function of $\p{\bm{q},\bm{p}}$, hence must be
defined on the phase manifold $M:=\mathbb{C}^4$. We can partially compactify $M$
into $\tilde M:=\mathbb{P}^2_{\mathbb{C}}\times\mathbb{C}^2$, where
$\mathbb{P}^2_{\mathbb{C}}$ stands for the compactification, by means of the
inverse stereographic projection $P^{-1}$, of the domain $\mathbb{C}^2$ for
$\p{q_1\,;\,p_1}=\p{r\,;\,p_r}$, whereas the second factor $\mathbb{C}^2$ is the
$\p{\theta\,;\,p_{\theta}}$-plane. Using $\left.P^{-1}\right|_{\overline{\Gamma}}$
in order to compactify the particular solution $\overline{\Gamma}=
\left\{\tilde{\bm{z}}\p{t}\right\}$, we obtain a Riemann sphere $\tilde\Gamma$
in $\mathbb{P}^2_{\mathbb{C}}$, whose immersion in $\mathbb{P}^2_{\mathbb{C}}
\times\mathbb{C}^2$ is contained in $\left\{\theta =0\right\}$; therefore
$\cos \theta$, $\sin \theta$ are holomorphic, hence meromorphic on a
neighbourhood of $\tilde\Gamma$ in $\tilde M$. Theorem \ref{moralesramis}
implies the absence of a complete set of first integrals which are meromorphic
on a neighbourhood of $\tilde\Gamma$ in the partial compactification
$\tilde{M}$. Since any rational function of $r,\theta,p_r, p_{\theta},
\cos \theta,\sin \theta$ must be meromorphic in a neighbourhood of
$\tilde\Gamma$ in $\tilde M$, we obtain the last claim in the previous
paragraph.
\end{enumerate}
\end{rems}

\section{Conclusion and perspectives} \label{conc}

In this paper, experimental and theoretical results concerning the \emph{Swinging
Atwood Machine} are presented. Equations of motion with two
pulleys are found generalizing those of previous studies. An experimental device for \emph{SAM} is
constructed and described in detail. Experiments are
conducted and the trajectories retrieved from computer video analysis closely match
those arising from numerical solution of the equations of motion.
Such comparisons seem to show that the motion is influenced by the
non-negligible dimension and the rotation of the pulleys; in particular, a
non-zero pulley radius leads to asymmetric pendulum trajectories. We conclude
that pulleys cannot be ignored in \emph{SAM} dynamics.

\medskip

Finally, after giving some numerical evidence of the lack of integrability of
\emph{SAM} in the absence of dissipation, a complete proof of this fact is given
using differential Galois theory and the necessary conditions following from
Morales-Ramis theory.

\medskip

Several perspectives of this work can be considered. First of all, other
experiments are currently conducted with $\mu_{\mathrm{exp}}=2.03$ and qualitative
preliminary results (not shown in this paper) indicate that the dynamics of
\emph{SAM} seems to be irregular -- chaotic behaviour is expected. Detailed
research will be performed in a future work. Secondly, the influence of
dissipative phenomena on \emph{SAM} dynamics should be studied; a possible procedure in such direction
is increasing air friction by coating the pendulum with different
materials, judiciously chosen so as to induce changes in drag force. Another
method could consist in placing the counterweight in media more viscous than
air, such as water or glycerin, and forcing it to evolve therein. Perhaps an
ultra fast camera of about 1000 images per second could be necessary to pick up
much more points and achieve a better resolution in pendulum trajectories.

\medskip

Concerning the integrability of \emph{SAM} without pulleys, that
is Hamiltonian ${\cal{H}}_w$ in \eqref{Hw} obtained from \eqref{H}
by skipping the contributions of $R$ and $I_p$, the results
summarized in Theorem \ref{previous}, along with the result shown
in the previous section for $\mu=1$, close the problem: the
pulley-less case is non-integrable for all values of $\mu\neq 3$.

\ack{The authors would like to thank the ``agr\'egation'' physics laboratory of
the ``Universit\'e Paul Sabatier (Toulouse, France)'' for technical support. 
Research by C. Sim\'o and S. Simon has been supported by grant MTM$2006-05849$/Consolider (Spain).
S. Simon is also grateful to the \emph{D\'epartement Maths Informatique} (\emph{Institut de Recherche XLIM-UMR CNRS $6172$},
\emph{Universit\'e de Limoges}) for a post-doc stay during which much of his contribution to this paper was completed.}

\bibliographystyle{plain}
\bibliography{SAMpulley}

\end{document}

%% file: Figure7a.tex
\begingroup
  \makeatletter
  \providecommand\color[2][]{%
    \GenericError{(gnuplot) \space\space\space\@spaces}{%
      Package color not loaded in conjunction with
      terminal option `colourtext'%
    }{See the gnuplot documentation for explanation.%
    }{Either use 'blacktext' in gnuplot or load the package
      color.sty in LaTeX.}%
    \renewcommand\color[2][]{}%
  }%
  \providecommand\includegraphics[2][]{%
    \GenericError{(gnuplot) \space\space\space\@spaces}{%
      Package graphicx or graphics not loaded%
    }{See the gnuplot documentation for explanation.%
    }{The gnuplot epslatex terminal needs graphicx.sty or graphics.sty.}%
    \renewcommand\includegraphics[2][]{}%
  }%
  \providecommand\rotatebox[2]{#2}%
  \@ifundefined{ifGPcolor}{%
    \newif\ifGPcolor
    \GPcolorfalse
  }{}%
  \@ifundefined{ifGPblacktext}{%
    \newif\ifGPblacktext
    \GPblacktexttrue
  }{}%
  \let\gplgaddtomacro\g@addto@macro
  \gdef\gplbacktext{}%
  \gdef\gplfronttext{}%
  \makeatother
  \ifGPblacktext
    \def\colorrgb#1{}%
    \def\colorgray#1{}%
  \else
    \ifGPcolor
      \def\colorrgb#1{\color[rgb]{#1}}%
      \def\colorgray#1{\color[gray]{#1}}%
      \expandafter\def\csname LTw\endcsname{\color{white}}%
      \expandafter\def\csname LTb\endcsname{\color{black}}%
      \expandafter\def\csname LTa\endcsname{\color{black}}%
      \expandafter\def\csname LT0\endcsname{\color[rgb]{1,0,0}}%
      \expandafter\def\csname LT1\endcsname{\color[rgb]{0,1,0}}%
      \expandafter\def\csname LT2\endcsname{\color[rgb]{0,0,1}}%
      \expandafter\def\csname LT3\endcsname{\color[rgb]{1,0,1}}%
      \expandafter\def\csname LT4\endcsname{\color[rgb]{0,1,1}}%
      \expandafter\def\csname LT5\endcsname{\color[rgb]{1,1,0}}%
      \expandafter\def\csname LT6\endcsname{\color[rgb]{0,0,0}}%
      \expandafter\def\csname LT7\endcsname{\color[rgb]{1,0.3,0}}%
      \expandafter\def\csname LT8\endcsname{\color[rgb]{0.5,0.5,0.5}}%
    \else
      \def\colorrgb#1{\color{black}}%
      \def\colorgray#1{\color[gray]{#1}}%
      \expandafter\def\csname LTw\endcsname{\color{white}}%
      \expandafter\def\csname LTb\endcsname{\color{black}}%
      \expandafter\def\csname LTa\endcsname{\color{black}}%
      \expandafter\def\csname LT0\endcsname{\color{black}}%
      \expandafter\def\csname LT1\endcsname{\color{black}}%
      \expandafter\def\csname LT2\endcsname{\color{black}}%
      \expandafter\def\csname LT3\endcsname{\color{black}}%
      \expandafter\def\csname LT4\endcsname{\color{black}}%
      \expandafter\def\csname LT5\endcsname{\color{black}}%
      \expandafter\def\csname LT6\endcsname{\color{black}}%
      \expandafter\def\csname LT7\endcsname{\color{black}}%
      \expandafter\def\csname LT8\endcsname{\color{black}}%
    \fi
  \fi
  \setlength{\unitlength}{0.0500bp}%
  \begin{picture}(7200.00,5040.00)%
    \gplgaddtomacro\gplbacktext{%
      \csname LTb\endcsname%
      \put(990,3915){\makebox(0,0)[r]{\strut{}-0.2}}%
      \put(990,2985){\makebox(0,0)[r]{\strut{} 0}}%
      \put(990,2055){\makebox(0,0)[r]{\strut{} 0.2}}%
      \put(990,1125){\makebox(0,0)[r]{\strut{} 0.4}}%
      \put(1122,440){\makebox(0,0){\strut{}-0.6}}%
      \put(2263,440){\makebox(0,0){\strut{}-0.4}}%
      \put(3404,440){\makebox(0,0){\strut{}-0.2}}%
      \put(4544,440){\makebox(0,0){\strut{} 0}}%
      \put(5685,440){\makebox(0,0){\strut{} 0.2}}%
      \put(6826,440){\makebox(0,0){\strut{} 0.4}}%
      \put(220,2520){\rotatebox{90}{\makebox(0,0){\strut{}$z\,\mathrm{(m)}$}}}%
      \put(3974,110){\makebox(0,0){\strut{}$y\,\mathrm{(m)}$}}%
      \put(3974,4710){\makebox(0,0){\strut{}(a)}}%
    }%
    \gplgaddtomacro\gplfronttext{%
    }%
    \gplbacktext
    \put(0,0){\includegraphics{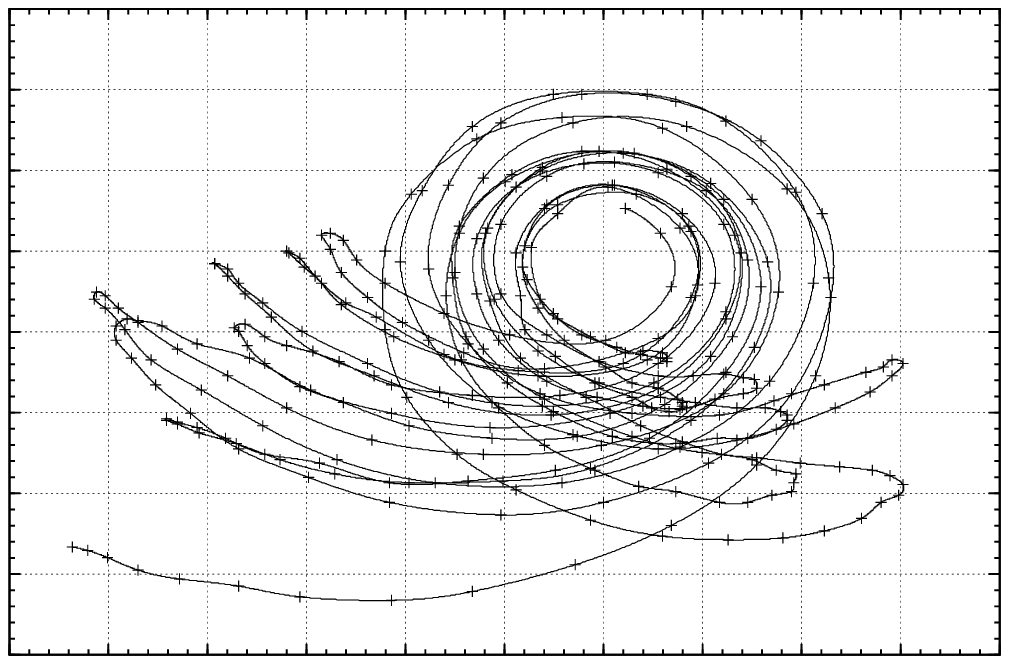}}%
    \put(4544,2985){\pscircle(0,0){0.25}}
    \gplfronttext
  \end{picture}%
\endgroup

%% file: Figure7b.tex
\begingroup
  \makeatletter
  \providecommand\color[2][]{%
    \GenericError{(gnuplot) \space\space\space\@spaces}{%
      Package color not loaded in conjunction with
      terminal option `colourtext'%
    }{See the gnuplot documentation for explanation.%
    }{Either use 'blacktext' in gnuplot or load the package
      color.sty in LaTeX.}%
    \renewcommand\color[2][]{}%
  }%
  \providecommand\includegraphics[2][]{%
    \GenericError{(gnuplot) \space\space\space\@spaces}{%
      Package graphicx or graphics not loaded%
    }{See the gnuplot documentation for explanation.%
    }{The gnuplot epslatex terminal needs graphicx.sty or graphics.sty.}%
    \renewcommand\includegraphics[2][]{}%
  }%
  \providecommand\rotatebox[2]{#2}%
  \@ifundefined{ifGPcolor}{%
    \newif\ifGPcolor
    \GPcolorfalse
  }{}%
  \@ifundefined{ifGPblacktext}{%
    \newif\ifGPblacktext
    \GPblacktexttrue
  }{}%
  \let\gplgaddtomacro\g@addto@macro
  \gdef\gplbacktext{}%
  \gdef\gplfronttext{}%
  \makeatother
  \ifGPblacktext
    \def\colorrgb#1{}%
    \def\colorgray#1{}%
  \else
    \ifGPcolor
      \def\colorrgb#1{\color[rgb]{#1}}%
      \def\colorgray#1{\color[gray]{#1}}%
      \expandafter\def\csname LTw\endcsname{\color{white}}%
      \expandafter\def\csname LTb\endcsname{\color{black}}%
      \expandafter\def\csname LTa\endcsname{\color{black}}%
      \expandafter\def\csname LT0\endcsname{\color[rgb]{1,0,0}}%
      \expandafter\def\csname LT1\endcsname{\color[rgb]{0,1,0}}%
      \expandafter\def\csname LT2\endcsname{\color[rgb]{0,0,1}}%
      \expandafter\def\csname LT3\endcsname{\color[rgb]{1,0,1}}%
      \expandafter\def\csname LT4\endcsname{\color[rgb]{0,1,1}}%
      \expandafter\def\csname LT5\endcsname{\color[rgb]{1,1,0}}%
      \expandafter\def\csname LT6\endcsname{\color[rgb]{0,0,0}}%
      \expandafter\def\csname LT7\endcsname{\color[rgb]{1,0.3,0}}%
      \expandafter\def\csname LT8\endcsname{\color[rgb]{0.5,0.5,0.5}}%
    \else
      \def\colorrgb#1{\color{black}}%
      \def\colorgray#1{\color[gray]{#1}}%
      \expandafter\def\csname LTw\endcsname{\color{white}}%
      \expandafter\def\csname LTb\endcsname{\color{black}}%
      \expandafter\def\csname LTa\endcsname{\color{black}}%
      \expandafter\def\csname LT0\endcsname{\color{black}}%
      \expandafter\def\csname LT1\endcsname{\color{black}}%
      \expandafter\def\csname LT2\endcsname{\color{black}}%
      \expandafter\def\csname LT3\endcsname{\color{black}}%
      \expandafter\def\csname LT4\endcsname{\color{black}}%
      \expandafter\def\csname LT5\endcsname{\color{black}}%
      \expandafter\def\csname LT6\endcsname{\color{black}}%
      \expandafter\def\csname LT7\endcsname{\color{black}}%
      \expandafter\def\csname LT8\endcsname{\color{black}}%
    \fi
  \fi
  \setlength{\unitlength}{0.0500bp}%
  \begin{picture}(7200.00,3024.00)%
    \gplgaddtomacro\gplbacktext{%
      \csname LTb\endcsname%
      \put(990,660){\makebox(0,0)[r]{\strut{} 0}}%
      \put(990,1147){\makebox(0,0)[r]{\strut{} 0.2}}%
      \put(990,1634){\makebox(0,0)[r]{\strut{} 0.4}}%
      \put(990,2121){\makebox(0,0)[r]{\strut{} 0.6}}%
      \put(1122,440){\makebox(0,0){\strut{} 0}}%
      \put(2263,440){\makebox(0,0){\strut{} 5}}%
      \put(3404,440){\makebox(0,0){\strut{} 10}}%
      \put(4544,440){\makebox(0,0){\strut{} 15}}%
      \put(5685,440){\makebox(0,0){\strut{} 20}}%
      \put(6826,440){\makebox(0,0){\strut{} 25}}%
      \put(220,1512){\rotatebox{90}{\makebox(0,0){\strut{}$r\,\mathrm{(m)}$}}}%
      \put(3974,110){\makebox(0,0){\strut{}$t\,\mathrm{(s)}$}}%
      \put(3974,2694){\makebox(0,0){\strut{}(b)}}%
    }%
    \gplgaddtomacro\gplfronttext{%
    }%
    \gplbacktext
    \put(0,0){\includegraphics{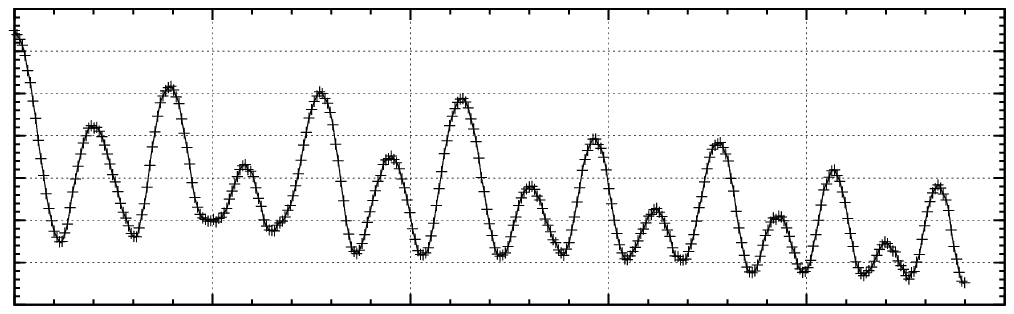}}%
    \gplfronttext
  \end{picture}%
\endgroup

%% file: Figure7c.tex
\begingroup
  \makeatletter
  \providecommand\color[2][]{%
    \GenericError{(gnuplot) \space\space\space\@spaces}{%
      Package color not loaded in conjunction with
      terminal option `colourtext'%
    }{See the gnuplot documentation for explanation.%
    }{Either use 'blacktext' in gnuplot or load the package
      color.sty in LaTeX.}%
    \renewcommand\color[2][]{}%
  }%
  \providecommand\includegraphics[2][]{%
    \GenericError{(gnuplot) \space\space\space\@spaces}{%
      Package graphicx or graphics not loaded%
    }{See the gnuplot documentation for explanation.%
    }{The gnuplot epslatex terminal needs graphicx.sty or graphics.sty.}%
    \renewcommand\includegraphics[2][]{}%
  }%
  \providecommand\rotatebox[2]{#2}%
  \@ifundefined{ifGPcolor}{%
    \newif\ifGPcolor
    \GPcolorfalse
  }{}%
  \@ifundefined{ifGPblacktext}{%
    \newif\ifGPblacktext
    \GPblacktexttrue
  }{}%
  \let\gplgaddtomacro\g@addto@macro
  \gdef\gplbacktext{}%
  \gdef\gplfronttext{}%
  \makeatother
  \ifGPblacktext
    \def\colorrgb#1{}%
    \def\colorgray#1{}%
  \else
    \ifGPcolor
      \def\colorrgb#1{\color[rgb]{#1}}%
      \def\colorgray#1{\color[gray]{#1}}%
      \expandafter\def\csname LTw\endcsname{\color{white}}%
      \expandafter\def\csname LTb\endcsname{\color{black}}%
      \expandafter\def\csname LTa\endcsname{\color{black}}%
      \expandafter\def\csname LT0\endcsname{\color[rgb]{1,0,0}}%
      \expandafter\def\csname LT1\endcsname{\color[rgb]{0,1,0}}%
      \expandafter\def\csname LT2\endcsname{\color[rgb]{0,0,1}}%
      \expandafter\def\csname LT3\endcsname{\color[rgb]{1,0,1}}%
      \expandafter\def\csname LT4\endcsname{\color[rgb]{0,1,1}}%
      \expandafter\def\csname LT5\endcsname{\color[rgb]{1,1,0}}%
      \expandafter\def\csname LT6\endcsname{\color[rgb]{0,0,0}}%
      \expandafter\def\csname LT7\endcsname{\color[rgb]{1,0.3,0}}%
      \expandafter\def\csname LT8\endcsname{\color[rgb]{0.5,0.5,0.5}}%
    \else
      \def\colorrgb#1{\color{black}}%
      \def\colorgray#1{\color[gray]{#1}}%
      \expandafter\def\csname LTw\endcsname{\color{white}}%
      \expandafter\def\csname LTb\endcsname{\color{black}}%
      \expandafter\def\csname LTa\endcsname{\color{black}}%
      \expandafter\def\csname LT0\endcsname{\color{black}}%
      \expandafter\def\csname LT1\endcsname{\color{black}}%
      \expandafter\def\csname LT2\endcsname{\color{black}}%
      \expandafter\def\csname LT3\endcsname{\color{black}}%
      \expandafter\def\csname LT4\endcsname{\color{black}}%
      \expandafter\def\csname LT5\endcsname{\color{black}}%
      \expandafter\def\csname LT6\endcsname{\color{black}}%
      \expandafter\def\csname LT7\endcsname{\color{black}}%
      \expandafter\def\csname LT8\endcsname{\color{black}}%
    \fi
  \fi
  \setlength{\unitlength}{0.0500bp}%
  \begin{picture}(7200.00,3024.00)%
    \gplgaddtomacro\gplbacktext{%
      \csname LTb\endcsname%
      \put(726,660){\makebox(0,0)[r]{\strut{}-8}}%
      \put(726,1342){\makebox(0,0)[r]{\strut{}-4}}%
      \put(726,2023){\makebox(0,0)[r]{\strut{} 0}}%
      \put(858,440){\makebox(0,0){\strut{} 0}}%
      \put(2052,440){\makebox(0,0){\strut{} 5}}%
      \put(3245,440){\makebox(0,0){\strut{} 10}}%
      \put(4439,440){\makebox(0,0){\strut{} 15}}%
      \put(5632,440){\makebox(0,0){\strut{} 20}}%
      \put(6826,440){\makebox(0,0){\strut{} 25}}%
      \put(220,1512){\rotatebox{90}{\makebox(0,0){\strut{}$\theta$ (rad)}}}%
      \put(3842,110){\makebox(0,0){\strut{}$t$ (s)}}%
      \put(3842,2694){\makebox(0,0){\strut{}(c)}}%
    }%
    \gplgaddtomacro\gplfronttext{%
    }%
    \gplbacktext
    \put(0,0){\includegraphics{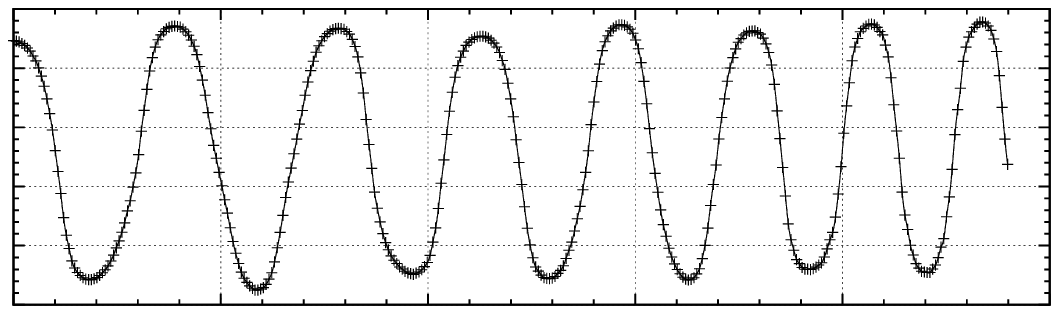}}%
    \gplfronttext
  \end{picture}%
\endgroup

%% file: Figure8.tex
\begingroup
  \makeatletter
  \providecommand\color[2][]{%
    \GenericError{(gnuplot) \space\space\space\@spaces}{%
      Package color not loaded in conjunction with
      terminal option `colourtext'%
    }{See the gnuplot documentation for explanation.%
    }{Either use 'blacktext' in gnuplot or load the package
      color.sty in LaTeX.}%
    \renewcommand\color[2][]{}%
  }%
  \providecommand\includegraphics[2][]{%
    \GenericError{(gnuplot) \space\space\space\@spaces}{%
      Package graphicx or graphics not loaded%
    }{See the gnuplot documentation for explanation.%
    }{The gnuplot epslatex terminal needs graphicx.sty or graphics.sty.}%
    \renewcommand\includegraphics[2][]{}%
  }%
  \providecommand\rotatebox[2]{#2}%
  \@ifundefined{ifGPcolor}{%
    \newif\ifGPcolor
    \GPcolorfalse
  }{}%
  \@ifundefined{ifGPblacktext}{%
    \newif\ifGPblacktext
    \GPblacktexttrue
  }{}%
  \let\gplgaddtomacro\g@addto@macro
  \gdef\gplbacktext{}%
  \gdef\gplfronttext{}%
  \makeatother
  \ifGPblacktext
    \def\colorrgb#1{}%
    \def\colorgray#1{}%
  \else
    \ifGPcolor
      \def\colorrgb#1{\color[rgb]{#1}}%
      \def\colorgray#1{\color[gray]{#1}}%
      \expandafter\def\csname LTw\endcsname{\color{white}}%
      \expandafter\def\csname LTb\endcsname{\color{black}}%
      \expandafter\def\csname LTa\endcsname{\color{black}}%
      \expandafter\def\csname LT0\endcsname{\color[rgb]{1,0,0}}%
      \expandafter\def\csname LT1\endcsname{\color[rgb]{0,1,0}}%
      \expandafter\def\csname LT2\endcsname{\color[rgb]{0,0,1}}%
      \expandafter\def\csname LT3\endcsname{\color[rgb]{1,0,1}}%
      \expandafter\def\csname LT4\endcsname{\color[rgb]{0,1,1}}%
      \expandafter\def\csname LT5\endcsname{\color[rgb]{1,1,0}}%
      \expandafter\def\csname LT6\endcsname{\color[rgb]{0,0,0}}%
      \expandafter\def\csname LT7\endcsname{\color[rgb]{1,0.3,0}}%
      \expandafter\def\csname LT8\endcsname{\color[rgb]{0.5,0.5,0.5}}%
    \else
      \def\colorrgb#1{\color{black}}%
      \def\colorgray#1{\color[gray]{#1}}%
      \expandafter\def\csname LTw\endcsname{\color{white}}%
      \expandafter\def\csname LTb\endcsname{\color{black}}%
      \expandafter\def\csname LTa\endcsname{\color{black}}%
      \expandafter\def\csname LT0\endcsname{\color{black}}%
      \expandafter\def\csname LT1\endcsname{\color{black}}%
      \expandafter\def\csname LT2\endcsname{\color{black}}%
      \expandafter\def\csname LT3\endcsname{\color{black}}%
      \expandafter\def\csname LT4\endcsname{\color{black}}%
      \expandafter\def\csname LT5\endcsname{\color{black}}%
      \expandafter\def\csname LT6\endcsname{\color{black}}%
      \expandafter\def\csname LT7\endcsname{\color{black}}%
      \expandafter\def\csname LT8\endcsname{\color{black}}%
    \fi
  \fi
  \setlength{\unitlength}{0.0500bp}%
  \begin{picture}(4320.00,3024.00)%
    \gplgaddtomacro\gplbacktext{%
      \csname LTb\endcsname%
      \put(1122,851){\makebox(0,0)[r]{\strut{}$0.8$}}%
      \put(1122,1233){\makebox(0,0)[r]{\strut{}$1$}}%
      \put(1122,1615){\makebox(0,0)[r]{\strut{}$1.2$}}%
      \put(1122,1996){\makebox(0,0)[r]{\strut{}$1.4$}}%
      \put(1122,2378){\makebox(0,0)[r]{\strut{}$1.6$}}%
      \put(1122,2760){\makebox(0,0)[r]{\strut{}$1.8$}}%
      \put(1254,440){\makebox(0,0){\strut{}$0$}}%
      \put(1792,440){\makebox(0,0){\strut{}$5$}}%
      \put(2331,440){\makebox(0,0){\strut{}$10$}}%
      \put(2869,440){\makebox(0,0){\strut{}$15$}}%
      \put(3408,440){\makebox(0,0){\strut{}$20$}}%
      \put(3946,440){\makebox(0,0){\strut{}$25$}}%
      \put(220,1710){\rotatebox{90}{\makebox(0,0){\strut{}${\cal{E}}_m  \ut{(J)}$}}}%
      \put(2600,110){\makebox(0,0){\strut{}$t  \ut{(s)}$}}%
    }%
    \gplgaddtomacro\gplfronttext{%
    }%
    \gplbacktext
    \put(0,0){\includegraphics{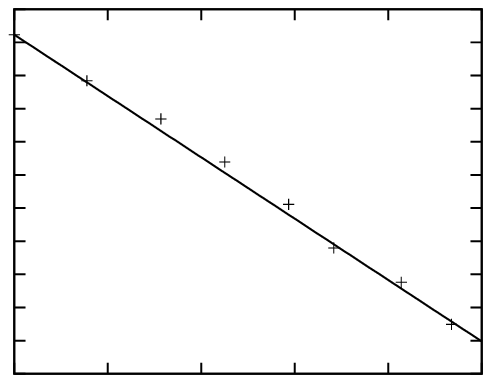}}%
    \gplfronttext
  \end{picture}%
\endgroup

%% file: Figure9a.tex
\begingroup
  \makeatletter
  \providecommand\color[2][]{%
    \GenericError{(gnuplot) \space\space\space\@spaces}{%
      Package color not loaded in conjunction with
      terminal option `colourtext'%
    }{See the gnuplot documentation for explanation.%
    }{Either use 'blacktext' in gnuplot or load the package
      color.sty in LaTeX.}%
    \renewcommand\color[2][]{}%
  }%
  \providecommand\includegraphics[2][]{%
    \GenericError{(gnuplot) \space\space\space\@spaces}{%
      Package graphicx or graphics not loaded%
    }{See the gnuplot documentation for explanation.%
    }{The gnuplot epslatex terminal needs graphicx.sty or graphics.sty.}%
    \renewcommand\includegraphics[2][]{}%
  }%
  \providecommand\rotatebox[2]{#2}%
  \@ifundefined{ifGPcolor}{%
    \newif\ifGPcolor
    \GPcolorfalse
  }{}%
  \@ifundefined{ifGPblacktext}{%
    \newif\ifGPblacktext
    \GPblacktexttrue
  }{}%
  \let\gplgaddtomacro\g@addto@macro
  \gdef\gplbacktext{}%
  \gdef\gplfronttext{}%
  \makeatother
  \ifGPblacktext
    \def\colorrgb#1{}%
    \def\colorgray#1{}%
  \else
    \ifGPcolor
      \def\colorrgb#1{\color[rgb]{#1}}%
      \def\colorgray#1{\color[gray]{#1}}%
      \expandafter\def\csname LTw\endcsname{\color{white}}%
      \expandafter\def\csname LTb\endcsname{\color{black}}%
      \expandafter\def\csname LTa\endcsname{\color{black}}%
      \expandafter\def\csname LT0\endcsname{\color[rgb]{1,0,0}}%
      \expandafter\def\csname LT1\endcsname{\color[rgb]{0,1,0}}%
      \expandafter\def\csname LT2\endcsname{\color[rgb]{0,0,1}}%
      \expandafter\def\csname LT3\endcsname{\color[rgb]{1,0,1}}%
      \expandafter\def\csname LT4\endcsname{\color[rgb]{0,1,1}}%
      \expandafter\def\csname LT5\endcsname{\color[rgb]{1,1,0}}%
      \expandafter\def\csname LT6\endcsname{\color[rgb]{0,0,0}}%
      \expandafter\def\csname LT7\endcsname{\color[rgb]{1,0.3,0}}%
      \expandafter\def\csname LT8\endcsname{\color[rgb]{0.5,0.5,0.5}}%
    \else
      \def\colorrgb#1{\color{black}}%
      \def\colorgray#1{\color[gray]{#1}}%
      \expandafter\def\csname LTw\endcsname{\color{white}}%
      \expandafter\def\csname LTb\endcsname{\color{black}}%
      \expandafter\def\csname LTa\endcsname{\color{black}}%
      \expandafter\def\csname LT0\endcsname{\color{black}}%
      \expandafter\def\csname LT1\endcsname{\color{black}}%
      \expandafter\def\csname LT2\endcsname{\color{black}}%
      \expandafter\def\csname LT3\endcsname{\color{black}}%
      \expandafter\def\csname LT4\endcsname{\color{black}}%
      \expandafter\def\csname LT5\endcsname{\color{black}}%
      \expandafter\def\csname LT6\endcsname{\color{black}}%
      \expandafter\def\csname LT7\endcsname{\color{black}}%
      \expandafter\def\csname LT8\endcsname{\color{black}}%
    \fi
  \fi
  \setlength{\unitlength}{0.0500bp}%
  \begin{picture}(7200.00,5040.00)%
    \gplgaddtomacro\gplbacktext{%
      \csname LTb\endcsname%
      \put(1122,4380){\makebox(0,0)[r]{\strut{} 0}}%
      \put(1122,3450){\makebox(0,0)[r]{\strut{} 0.1}}%
      \put(1122,2520){\makebox(0,0)[r]{\strut{} 0.2}}%
      \put(1122,1590){\makebox(0,0)[r]{\strut{} 0.3}}%
       \put(1122,660){\makebox(0,0)[r]{\strut{} 0.4}}%
      \csname LTb\endcsname%
      \csname LTb\endcsname%
      \put(1761,440){\makebox(0,0){\strut{}-0.5}}%
      \csname LTb\endcsname%
      \put(2267,440){\makebox(0,0){\strut{}-0.4}}%
      \csname LTb\endcsname%
      \put(2774,440){\makebox(0,0){\strut{}-0.3}}%
      \csname LTb\endcsname%
      \put(3280,440){\makebox(0,0){\strut{}-0.2}}%
      \csname LTb\endcsname%
      \put(3787,440){\makebox(0,0){\strut{}-0.1}}%
      \csname LTb\endcsname%
      \put(4293,440){\makebox(0,0){\strut{} 0}}%
      \csname LTb\endcsname%
      \put(4800,440){\makebox(0,0){\strut{} 0.1}}%
      \csname LTb\endcsname%
      \put(5306,440){\makebox(0,0){\strut{} 0.2}}%
      \csname LTb\endcsname%
      \put(5813,440){\makebox(0,0){\strut{} 0.3}}%
      \csname LTb\endcsname%
      \put(6319,440){\makebox(0,0){\strut{} 0.4}}%
      \csname LTb\endcsname%
      \put(220,2520){\rotatebox{90}{\makebox(0,0){\strut{}$z\,\mathrm{(m)}$}}}%
      \put(4040,110){\makebox(0,0){\strut{}$y\,\mathrm{(m)}$}}%
      \put(4040,4710){\makebox(0,0){\strut{}(a)}}%
    }%
    \gplgaddtomacro\gplfronttext{%
    }%
    \gplbacktext
    \put(0,0){\includegraphics{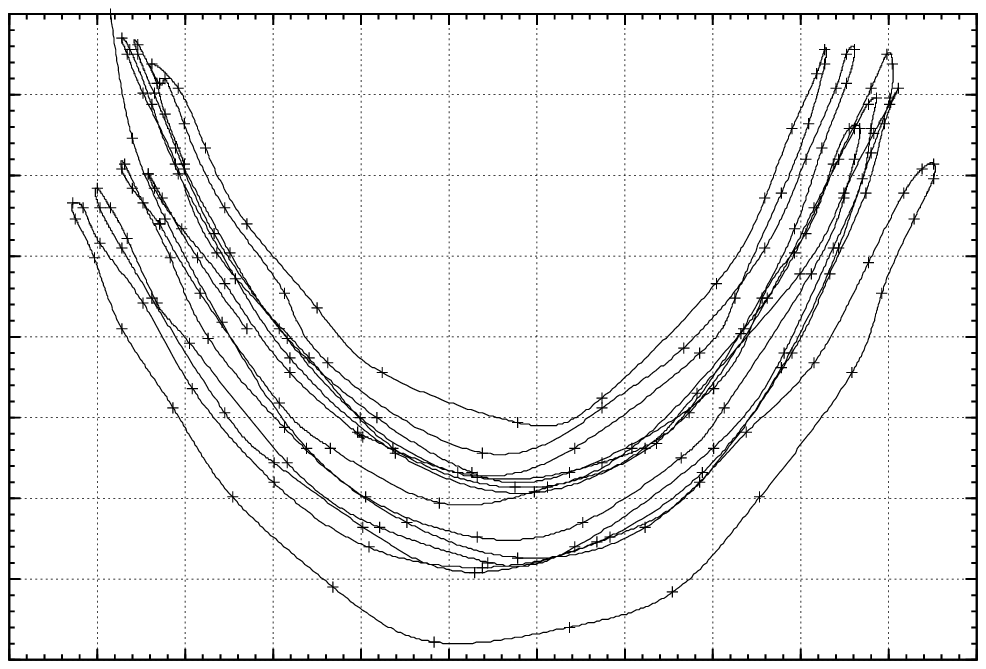}}%
    \gplfronttext
  \end{picture}%
\endgroup

%% file: Figure9b.tex
\begingroup
  \makeatletter
  \providecommand\color[2][]{%
    \GenericError{(gnuplot) \space\space\space\@spaces}{%
      Package color not loaded in conjunction with
      terminal option `colourtext'%
    }{See the gnuplot documentation for explanation.%
    }{Either use 'blacktext' in gnuplot or load the package
      color.sty in LaTeX.}%
    \renewcommand\color[2][]{}%
  }%
  \providecommand\includegraphics[2][]{%
    \GenericError{(gnuplot) \space\space\space\@spaces}{%
      Package graphicx or graphics not loaded%
    }{See the gnuplot documentation for explanation.%
    }{The gnuplot epslatex terminal needs graphicx.sty or graphics.sty.}%
    \renewcommand\includegraphics[2][]{}%
  }%
  \providecommand\rotatebox[2]{#2}%
  \@ifundefined{ifGPcolor}{%
    \newif\ifGPcolor
    \GPcolorfalse
  }{}%
  \@ifundefined{ifGPblacktext}{%
    \newif\ifGPblacktext
    \GPblacktexttrue
  }{}%
  \let\gplgaddtomacro\g@addto@macro
  \gdef\gplbacktext{}%
  \gdef\gplfronttext{}%
  \makeatother
  \ifGPblacktext
    \def\colorrgb#1{}%
    \def\colorgray#1{}%
  \else
    \ifGPcolor
      \def\colorrgb#1{\color[rgb]{#1}}%
      \def\colorgray#1{\color[gray]{#1}}%
      \expandafter\def\csname LTw\endcsname{\color{white}}%
      \expandafter\def\csname LTb\endcsname{\color{black}}%
      \expandafter\def\csname LTa\endcsname{\color{black}}%
      \expandafter\def\csname LT0\endcsname{\color[rgb]{1,0,0}}%
      \expandafter\def\csname LT1\endcsname{\color[rgb]{0,1,0}}%
      \expandafter\def\csname LT2\endcsname{\color[rgb]{0,0,1}}%
      \expandafter\def\csname LT3\endcsname{\color[rgb]{1,0,1}}%
      \expandafter\def\csname LT4\endcsname{\color[rgb]{0,1,1}}%
      \expandafter\def\csname LT5\endcsname{\color[rgb]{1,1,0}}%
      \expandafter\def\csname LT6\endcsname{\color[rgb]{0,0,0}}%
      \expandafter\def\csname LT7\endcsname{\color[rgb]{1,0.3,0}}%
      \expandafter\def\csname LT8\endcsname{\color[rgb]{0.5,0.5,0.5}}%
    \else
      \def\colorrgb#1{\color{black}}%
      \def\colorgray#1{\color[gray]{#1}}%
      \expandafter\def\csname LTw\endcsname{\color{white}}%
      \expandafter\def\csname LTb\endcsname{\color{black}}%
      \expandafter\def\csname LTa\endcsname{\color{black}}%
      \expandafter\def\csname LT0\endcsname{\color{black}}%
      \expandafter\def\csname LT1\endcsname{\color{black}}%
      \expandafter\def\csname LT2\endcsname{\color{black}}%
      \expandafter\def\csname LT3\endcsname{\color{black}}%
      \expandafter\def\csname LT4\endcsname{\color{black}}%
      \expandafter\def\csname LT5\endcsname{\color{black}}%
      \expandafter\def\csname LT6\endcsname{\color{black}}%
      \expandafter\def\csname LT7\endcsname{\color{black}}%
      \expandafter\def\csname LT8\endcsname{\color{black}}%
    \fi
  \fi
  \setlength{\unitlength}{0.0500bp}%
  \begin{picture}(7200.00,3024.00)%
    \gplgaddtomacro\gplbacktext{%
      \csname LTb\endcsname%
      \put(1122,660){\makebox(0,0)[r]{\strut{} 0.2}}%
      \csname LTb\endcsname%
      \csname LTb\endcsname%
      \put(1122,1086){\makebox(0,0)[r]{\strut{} 0.3}}%
      \csname LTb\endcsname%
      \csname LTb\endcsname%
      \put(1122,1512){\makebox(0,0)[r]{\strut{} 0.4}}%
      \csname LTb\endcsname%
      \csname LTb\endcsname%
      \put(1122,1938){\makebox(0,0)[r]{\strut{} 0.5}}%
      \csname LTb\endcsname%
      \csname LTb\endcsname%
      \put(1122,2364){\makebox(0,0)[r]{\strut{} 0.6}}%
      \csname LTb\endcsname%
      \put(1254,440){\makebox(0,0){\strut{} 0}}%
      \csname LTb\endcsname%
      \put(1951,440){\makebox(0,0){\strut{} 1}}%
      \csname LTb\endcsname%
      \put(2647,440){\makebox(0,0){\strut{} 2}}%
      \csname LTb\endcsname%
      \put(3344,440){\makebox(0,0){\strut{} 3}}%
      \csname LTb\endcsname%
      \put(4040,440){\makebox(0,0){\strut{} 4}}%
      \csname LTb\endcsname%
      \put(4737,440){\makebox(0,0){\strut{} 5}}%
      \csname LTb\endcsname%
      \put(5433,440){\makebox(0,0){\strut{} 6}}%
      \csname LTb\endcsname%
      \put(6130,440){\makebox(0,0){\strut{} 7}}%
      \csname LTb\endcsname%
      \put(6826,440){\makebox(0,0){\strut{} 8}}%
      \put(220,1512){\rotatebox{90}{\makebox(0,0){\strut{}$r\,\mathrm{(m)}$}}}%
      \put(4040,110){\makebox(0,0){\strut{}$t\,\mathrm{(s)}$}}%
      \put(4040,2694){\makebox(0,0){\strut{}(b)}}%
    }%
    \gplgaddtomacro\gplfronttext{%
    }%
    \gplbacktext
    \put(0,0){\includegraphics{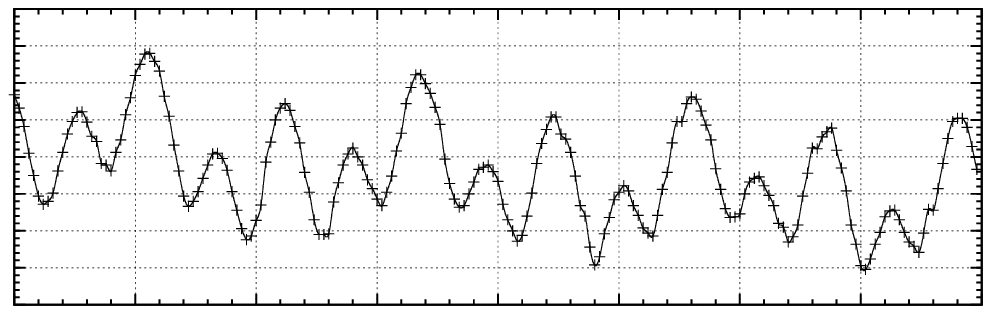}}%
    \gplfronttext
  \end{picture}%
\endgroup

%% file: Figure9c.tex
\begingroup
  \makeatletter
  \providecommand\color[2][]{%
    \GenericError{(gnuplot) \space\space\space\@spaces}{%
      Package color not loaded in conjunction with
      terminal option `colourtext'%
    }{See the gnuplot documentation for explanation.%
    }{Either use 'blacktext' in gnuplot or load the package
      color.sty in LaTeX.}%
    \renewcommand\color[2][]{}%
  }%
  \providecommand\includegraphics[2][]{%
    \GenericError{(gnuplot) \space\space\space\@spaces}{%
      Package graphicx or graphics not loaded%
    }{See the gnuplot documentation for explanation.%
    }{The gnuplot epslatex terminal needs graphicx.sty or graphics.sty.}%
    \renewcommand\includegraphics[2][]{}%
  }%
  \providecommand\rotatebox[2]{#2}%
  \@ifundefined{ifGPcolor}{%
    \newif\ifGPcolor
    \GPcolorfalse
  }{}%
  \@ifundefined{ifGPblacktext}{%
    \newif\ifGPblacktext
    \GPblacktexttrue
  }{}%
  \let\gplgaddtomacro\g@addto@macro
  \gdef\gplbacktext{}%
  \gdef\gplfronttext{}%
  \makeatother
  \ifGPblacktext
    \def\colorrgb#1{}%
    \def\colorgray#1{}%
  \else
    \ifGPcolor
      \def\colorrgb#1{\color[rgb]{#1}}%
      \def\colorgray#1{\color[gray]{#1}}%
      \expandafter\def\csname LTw\endcsname{\color{white}}%
      \expandafter\def\csname LTb\endcsname{\color{black}}%
      \expandafter\def\csname LTa\endcsname{\color{black}}%
      \expandafter\def\csname LT0\endcsname{\color[rgb]{1,0,0}}%
      \expandafter\def\csname LT1\endcsname{\color[rgb]{0,1,0}}%
      \expandafter\def\csname LT2\endcsname{\color[rgb]{0,0,1}}%
      \expandafter\def\csname LT3\endcsname{\color[rgb]{1,0,1}}%
      \expandafter\def\csname LT4\endcsname{\color[rgb]{0,1,1}}%
      \expandafter\def\csname LT5\endcsname{\color[rgb]{1,1,0}}%
      \expandafter\def\csname LT6\endcsname{\color[rgb]{0,0,0}}%
      \expandafter\def\csname LT7\endcsname{\color[rgb]{1,0.3,0}}%
      \expandafter\def\csname LT8\endcsname{\color[rgb]{0.5,0.5,0.5}}%
    \else
      \def\colorrgb#1{\color{black}}%
      \def\colorgray#1{\color[gray]{#1}}%
      \expandafter\def\csname LTw\endcsname{\color{white}}%
      \expandafter\def\csname LTb\endcsname{\color{black}}%
      \expandafter\def\csname LTa\endcsname{\color{black}}%
      \expandafter\def\csname LT0\endcsname{\color{black}}%
      \expandafter\def\csname LT1\endcsname{\color{black}}%
      \expandafter\def\csname LT2\endcsname{\color{black}}%
      \expandafter\def\csname LT3\endcsname{\color{black}}%
      \expandafter\def\csname LT4\endcsname{\color{black}}%
      \expandafter\def\csname LT5\endcsname{\color{black}}%
      \expandafter\def\csname LT6\endcsname{\color{black}}%
      \expandafter\def\csname LT7\endcsname{\color{black}}%
      \expandafter\def\csname LT8\endcsname{\color{black}}%
    \fi
  \fi
  \setlength{\unitlength}{0.0500bp}%
  \begin{picture}(7200.00,3024.00)%
    \gplgaddtomacro\gplbacktext{%
      \csname LTb\endcsname%
      \put(990,660){\makebox(0,0)[r]{\strut{}-2}}%
      \csname LTb\endcsname%
      \csname LTb\endcsname%
      \put(990,1086){\makebox(0,0)[r]{\strut{}-1}}%
      \csname LTb\endcsname%
      \csname LTb\endcsname%
      \put(990,1512){\makebox(0,0)[r]{\strut{} 0}}%
      \csname LTb\endcsname%
      \csname LTb\endcsname%
      \put(990,1938){\makebox(0,0)[r]{\strut{} 1}}%
      \csname LTb\endcsname%
      \csname LTb\endcsname%
      \put(990,2364){\makebox(0,0)[r]{\strut{} 2}}%
      \csname LTb\endcsname%
      \put(1122,440){\makebox(0,0){\strut{} 0}}%
      \csname LTb\endcsname%
      \put(1835,440){\makebox(0,0){\strut{} 1}}%
      \csname LTb\endcsname%
      \put(2548,440){\makebox(0,0){\strut{} 2}}%
      \csname LTb\endcsname%
      \put(3261,440){\makebox(0,0){\strut{} 3}}%
      \csname LTb\endcsname%
      \put(3974,440){\makebox(0,0){\strut{} 4}}%
      \csname LTb\endcsname%
      \put(4687,440){\makebox(0,0){\strut{} 5}}%
      \csname LTb\endcsname%
      \put(5400,440){\makebox(0,0){\strut{} 6}}%
      \csname LTb\endcsname%
      \put(6113,440){\makebox(0,0){\strut{} 7}}%
      \csname LTb\endcsname%
      \put(6826,440){\makebox(0,0){\strut{} 8}}%
      \put(220,1512){\rotatebox{90}{\makebox(0,0){\strut{}$\theta$ (rad)}}}%
      \put(3974,110){\makebox(0,0){\strut{}$t$ (s)}}%
      \put(3974,2694){\makebox(0,0){\strut{}(c)}}%
    }%
    \gplgaddtomacro\gplfronttext{%
    }%
    \gplbacktext
    \put(0,0){\includegraphics{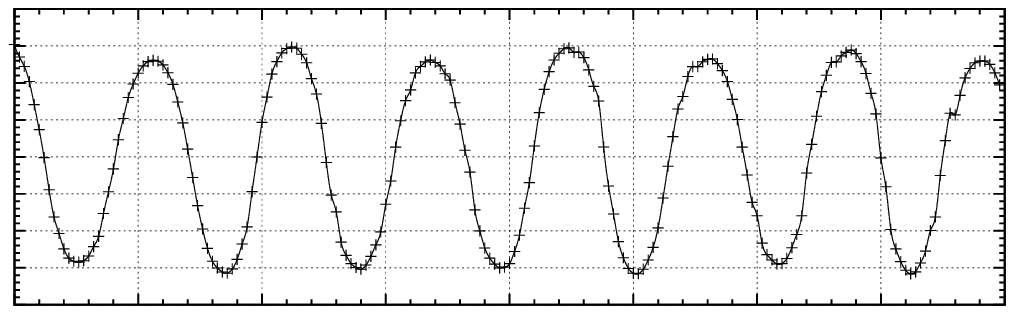}}%
    \gplfronttext
  \end{picture}%
\endgroup

%% file: Figure10.tex
\begingroup
  \makeatletter
  \providecommand\color[2][]{%
    \GenericError{(gnuplot) \space\space\space\@spaces}{%
      Package color not loaded in conjunction with
      terminal option `colourtext'%
    }{See the gnuplot documentation for explanation.%
    }{Either use 'blacktext' in gnuplot or load the package
      color.sty in LaTeX.}%
    \renewcommand\color[2][]{}%
  }%
  \providecommand\includegraphics[2][]{%
    \GenericError{(gnuplot) \space\space\space\@spaces}{%
      Package graphicx or graphics not loaded%
    }{See the gnuplot documentation for explanation.%
    }{The gnuplot epslatex terminal needs graphicx.sty or graphics.sty.}%
    \renewcommand\includegraphics[2][]{}%
  }%
  \providecommand\rotatebox[2]{#2}%
  \@ifundefined{ifGPcolor}{%
    \newif\ifGPcolor
    \GPcolorfalse
  }{}%
  \@ifundefined{ifGPblacktext}{%
    \newif\ifGPblacktext
    \GPblacktexttrue
  }{}%
  \let\gplgaddtomacro\g@addto@macro
  \gdef\gplbacktext{}%
  \gdef\gplfronttext{}%
  \makeatother
  \ifGPblacktext
    \def\colorrgb#1{}%
    \def\colorgray#1{}%
  \else
    \ifGPcolor
      \def\colorrgb#1{\color[rgb]{#1}}%
      \def\colorgray#1{\color[gray]{#1}}%
      \expandafter\def\csname LTw\endcsname{\color{white}}%
      \expandafter\def\csname LTb\endcsname{\color{black}}%
      \expandafter\def\csname LTa\endcsname{\color{black}}%
      \expandafter\def\csname LT0\endcsname{\color[rgb]{1,0,0}}%
      \expandafter\def\csname LT1\endcsname{\color[rgb]{0,1,0}}%
      \expandafter\def\csname LT2\endcsname{\color[rgb]{0,0,1}}%
      \expandafter\def\csname LT3\endcsname{\color[rgb]{1,0,1}}%
      \expandafter\def\csname LT4\endcsname{\color[rgb]{0,1,1}}%
      \expandafter\def\csname LT5\endcsname{\color[rgb]{1,1,0}}%
      \expandafter\def\csname LT6\endcsname{\color[rgb]{0,0,0}}%
      \expandafter\def\csname LT7\endcsname{\color[rgb]{1,0.3,0}}%
      \expandafter\def\csname LT8\endcsname{\color[rgb]{0.5,0.5,0.5}}%
    \else
      \def\colorrgb#1{\color{black}}%
      \def\colorgray#1{\color[gray]{#1}}%
      \expandafter\def\csname LTw\endcsname{\color{white}}%
      \expandafter\def\csname LTb\endcsname{\color{black}}%
      \expandafter\def\csname LTa\endcsname{\color{black}}%
      \expandafter\def\csname LT0\endcsname{\color{black}}%
      \expandafter\def\csname LT1\endcsname{\color{black}}%
      \expandafter\def\csname LT2\endcsname{\color{black}}%
      \expandafter\def\csname LT3\endcsname{\color{black}}%
      \expandafter\def\csname LT4\endcsname{\color{black}}%
      \expandafter\def\csname LT5\endcsname{\color{black}}%
      \expandafter\def\csname LT6\endcsname{\color{black}}%
      \expandafter\def\csname LT7\endcsname{\color{black}}%
      \expandafter\def\csname LT8\endcsname{\color{black}}%
    \fi
  \fi
  \setlength{\unitlength}{0.0500bp}%
  \begin{picture}(4320.00,3024.00)%
    \gplgaddtomacro\gplbacktext{%
      \csname LTb\endcsname%
      \put(1254,851){\makebox(0,0)[r]{\strut{}$0.56$}}%
      \put(1254,1233){\makebox(0,0)[r]{\strut{}$0.6$}}%
      \put(1254,1615){\makebox(0,0)[r]{\strut{}$0.64$}}%
      \put(1254,1996){\makebox(0,0)[r]{\strut{}$0.68$}}%
      \put(1254,2378){\makebox(0,0)[r]{\strut{}$0.72$}}%
      \put(1254,2760){\makebox(0,0)[r]{\strut{}$0.76$}}%
      \put(1386,440){\makebox(0,0){\strut{}$0$}}%
      \put(1706,440){\makebox(0,0){\strut{}$1$}}%
      \put(2026,440){\makebox(0,0){\strut{}$2$}}%
      \put(2346,440){\makebox(0,0){\strut{}$3$}}%
      \put(2666,440){\makebox(0,0){\strut{}$4$}}%
      \put(2986,440){\makebox(0,0){\strut{}$5$}}%
      \put(3306,440){\makebox(0,0){\strut{}$6$}}%
      \put(3626,440){\makebox(0,0){\strut{}$7$}}%
      \put(3946,440){\makebox(0,0){\strut{}$8$}}%
      \put(220,1710){\rotatebox{90}{\makebox(0,0){\strut{}${\cal{E}}_m  \ut{(J)}$}}}%
      \put(2666,110){\makebox(0,0){\strut{}$t  \ut{(s)}$}}%
    }%
    \gplgaddtomacro\gplfronttext{%
    }%
    \gplbacktext
    \put(0,0){\includegraphics{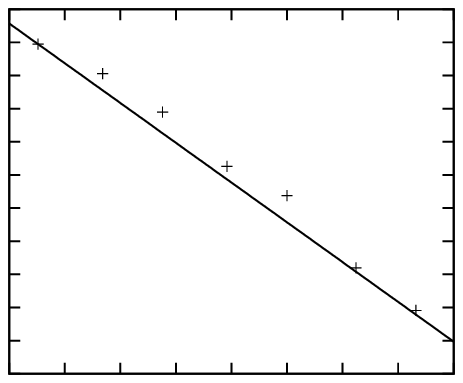}}%
    \gplfronttext
  \end{picture}%
\endgroup

%% file: Figure11a.tex
\begingroup
  \makeatletter
  \providecommand\color[2][]{%
    \GenericError{(gnuplot) \space\space\space\@spaces}{%
      Package color not loaded in conjunction with
      terminal option `colourtext'%
    }{See the gnuplot documentation for explanation.%
    }{Either use 'blacktext' in gnuplot or load the package
      color.sty in LaTeX.}%
    \renewcommand\color[2][]{}%
  }%
  \providecommand\includegraphics[2][]{%
    \GenericError{(gnuplot) \space\space\space\@spaces}{%
      Package graphicx or graphics not loaded%
    }{See the gnuplot documentation for explanation.%
    }{The gnuplot epslatex terminal needs graphicx.sty or graphics.sty.}%
    \renewcommand\includegraphics[2][]{}%
  }%
  \providecommand\rotatebox[2]{#2}%
  \@ifundefined{ifGPcolor}{%
    \newif\ifGPcolor
    \GPcolorfalse
  }{}%
  \@ifundefined{ifGPblacktext}{%
    \newif\ifGPblacktext
    \GPblacktexttrue
  }{}%
  \let\gplgaddtomacro\g@addto@macro
  \gdef\gplbacktext{}%
  \gdef\gplfronttext{}%
  \makeatother
  \ifGPblacktext
    \def\colorrgb#1{}%
    \def\colorgray#1{}%
  \else
    \ifGPcolor
      \def\colorrgb#1{\color[rgb]{#1}}%
      \def\colorgray#1{\color[gray]{#1}}%
      \expandafter\def\csname LTw\endcsname{\color{white}}%
      \expandafter\def\csname LTb\endcsname{\color{black}}%
      \expandafter\def\csname LTa\endcsname{\color{black}}%
      \expandafter\def\csname LT0\endcsname{\color[rgb]{1,0,0}}%
      \expandafter\def\csname LT1\endcsname{\color[rgb]{0,1,0}}%
      \expandafter\def\csname LT2\endcsname{\color[rgb]{0,0,1}}%
      \expandafter\def\csname LT3\endcsname{\color[rgb]{1,0,1}}%
      \expandafter\def\csname LT4\endcsname{\color[rgb]{0,1,1}}%
      \expandafter\def\csname LT5\endcsname{\color[rgb]{1,1,0}}%
      \expandafter\def\csname LT6\endcsname{\color[rgb]{0,0,0}}%
      \expandafter\def\csname LT7\endcsname{\color[rgb]{1,0.3,0}}%
      \expandafter\def\csname LT8\endcsname{\color[rgb]{0.5,0.5,0.5}}%
    \else
      \def\colorrgb#1{\color{black}}%
      \def\colorgray#1{\color[gray]{#1}}%
      \expandafter\def\csname LTw\endcsname{\color{white}}%
      \expandafter\def\csname LTb\endcsname{\color{black}}%
      \expandafter\def\csname LTa\endcsname{\color{black}}%
      \expandafter\def\csname LT0\endcsname{\color{black}}%
      \expandafter\def\csname LT1\endcsname{\color{black}}%
      \expandafter\def\csname LT2\endcsname{\color{black}}%
      \expandafter\def\csname LT3\endcsname{\color{black}}%
      \expandafter\def\csname LT4\endcsname{\color{black}}%
      \expandafter\def\csname LT5\endcsname{\color{black}}%
      \expandafter\def\csname LT6\endcsname{\color{black}}%
      \expandafter\def\csname LT7\endcsname{\color{black}}%
      \expandafter\def\csname LT8\endcsname{\color{black}}%
    \fi
  \fi
  \setlength{\unitlength}{0.0500bp}%
  \begin{picture}(7200.00,5040.00)%
    \gplgaddtomacro\gplbacktext{%
      \csname LTb\endcsname%
      \put(990,4094){\makebox(0,0)[r]{\strut{} 0}}%
      \put(990,2949){\makebox(0,0)[r]{\strut{} 0.4}}%
      \put(990,1805){\makebox(0,0)[r]{\strut{} 0.8}}%
      \put(990,660){\makebox(0,0)[r]{\strut{} 1.2}}%
      \put(1122,440){\makebox(0,0){\strut{}-0.2}}%
      \put(2548,440){\makebox(0,0){\strut{}-0.1}}%
      \put(3974,440){\makebox(0,0){\strut{} 0}}%
      \put(5400,440){\makebox(0,0){\strut{} 0.1}}%
      \put(6826,440){\makebox(0,0){\strut{} 0.2}}%
      \put(220,2520){\rotatebox{90}{\makebox(0,0){\strut{}$z\,\mathrm{(m)}$}}}%
      \put(3974,110){\makebox(0,0){\strut{}$y\,\mathrm{(m)}$}}%
      \put(3974,4710){\makebox(0,0){\strut{}(a)}}%
    }%
    \gplgaddtomacro\gplfronttext{%
    }%
    \gplbacktext
    \put(0,0){\includegraphics{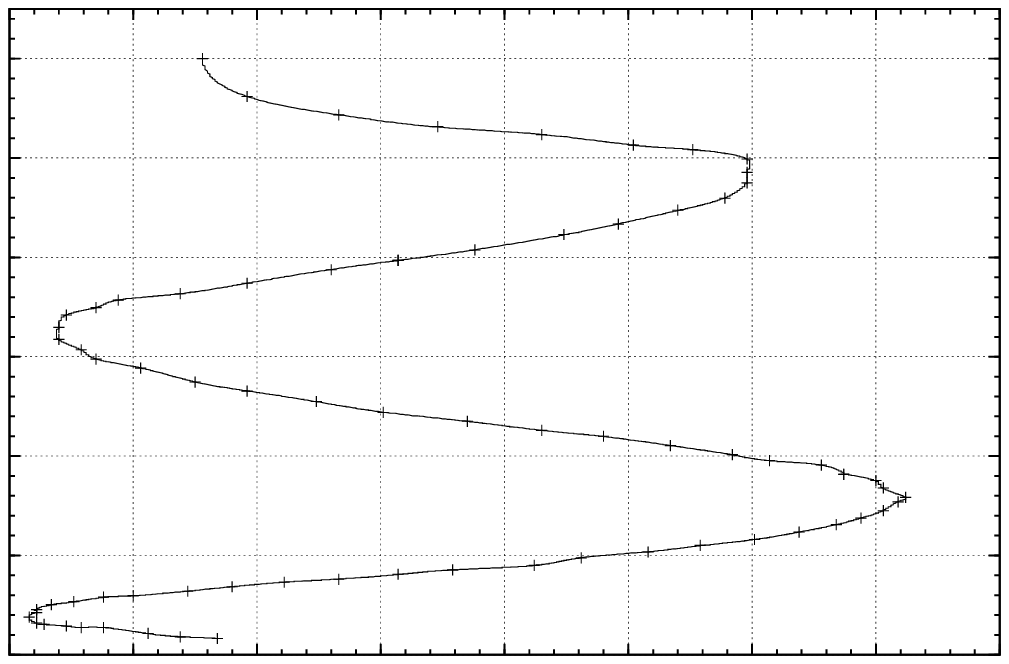}}%
    \gplfronttext
  \end{picture}%
\endgroup

%% file: Figure11b.tex
\begingroup
  \makeatletter
  \providecommand\color[2][]{%
    \GenericError{(gnuplot) \space\space\space\@spaces}{%
      Package color not loaded in conjunction with
      terminal option `colourtext'%
    }{See the gnuplot documentation for explanation.%
    }{Either use 'blacktext' in gnuplot or load the package
      color.sty in LaTeX.}%
    \renewcommand\color[2][]{}%
  }%
  \providecommand\includegraphics[2][]{%
    \GenericError{(gnuplot) \space\space\space\@spaces}{%
      Package graphicx or graphics not loaded%
    }{See the gnuplot documentation for explanation.%
    }{The gnuplot epslatex terminal needs graphicx.sty or graphics.sty.}%
    \renewcommand\includegraphics[2][]{}%
  }%
  \providecommand\rotatebox[2]{#2}%
  \@ifundefined{ifGPcolor}{%
    \newif\ifGPcolor
    \GPcolorfalse
  }{}%
  \@ifundefined{ifGPblacktext}{%
    \newif\ifGPblacktext
    \GPblacktexttrue
  }{}%
  \let\gplgaddtomacro\g@addto@macro
  \gdef\gplbacktext{}%
  \gdef\gplfronttext{}%
  \makeatother
  \ifGPblacktext
    \def\colorrgb#1{}%
    \def\colorgray#1{}%
  \else
    \ifGPcolor
      \def\colorrgb#1{\color[rgb]{#1}}%
      \def\colorgray#1{\color[gray]{#1}}%
      \expandafter\def\csname LTw\endcsname{\color{white}}%
      \expandafter\def\csname LTb\endcsname{\color{black}}%
      \expandafter\def\csname LTa\endcsname{\color{black}}%
      \expandafter\def\csname LT0\endcsname{\color[rgb]{1,0,0}}%
      \expandafter\def\csname LT1\endcsname{\color[rgb]{0,1,0}}%
      \expandafter\def\csname LT2\endcsname{\color[rgb]{0,0,1}}%
      \expandafter\def\csname LT3\endcsname{\color[rgb]{1,0,1}}%
      \expandafter\def\csname LT4\endcsname{\color[rgb]{0,1,1}}%
      \expandafter\def\csname LT5\endcsname{\color[rgb]{1,1,0}}%
      \expandafter\def\csname LT6\endcsname{\color[rgb]{0,0,0}}%
      \expandafter\def\csname LT7\endcsname{\color[rgb]{1,0.3,0}}%
      \expandafter\def\csname LT8\endcsname{\color[rgb]{0.5,0.5,0.5}}%
    \else
      \def\colorrgb#1{\color{black}}%
      \def\colorgray#1{\color[gray]{#1}}%
      \expandafter\def\csname LTw\endcsname{\color{white}}%
      \expandafter\def\csname LTb\endcsname{\color{black}}%
      \expandafter\def\csname LTa\endcsname{\color{black}}%
      \expandafter\def\csname LT0\endcsname{\color{black}}%
      \expandafter\def\csname LT1\endcsname{\color{black}}%
      \expandafter\def\csname LT2\endcsname{\color{black}}%
      \expandafter\def\csname LT3\endcsname{\color{black}}%
      \expandafter\def\csname LT4\endcsname{\color{black}}%
      \expandafter\def\csname LT5\endcsname{\color{black}}%
      \expandafter\def\csname LT6\endcsname{\color{black}}%
      \expandafter\def\csname LT7\endcsname{\color{black}}%
      \expandafter\def\csname LT8\endcsname{\color{black}}%
    \fi
  \fi
  \setlength{\unitlength}{0.0500bp}%
  \begin{picture}(7200.00,3024.00)%
    \gplgaddtomacro\gplbacktext{%
      \csname LTb\endcsname%
      \put(990,660){\makebox(0,0)[r]{\strut{} 0}}%
      \put(990,1184){\makebox(0,0)[r]{\strut{} 0.4}}%
      \put(990,1709){\makebox(0,0)[r]{\strut{} 0.8}}%
      \put(990,2233){\makebox(0,0)[r]{\strut{} 1.2}}%
      \csname LTb\endcsname%
      \put(1122,440){\makebox(0,0){\strut{} 0}}%
      \csname LTb\endcsname%
      \put(2073,440){\makebox(0,0){\strut{} 0.5}}%
      \csname LTb\endcsname%
      \put(3023,440){\makebox(0,0){\strut{} 1}}%
      \csname LTb\endcsname%
      \put(3974,440){\makebox(0,0){\strut{} 1.5}}%
      \csname LTb\endcsname%
      \put(4925,440){\makebox(0,0){\strut{} 2}}%
      \csname LTb\endcsname%
      \put(5875,440){\makebox(0,0){\strut{} 2.5}}%
      \csname LTb\endcsname%
      \put(6826,440){\makebox(0,0){\strut{} 3}}%
      \put(220,1512){\rotatebox{90}{\makebox(0,0){\strut{}$r\ut{(m)}$}}}%
      \put(3974,110){\makebox(0,0){\strut{}$t\ut{(s)}$}}%
      \put(3974,2694){\makebox(0,0){\strut{}(b)}}%
    }%
    \gplgaddtomacro\gplfronttext{%
    }%
    \gplbacktext
    \put(0,0){\includegraphics{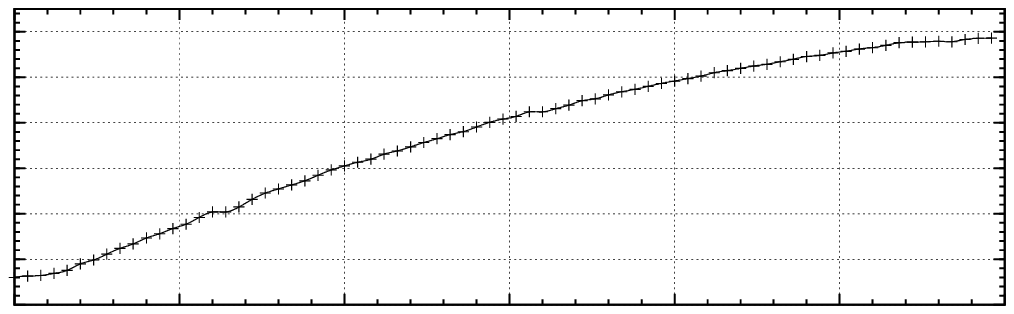}}%
    \gplfronttext
  \end{picture}%
\endgroup

%% file: Figure11c.tex
\begingroup
  \makeatletter
  \providecommand\color[2][]{%
    \GenericError{(gnuplot) \space\space\space\@spaces}{%
      Package color not loaded in conjunction with
      terminal option `colourtext'%
    }{See the gnuplot documentation for explanation.%
    }{Either use 'blacktext' in gnuplot or load the package
      color.sty in LaTeX.}%
    \renewcommand\color[2][]{}%
  }%
  \providecommand\includegraphics[2][]{%
    \GenericError{(gnuplot) \space\space\space\@spaces}{%
      Package graphicx or graphics not loaded%
    }{See the gnuplot documentation for explanation.%
    }{The gnuplot epslatex terminal needs graphicx.sty or graphics.sty.}%
    \renewcommand\includegraphics[2][]{}%
  }%
  \providecommand\rotatebox[2]{#2}%
  \@ifundefined{ifGPcolor}{%
    \newif\ifGPcolor
    \GPcolorfalse
  }{}%
  \@ifundefined{ifGPblacktext}{%
    \newif\ifGPblacktext
    \GPblacktexttrue
  }{}%
  \let\gplgaddtomacro\g@addto@macro
  \gdef\gplbacktext{}%
  \gdef\gplfronttext{}%
  \makeatother
  \ifGPblacktext
    \def\colorrgb#1{}%
    \def\colorgray#1{}%
  \else
    \ifGPcolor
      \def\colorrgb#1{\color[rgb]{#1}}%
      \def\colorgray#1{\color[gray]{#1}}%
      \expandafter\def\csname LTw\endcsname{\color{white}}%
      \expandafter\def\csname LTb\endcsname{\color{black}}%
      \expandafter\def\csname LTa\endcsname{\color{black}}%
      \expandafter\def\csname LT0\endcsname{\color[rgb]{1,0,0}}%
      \expandafter\def\csname LT1\endcsname{\color[rgb]{0,1,0}}%
      \expandafter\def\csname LT2\endcsname{\color[rgb]{0,0,1}}%
      \expandafter\def\csname LT3\endcsname{\color[rgb]{1,0,1}}%
      \expandafter\def\csname LT4\endcsname{\color[rgb]{0,1,1}}%
      \expandafter\def\csname LT5\endcsname{\color[rgb]{1,1,0}}%
      \expandafter\def\csname LT6\endcsname{\color[rgb]{0,0,0}}%
      \expandafter\def\csname LT7\endcsname{\color[rgb]{1,0.3,0}}%
      \expandafter\def\csname LT8\endcsname{\color[rgb]{0.5,0.5,0.5}}%
    \else
      \def\colorrgb#1{\color{black}}%
      \def\colorgray#1{\color[gray]{#1}}%
      \expandafter\def\csname LTw\endcsname{\color{white}}%
      \expandafter\def\csname LTb\endcsname{\color{black}}%
      \expandafter\def\csname LTa\endcsname{\color{black}}%
      \expandafter\def\csname LT0\endcsname{\color{black}}%
      \expandafter\def\csname LT1\endcsname{\color{black}}%
      \expandafter\def\csname LT2\endcsname{\color{black}}%
      \expandafter\def\csname LT3\endcsname{\color{black}}%
      \expandafter\def\csname LT4\endcsname{\color{black}}%
      \expandafter\def\csname LT5\endcsname{\color{black}}%
      \expandafter\def\csname LT6\endcsname{\color{black}}%
      \expandafter\def\csname LT7\endcsname{\color{black}}%
      \expandafter\def\csname LT8\endcsname{\color{black}}%
    \fi
  \fi
  \setlength{\unitlength}{0.0500bp}%
  \begin{picture}(7200.00,3024.00)%
    \gplgaddtomacro\gplbacktext{%
      \csname LTb\endcsname%
      \put(990,660){\makebox(0,0)[r]{\strut{}-1}}%
      \put(990,1228){\makebox(0,0)[r]{\strut{} 0}}%
      \put(990,1796){\makebox(0,0)[r]{\strut{} 1}}%
      \put(990,2364){\makebox(0,0)[r]{\strut{} 2}}%
      \csname LTb\endcsname%
      \put(1122,440){\makebox(0,0){\strut{} 0}}%
      \csname LTb\endcsname%
      \put(2073,440){\makebox(0,0){\strut{} 0.5}}%
      \csname LTb\endcsname%
      \put(3023,440){\makebox(0,0){\strut{} 1}}%
      \csname LTb\endcsname%
      \put(3974,440){\makebox(0,0){\strut{} 1.5}}%
      \csname LTb\endcsname%
      \put(4925,440){\makebox(0,0){\strut{} 2}}%
      \csname LTb\endcsname%
      \put(5875,440){\makebox(0,0){\strut{} 2.5}}%
      \csname LTb\endcsname%
      \put(6826,440){\makebox(0,0){\strut{} 3}}%
      \put(220,1512){\rotatebox{90}{\makebox(0,0){\strut{}$\theta$ (rad)}}}%
      \put(3974,110){\makebox(0,0){\strut{}$t$ (s)}}%
      \put(3974,2694){\makebox(0,0){\strut{}(c)}}%
    }%
    \gplgaddtomacro\gplfronttext{%
    }%
    \gplbacktext
    \put(0,0){\includegraphics{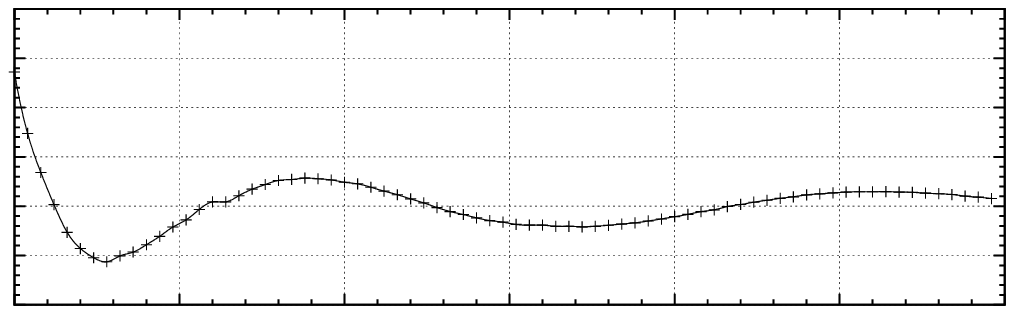}}%
    \gplfronttext
  \end{picture}%
\endgroup

%% file: Figure12a.tex
\begingroup
  \makeatletter
  \providecommand\color[2][]{%
    \GenericError{(gnuplot) \space\space\space\@spaces}{%
      Package color not loaded in conjunction with
      terminal option `colourtext'%
    }{See the gnuplot documentation for explanation.%
    }{Either use 'blacktext' in gnuplot or load the package
      color.sty in LaTeX.}%
    \renewcommand\color[2][]{}%
  }%
  \providecommand\includegraphics[2][]{%
    \GenericError{(gnuplot) \space\space\space\@spaces}{%
      Package graphicx or graphics not loaded%
    }{See the gnuplot documentation for explanation.%
    }{The gnuplot epslatex terminal needs graphicx.sty or graphics.sty.}%
    \renewcommand\includegraphics[2][]{}%
  }%
  \providecommand\rotatebox[2]{#2}%
  \@ifundefined{ifGPcolor}{%
    \newif\ifGPcolor
    \GPcolorfalse
  }{}%
  \@ifundefined{ifGPblacktext}{%
    \newif\ifGPblacktext
    \GPblacktexttrue
  }{}%
  \let\gplgaddtomacro\g@addto@macro
  \gdef\gplbacktext{}%
  \gdef\gplfronttext{}%
  \makeatother
  \ifGPblacktext
    \def\colorrgb#1{}%
    \def\colorgray#1{}%
  \else
    \ifGPcolor
      \def\colorrgb#1{\color[rgb]{#1}}%
      \def\colorgray#1{\color[gray]{#1}}%
      \expandafter\def\csname LTw\endcsname{\color{white}}%
      \expandafter\def\csname LTb\endcsname{\color{black}}%
      \expandafter\def\csname LTa\endcsname{\color{black}}%
      \expandafter\def\csname LT0\endcsname{\color[rgb]{1,0,0}}%
      \expandafter\def\csname LT1\endcsname{\color[rgb]{0,1,0}}%
      \expandafter\def\csname LT2\endcsname{\color[rgb]{0,0,1}}%
      \expandafter\def\csname LT3\endcsname{\color[rgb]{1,0,1}}%
      \expandafter\def\csname LT4\endcsname{\color[rgb]{0,1,1}}%
      \expandafter\def\csname LT5\endcsname{\color[rgb]{1,1,0}}%
      \expandafter\def\csname LT6\endcsname{\color[rgb]{0,0,0}}%
      \expandafter\def\csname LT7\endcsname{\color[rgb]{1,0.3,0}}%
      \expandafter\def\csname LT8\endcsname{\color[rgb]{0.5,0.5,0.5}}%
    \else
      \def\colorrgb#1{\color{black}}%
      \def\colorgray#1{\color[gray]{#1}}%
      \expandafter\def\csname LTw\endcsname{\color{white}}%
      \expandafter\def\csname LTb\endcsname{\color{black}}%
      \expandafter\def\csname LTa\endcsname{\color{black}}%
      \expandafter\def\csname LT0\endcsname{\color{black}}%
      \expandafter\def\csname LT1\endcsname{\color{black}}%
      \expandafter\def\csname LT2\endcsname{\color{black}}%
      \expandafter\def\csname LT3\endcsname{\color{black}}%
      \expandafter\def\csname LT4\endcsname{\color{black}}%
      \expandafter\def\csname LT5\endcsname{\color{black}}%
      \expandafter\def\csname LT6\endcsname{\color{black}}%
      \expandafter\def\csname LT7\endcsname{\color{black}}%
      \expandafter\def\csname LT8\endcsname{\color{black}}%
    \fi
  \fi
  \setlength{\unitlength}{0.0500bp}%
  \begin{picture}(5760.00,4032.00)%
    \gplgaddtomacro\gplbacktext{%
      \csname LTb\endcsname%
      \put(990,3033){\makebox(0,0)[r]{\strut{}-0.2}}%
      \put(990,2355){\makebox(0,0)[r]{\strut{} 0}}%
      \put(990,1677){\makebox(0,0)[r]{\strut{} 0.2}}%
      \put(990,999){\makebox(0,0)[r]{\strut{} 0.4}}%
      \put(1122,440){\makebox(0,0){\strut{}-0.6}}%
      \put(1975,440){\makebox(0,0){\strut{}-0.4}}%
      \put(2828,440){\makebox(0,0){\strut{}-0.2}}%
      \put(3680,440){\makebox(0,0){\strut{} 0}}%
      \put(4533,440){\makebox(0,0){\strut{} 0.2}}%
      \put(5386,440){\makebox(0,0){\strut{} 0.4}}%
      \put(220,2016){\rotatebox{90}{\makebox(0,0){\strut{}$z\,\mathrm{(m)}$}}}%
      \put(3254,110){\makebox(0,0){\strut{}$y\,\mathrm{(m)}$}}%
      \put(3254,3702){\makebox(0,0){\strut{}(a)}}%
    }%
    \gplgaddtomacro\gplfronttext{%
    }%
    \gplbacktext
    \put(0,0){\includegraphics{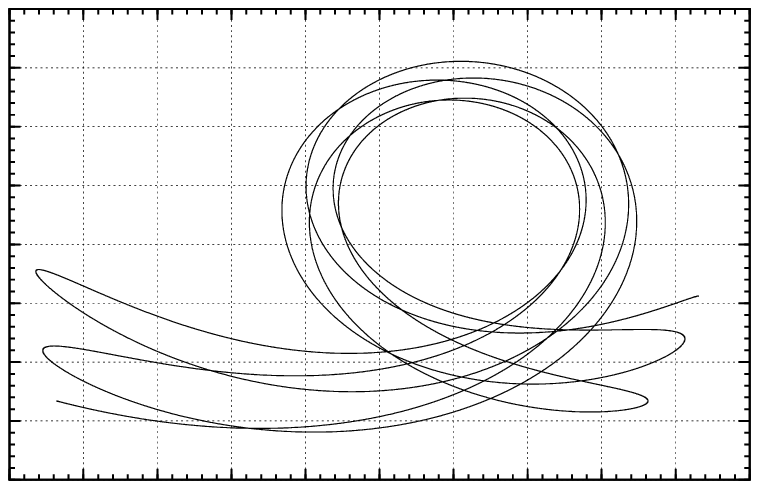}}%
    \put(3680,2355){\pscircle(0,0){0.25}}
    \gplfronttext
  \end{picture}%
\endgroup

%% file: Figure12b.tex
\begingroup
  \makeatletter
  \providecommand\color[2][]{%
    \GenericError{(gnuplot) \space\space\space\@spaces}{%
      Package color not loaded in conjunction with
      terminal option `colourtext'%
    }{See the gnuplot documentation for explanation.%
    }{Either use 'blacktext' in gnuplot or load the package
      color.sty in LaTeX.}%
    \renewcommand\color[2][]{}%
  }%
  \providecommand\includegraphics[2][]{%
    \GenericError{(gnuplot) \space\space\space\@spaces}{%
      Package graphicx or graphics not loaded%
    }{See the gnuplot documentation for explanation.%
    }{The gnuplot epslatex terminal needs graphicx.sty or graphics.sty.}%
    \renewcommand\includegraphics[2][]{}%
  }%
  \providecommand\rotatebox[2]{#2}%
  \@ifundefined{ifGPcolor}{%
    \newif\ifGPcolor
    \GPcolorfalse
  }{}%
  \@ifundefined{ifGPblacktext}{%
    \newif\ifGPblacktext
    \GPblacktexttrue
  }{}%
  \let\gplgaddtomacro\g@addto@macro
  \gdef\gplbacktext{}%
  \gdef\gplfronttext{}%
  \makeatother
  \ifGPblacktext
    \def\colorrgb#1{}%
    \def\colorgray#1{}%
  \else
    \ifGPcolor
      \def\colorrgb#1{\color[rgb]{#1}}%
      \def\colorgray#1{\color[gray]{#1}}%
      \expandafter\def\csname LTw\endcsname{\color{white}}%
      \expandafter\def\csname LTb\endcsname{\color{black}}%
      \expandafter\def\csname LTa\endcsname{\color{black}}%
      \expandafter\def\csname LT0\endcsname{\color[rgb]{1,0,0}}%
      \expandafter\def\csname LT1\endcsname{\color[rgb]{0,1,0}}%
      \expandafter\def\csname LT2\endcsname{\color[rgb]{0,0,1}}%
      \expandafter\def\csname LT3\endcsname{\color[rgb]{1,0,1}}%
      \expandafter\def\csname LT4\endcsname{\color[rgb]{0,1,1}}%
      \expandafter\def\csname LT5\endcsname{\color[rgb]{1,1,0}}%
      \expandafter\def\csname LT6\endcsname{\color[rgb]{0,0,0}}%
      \expandafter\def\csname LT7\endcsname{\color[rgb]{1,0.3,0}}%
      \expandafter\def\csname LT8\endcsname{\color[rgb]{0.5,0.5,0.5}}%
    \else
      \def\colorrgb#1{\color{black}}%
      \def\colorgray#1{\color[gray]{#1}}%
      \expandafter\def\csname LTw\endcsname{\color{white}}%
      \expandafter\def\csname LTb\endcsname{\color{black}}%
      \expandafter\def\csname LTa\endcsname{\color{black}}%
      \expandafter\def\csname LT0\endcsname{\color{black}}%
      \expandafter\def\csname LT1\endcsname{\color{black}}%
      \expandafter\def\csname LT2\endcsname{\color{black}}%
      \expandafter\def\csname LT3\endcsname{\color{black}}%
      \expandafter\def\csname LT4\endcsname{\color{black}}%
      \expandafter\def\csname LT5\endcsname{\color{black}}%
      \expandafter\def\csname LT6\endcsname{\color{black}}%
      \expandafter\def\csname LT7\endcsname{\color{black}}%
      \expandafter\def\csname LT8\endcsname{\color{black}}%
    \fi
  \fi
  \setlength{\unitlength}{0.0500bp}%
  \begin{picture}(5760.00,4032.00)%
    \gplgaddtomacro\gplbacktext{%
      \csname LTb\endcsname%
      \put(990,2920){\makebox(0,0)[r]{\strut{} 0}}%
      \put(990,2016){\makebox(0,0)[r]{\strut{} 0.2}}%
      \put(990,1112){\makebox(0,0)[r]{\strut{} 0.4}}%
      \put(1122,440){\makebox(0,0){\strut{}-0.6}}%
      \csname LTb\endcsname%
      \put(1925,440){\makebox(0,0){\strut{}-0.4}}%
      \csname LTb\endcsname%
      \put(2729,440){\makebox(0,0){\strut{}-0.2}}%
      \csname LTb\endcsname%
      \put(3532,440){\makebox(0,0){\strut{} 0}}%
      \csname LTb\endcsname%
      \put(4335,440){\makebox(0,0){\strut{} 0.2}}%
      \csname LTb\endcsname%
      \put(5138,440){\makebox(0,0){\strut{} 0.4}}%
      \put(220,2016){\rotatebox{90}{\makebox(0,0){\strut{}$z\,\mathrm{(m)}$}}}%
      \put(5759,2016){\rotatebox{90}{\makebox(0,0){\strut{}}}}%
      \put(3331,110){\makebox(0,0){\strut{}$y\,\mathrm{(m)}$}}%
      \put(3331,3702){\makebox(0,0){\strut{}(b)}}%
      \put(3331,3261){\makebox(0,0){\strut{}}}%
      \put(330,110){\makebox(0,0)[l]{\strut{}}}%
    }%
    \gplgaddtomacro\gplfronttext{%
    }%
    \gplbacktext
    \put(0,0){\includegraphics{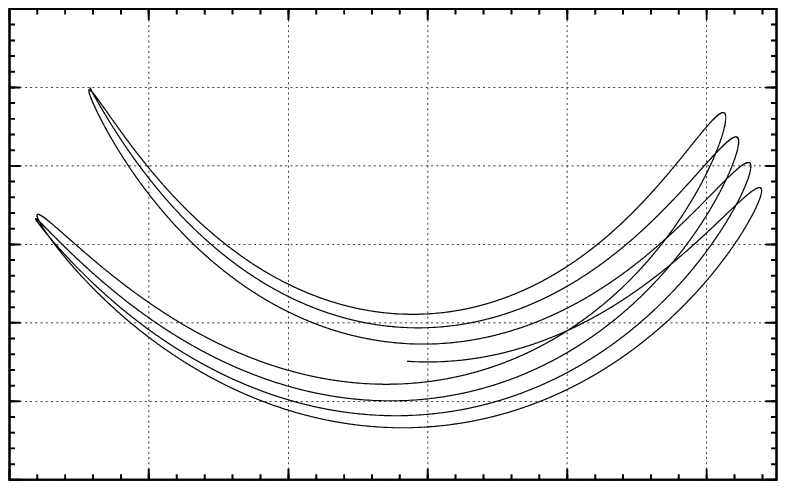}}%
    \gplfronttext
  \end{picture}%
\endgroup

%% file: Figure12c.tex
\begingroup
  \makeatletter
  \providecommand\color[2][]{%
    \GenericError{(gnuplot) \space\space\space\@spaces}{%
      Package color not loaded in conjunction with
      terminal option `colourtext'%
    }{See the gnuplot documentation for explanation.%
    }{Either use 'blacktext' in gnuplot or load the package
      color.sty in LaTeX.}%
    \renewcommand\color[2][]{}%
  }%
  \providecommand\includegraphics[2][]{%
    \GenericError{(gnuplot) \space\space\space\@spaces}{%
      Package graphicx or graphics not loaded%
    }{See the gnuplot documentation for explanation.%
    }{The gnuplot epslatex terminal needs graphicx.sty or graphics.sty.}%
    \renewcommand\includegraphics[2][]{}%
  }%
  \providecommand\rotatebox[2]{#2}%
  \@ifundefined{ifGPcolor}{%
    \newif\ifGPcolor
    \GPcolorfalse
  }{}%
  \@ifundefined{ifGPblacktext}{%
    \newif\ifGPblacktext
    \GPblacktexttrue
  }{}%
  \let\gplgaddtomacro\g@addto@macro
  \gdef\gplbacktext{}%
  \gdef\gplfronttext{}%
  \makeatother
  \ifGPblacktext
    \def\colorrgb#1{}%
    \def\colorgray#1{}%
  \else
    \ifGPcolor
      \def\colorrgb#1{\color[rgb]{#1}}%
      \def\colorgray#1{\color[gray]{#1}}%
      \expandafter\def\csname LTw\endcsname{\color{white}}%
      \expandafter\def\csname LTb\endcsname{\color{black}}%
      \expandafter\def\csname LTa\endcsname{\color{black}}%
      \expandafter\def\csname LT0\endcsname{\color[rgb]{1,0,0}}%
      \expandafter\def\csname LT1\endcsname{\color[rgb]{0,1,0}}%
      \expandafter\def\csname LT2\endcsname{\color[rgb]{0,0,1}}%
      \expandafter\def\csname LT3\endcsname{\color[rgb]{1,0,1}}%
      \expandafter\def\csname LT4\endcsname{\color[rgb]{0,1,1}}%
      \expandafter\def\csname LT5\endcsname{\color[rgb]{1,1,0}}%
      \expandafter\def\csname LT6\endcsname{\color[rgb]{0,0,0}}%
      \expandafter\def\csname LT7\endcsname{\color[rgb]{1,0.3,0}}%
      \expandafter\def\csname LT8\endcsname{\color[rgb]{0.5,0.5,0.5}}%
    \else
      \def\colorrgb#1{\color{black}}%
      \def\colorgray#1{\color[gray]{#1}}%
      \expandafter\def\csname LTw\endcsname{\color{white}}%
      \expandafter\def\csname LTb\endcsname{\color{black}}%
      \expandafter\def\csname LTa\endcsname{\color{black}}%
      \expandafter\def\csname LT0\endcsname{\color{black}}%
      \expandafter\def\csname LT1\endcsname{\color{black}}%
      \expandafter\def\csname LT2\endcsname{\color{black}}%
      \expandafter\def\csname LT3\endcsname{\color{black}}%
      \expandafter\def\csname LT4\endcsname{\color{black}}%
      \expandafter\def\csname LT5\endcsname{\color{black}}%
      \expandafter\def\csname LT6\endcsname{\color{black}}%
      \expandafter\def\csname LT7\endcsname{\color{black}}%
      \expandafter\def\csname LT8\endcsname{\color{black}}%
    \fi
  \fi
  \setlength{\unitlength}{0.0500bp}%
  \begin{picture}(5760.00,4032.00)%
    \gplgaddtomacro\gplbacktext{%
      \csname LTb\endcsname%
      \put(990,3191){\makebox(0,0)[r]{\strut{} 0}}%
      \put(990,2468){\makebox(0,0)[r]{\strut{} 0.4}}%
      \put(990,1745){\makebox(0,0)[r]{\strut{} 0.8}}%
      \put(990,1022){\makebox(0,0)[r]{\strut{} 1.2}}%
      \put(1564,440){\makebox(0,0){\strut{}-0.2}}%
      \csname LTb\endcsname%
      \put(2447,440){\makebox(0,0){\strut{}-0.1}}%
      \csname LTb\endcsname%
      \put(3331,440){\makebox(0,0){\strut{} 0}}%
      \csname LTb\endcsname%
      \put(4215,440){\makebox(0,0){\strut{} 0.1}}%
      \csname LTb\endcsname%
      \put(5098,440){\makebox(0,0){\strut{} 0.2}}%
      \put(220,2016){\rotatebox{90}{\makebox(0,0){\strut{}$z\,\mathrm{(m)}$}}}%
      \put(5759,2016){\rotatebox{90}{\makebox(0,0){\strut{}}}}%
      \put(3331,110){\makebox(0,0){\strut{}$y\,\mathrm{(m)}$}}%
      \put(3331,3702){\makebox(0,0){\strut{}(c)}}%
      \put(3331,3261){\makebox(0,0){\strut{}}}%
      \put(330,110){\makebox(0,0)[l]{\strut{}}}%
    }%
    \gplgaddtomacro\gplfronttext{%
    }%
    \gplbacktext
    \put(0,0){\includegraphics{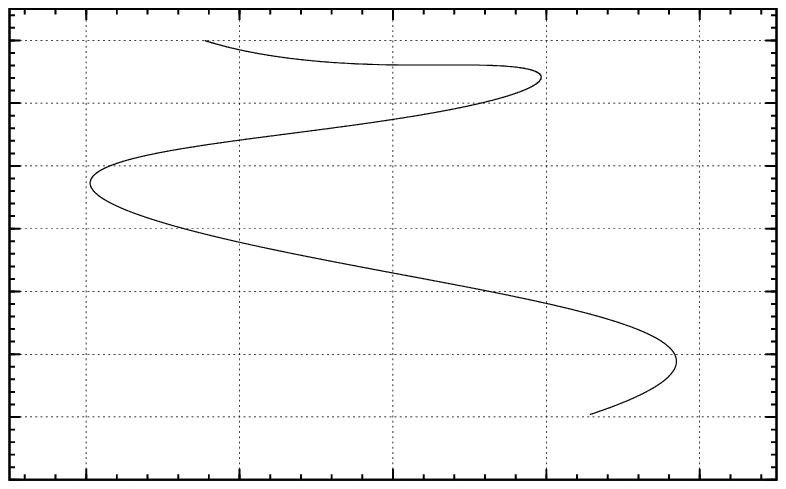}}%
    \gplfronttext
  \end{picture}%
\endgroup

%% file: Figure13a.tex
\begingroup
  \makeatletter
  \providecommand\color[2][]{%
    \GenericError{(gnuplot) \space\space\space\@spaces}{%
      Package color not loaded in conjunction with
      terminal option `colourtext'%
    }{See the gnuplot documentation for explanation.%
    }{Either use 'blacktext' in gnuplot or load the package
      color.sty in LaTeX.}%
    \renewcommand\color[2][]{}%
  }%
  \providecommand\includegraphics[2][]{%
    \GenericError{(gnuplot) \space\space\space\@spaces}{%
      Package graphicx or graphics not loaded%
    }{See the gnuplot documentation for explanation.%
    }{The gnuplot epslatex terminal needs graphicx.sty or graphics.sty.}%
    \renewcommand\includegraphics[2][]{}%
  }%
  \providecommand\rotatebox[2]{#2}%
  \@ifundefined{ifGPcolor}{%
    \newif\ifGPcolor
    \GPcolorfalse
  }{}%
  \@ifundefined{ifGPblacktext}{%
    \newif\ifGPblacktext
    \GPblacktexttrue
  }{}%
  \let\gplgaddtomacro\g@addto@macro
  \gdef\gplbacktext{}%
  \gdef\gplfronttext{}%
  \makeatother
  \ifGPblacktext
    \def\colorrgb#1{}%
    \def\colorgray#1{}%
  \else
    \ifGPcolor
      \def\colorrgb#1{\color[rgb]{#1}}%
      \def\colorgray#1{\color[gray]{#1}}%
      \expandafter\def\csname LTw\endcsname{\color{white}}%
      \expandafter\def\csname LTb\endcsname{\color{black}}%
      \expandafter\def\csname LTa\endcsname{\color{black}}%
      \expandafter\def\csname LT0\endcsname{\color[rgb]{1,0,0}}%
      \expandafter\def\csname LT1\endcsname{\color[rgb]{0,1,0}}%
      \expandafter\def\csname LT2\endcsname{\color[rgb]{0,0,1}}%
      \expandafter\def\csname LT3\endcsname{\color[rgb]{1,0,1}}%
      \expandafter\def\csname LT4\endcsname{\color[rgb]{0,1,1}}%
      \expandafter\def\csname LT5\endcsname{\color[rgb]{1,1,0}}%
      \expandafter\def\csname LT6\endcsname{\color[rgb]{0,0,0}}%
      \expandafter\def\csname LT7\endcsname{\color[rgb]{1,0.3,0}}%
      \expandafter\def\csname LT8\endcsname{\color[rgb]{0.5,0.5,0.5}}%
    \else
      \def\colorrgb#1{\color{black}}%
      \def\colorgray#1{\color[gray]{#1}}%
      \expandafter\def\csname LTw\endcsname{\color{white}}%
      \expandafter\def\csname LTb\endcsname{\color{black}}%
      \expandafter\def\csname LTa\endcsname{\color{black}}%
      \expandafter\def\csname LT0\endcsname{\color{black}}%
      \expandafter\def\csname LT1\endcsname{\color{black}}%
      \expandafter\def\csname LT2\endcsname{\color{black}}%
      \expandafter\def\csname LT3\endcsname{\color{black}}%
      \expandafter\def\csname LT4\endcsname{\color{black}}%
      \expandafter\def\csname LT5\endcsname{\color{black}}%
      \expandafter\def\csname LT6\endcsname{\color{black}}%
      \expandafter\def\csname LT7\endcsname{\color{black}}%
      \expandafter\def\csname LT8\endcsname{\color{black}}%
    \fi
  \fi
  \setlength{\unitlength}{0.0500bp}%
  \begin{picture}(3600.00,2520.00)%
    \gplgaddtomacro\gplbacktext{%
      \csname LTb\endcsname%
      \csname LTb\endcsname%
      \put(990,1727){\makebox(0,0)[r]{\strut{}-0.2}}%
      \csname LTb\endcsname%
      \csname LTb\endcsname%
      \put(990,1460){\makebox(0,0)[r]{\strut{} 0}}%
      \csname LTb\endcsname%
      \csname LTb\endcsname%
      \put(990,1193){\makebox(0,0)[r]{\strut{} 0.2}}%
      \csname LTb\endcsname%
      \csname LTb\endcsname%
      \put(990,927){\makebox(0,0)[r]{\strut{} 0.4}}%
      \csname LTb\endcsname%
      \csname LTb\endcsname%
      \csname LTb\endcsname%
      \put(1409,440){\makebox(0,0){\strut{}-0.4}}%
      \csname LTb\endcsname%
      \put(1791,440){\makebox(0,0){\strut{}-0.2}}%
      \csname LTb\endcsname%
      \put(2174,440){\makebox(0,0){\strut{} 0}}%
      \csname LTb\endcsname%
      \put(2557,440){\makebox(0,0){\strut{} 0.2}}%
      \csname LTb\endcsname%
      \put(2939,440){\makebox(0,0){\strut{} 0.4}}%
      \put(220,1260){\rotatebox{90}{\makebox(0,0){\strut{}$z$ (m)}}}%
      \put(2174,110){\makebox(0,0){\strut{}$y$ (m)}}%
      \put(2174,2190){\makebox(0,0){\strut{}(a)}}%
    }%
    \gplgaddtomacro\gplfronttext{%
    }%
    \gplbacktext
    \put(0,0){\includegraphics{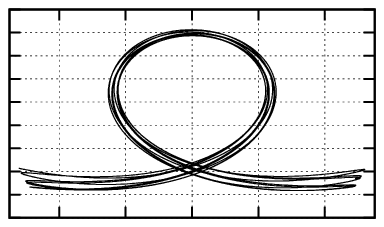}}%
    \gplfronttext
  \end{picture}%
\endgroup

%% file: Figure13b.tex
\begingroup
  \makeatletter
  \providecommand\color[2][]{%
    \GenericError{(gnuplot) \space\space\space\@spaces}{%
      Package color not loaded in conjunction with
      terminal option `colourtext'%
    }{See the gnuplot documentation for explanation.%
    }{Either use 'blacktext' in gnuplot or load the package
      color.sty in LaTeX.}%
    \renewcommand\color[2][]{}%
  }%
  \providecommand\includegraphics[2][]{%
    \GenericError{(gnuplot) \space\space\space\@spaces}{%
      Package graphicx or graphics not loaded%
    }{See the gnuplot documentation for explanation.%
    }{The gnuplot epslatex terminal needs graphicx.sty or graphics.sty.}%
    \renewcommand\includegraphics[2][]{}%
  }%
  \providecommand\rotatebox[2]{#2}%
  \@ifundefined{ifGPcolor}{%
    \newif\ifGPcolor
    \GPcolorfalse
  }{}%
  \@ifundefined{ifGPblacktext}{%
    \newif\ifGPblacktext
    \GPblacktexttrue
  }{}%
  \let\gplgaddtomacro\g@addto@macro
  \gdef\gplbacktext{}%
  \gdef\gplfronttext{}%
  \makeatother
  \ifGPblacktext
    \def\colorrgb#1{}%
    \def\colorgray#1{}%
  \else
    \ifGPcolor
      \def\colorrgb#1{\color[rgb]{#1}}%
      \def\colorgray#1{\color[gray]{#1}}%
      \expandafter\def\csname LTw\endcsname{\color{white}}%
      \expandafter\def\csname LTb\endcsname{\color{black}}%
      \expandafter\def\csname LTa\endcsname{\color{black}}%
      \expandafter\def\csname LT0\endcsname{\color[rgb]{1,0,0}}%
      \expandafter\def\csname LT1\endcsname{\color[rgb]{0,1,0}}%
      \expandafter\def\csname LT2\endcsname{\color[rgb]{0,0,1}}%
      \expandafter\def\csname LT3\endcsname{\color[rgb]{1,0,1}}%
      \expandafter\def\csname LT4\endcsname{\color[rgb]{0,1,1}}%
      \expandafter\def\csname LT5\endcsname{\color[rgb]{1,1,0}}%
      \expandafter\def\csname LT6\endcsname{\color[rgb]{0,0,0}}%
      \expandafter\def\csname LT7\endcsname{\color[rgb]{1,0.3,0}}%
      \expandafter\def\csname LT8\endcsname{\color[rgb]{0.5,0.5,0.5}}%
    \else
      \def\colorrgb#1{\color{black}}%
      \def\colorgray#1{\color[gray]{#1}}%
      \expandafter\def\csname LTw\endcsname{\color{white}}%
      \expandafter\def\csname LTb\endcsname{\color{black}}%
      \expandafter\def\csname LTa\endcsname{\color{black}}%
      \expandafter\def\csname LT0\endcsname{\color{black}}%
      \expandafter\def\csname LT1\endcsname{\color{black}}%
      \expandafter\def\csname LT2\endcsname{\color{black}}%
      \expandafter\def\csname LT3\endcsname{\color{black}}%
      \expandafter\def\csname LT4\endcsname{\color{black}}%
      \expandafter\def\csname LT5\endcsname{\color{black}}%
      \expandafter\def\csname LT6\endcsname{\color{black}}%
      \expandafter\def\csname LT7\endcsname{\color{black}}%
      \expandafter\def\csname LT8\endcsname{\color{black}}%
    \fi
  \fi
  \setlength{\unitlength}{0.0500bp}%
  \begin{picture}(3600.00,2520.00)%
    \gplgaddtomacro\gplbacktext{%
      \csname LTb\endcsname%
      \csname LTb\endcsname%
      \put(990,1727){\makebox(0,0)[r]{\strut{}-0.2}}%
      \csname LTb\endcsname%
      \csname LTb\endcsname%
      \put(990,1460){\makebox(0,0)[r]{\strut{} 0}}%
      \csname LTb\endcsname%
      \csname LTb\endcsname%
      \put(990,1193){\makebox(0,0)[r]{\strut{} 0.2}}%
      \csname LTb\endcsname%
      \csname LTb\endcsname%
      \put(990,927){\makebox(0,0)[r]{\strut{} 0.4}}%
      \csname LTb\endcsname%
      \csname LTb\endcsname%
      \csname LTb\endcsname%
      \csname LTb\endcsname%
      \put(1543,440){\makebox(0,0){\strut{}-0.4}}%
      \csname LTb\endcsname%
      \put(1964,440){\makebox(0,0){\strut{}-0.2}}%
      \csname LTb\endcsname%
      \put(2384,440){\makebox(0,0){\strut{} 0}}%
      \csname LTb\endcsname%
      \put(2805,440){\makebox(0,0){\strut{} 0.2}}%
      \csname LTb\endcsname%
      \put(220,1260){\rotatebox{90}{\makebox(0,0){\strut{}$z$ (m)}}}%
      \put(2174,110){\makebox(0,0){\strut{}$y$ (m)}}%
      \put(2174,2190){\makebox(0,0){\strut{}(b)}}%
    }%
    \gplgaddtomacro\gplfronttext{%
    }%
    \gplbacktext
    \put(0,0){\includegraphics{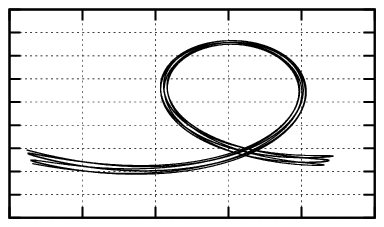}}%
    \gplfronttext
  \end{picture}%
\endgroup

%% file: Figure13c.tex
\begingroup
  \makeatletter
  \providecommand\color[2][]{%
    \GenericError{(gnuplot) \space\space\space\@spaces}{%
      Package color not loaded in conjunction with
      terminal option `colourtext'%
    }{See the gnuplot documentation for explanation.%
    }{Either use 'blacktext' in gnuplot or load the package
      color.sty in LaTeX.}%
    \renewcommand\color[2][]{}%
  }%
  \providecommand\includegraphics[2][]{%
    \GenericError{(gnuplot) \space\space\space\@spaces}{%
      Package graphicx or graphics not loaded%
    }{See the gnuplot documentation for explanation.%
    }{The gnuplot epslatex terminal needs graphicx.sty or graphics.sty.}%
    \renewcommand\includegraphics[2][]{}%
  }%
  \providecommand\rotatebox[2]{#2}%
  \@ifundefined{ifGPcolor}{%
    \newif\ifGPcolor
    \GPcolorfalse
  }{}%
  \@ifundefined{ifGPblacktext}{%
    \newif\ifGPblacktext
    \GPblacktexttrue
  }{}%
  \let\gplgaddtomacro\g@addto@macro
  \gdef\gplbacktext{}%
  \gdef\gplfronttext{}%
  \makeatother
  \ifGPblacktext
    \def\colorrgb#1{}%
    \def\colorgray#1{}%
  \else
    \ifGPcolor
      \def\colorrgb#1{\color[rgb]{#1}}%
      \def\colorgray#1{\color[gray]{#1}}%
      \expandafter\def\csname LTw\endcsname{\color{white}}%
      \expandafter\def\csname LTb\endcsname{\color{black}}%
      \expandafter\def\csname LTa\endcsname{\color{black}}%
      \expandafter\def\csname LT0\endcsname{\color[rgb]{1,0,0}}%
      \expandafter\def\csname LT1\endcsname{\color[rgb]{0,1,0}}%
      \expandafter\def\csname LT2\endcsname{\color[rgb]{0,0,1}}%
      \expandafter\def\csname LT3\endcsname{\color[rgb]{1,0,1}}%
      \expandafter\def\csname LT4\endcsname{\color[rgb]{0,1,1}}%
      \expandafter\def\csname LT5\endcsname{\color[rgb]{1,1,0}}%
      \expandafter\def\csname LT6\endcsname{\color[rgb]{0,0,0}}%
      \expandafter\def\csname LT7\endcsname{\color[rgb]{1,0.3,0}}%
      \expandafter\def\csname LT8\endcsname{\color[rgb]{0.5,0.5,0.5}}%
    \else
      \def\colorrgb#1{\color{black}}%
      \def\colorgray#1{\color[gray]{#1}}%
      \expandafter\def\csname LTw\endcsname{\color{white}}%
      \expandafter\def\csname LTb\endcsname{\color{black}}%
      \expandafter\def\csname LTa\endcsname{\color{black}}%
      \expandafter\def\csname LT0\endcsname{\color{black}}%
      \expandafter\def\csname LT1\endcsname{\color{black}}%
      \expandafter\def\csname LT2\endcsname{\color{black}}%
      \expandafter\def\csname LT3\endcsname{\color{black}}%
      \expandafter\def\csname LT4\endcsname{\color{black}}%
      \expandafter\def\csname LT5\endcsname{\color{black}}%
      \expandafter\def\csname LT6\endcsname{\color{black}}%
      \expandafter\def\csname LT7\endcsname{\color{black}}%
      \expandafter\def\csname LT8\endcsname{\color{black}}%
    \fi
  \fi
  \setlength{\unitlength}{0.0500bp}%
  \begin{picture}(3600.00,2520.00)%
    \gplgaddtomacro\gplbacktext{%
      \csname LTb\endcsname%
      \csname LTb\endcsname%
      \put(990,1727){\makebox(0,0)[r]{\strut{}-0.2}}%
      \csname LTb\endcsname%
      \csname LTb\endcsname%
      \put(990,1460){\makebox(0,0)[r]{\strut{} 0}}%
      \csname LTb\endcsname%
      \csname LTb\endcsname%
      \put(990,1193){\makebox(0,0)[r]{\strut{} 0.2}}%
      \csname LTb\endcsname%
      \csname LTb\endcsname%
      \put(990,927){\makebox(0,0)[r]{\strut{} 0.4}}%
      \csname LTb\endcsname%
      \csname LTb\endcsname%
      \csname LTb\endcsname%
      \put(1122,440){\makebox(0,0){\strut{}-0.6}}%
      \csname LTb\endcsname%
      \csname LTb\endcsname%
      \put(1590,440){\makebox(0,0){\strut{}-0.4}}%
      \csname LTb\endcsname%
      \csname LTb\endcsname%
      \put(2057,440){\makebox(0,0){\strut{}-0.2}}%
      \csname LTb\endcsname%
      \csname LTb\endcsname%
      \put(2525,440){\makebox(0,0){\strut{} 0}}%
      \csname LTb\endcsname%
      \csname LTb\endcsname%
      \put(2992,440){\makebox(0,0){\strut{} 0.2}}%
      \csname LTb\endcsname%
      \put(220,1260){\rotatebox{90}{\makebox(0,0){\strut{}$z$ (m)}}}%
      \put(2174,110){\makebox(0,0){\strut{}$y$ (m)}}%
      \put(2174,2190){\makebox(0,0){\strut{}(c)}}%
    }%
    \gplgaddtomacro\gplfronttext{%
    }%
    \gplbacktext
    \put(0,0){\includegraphics{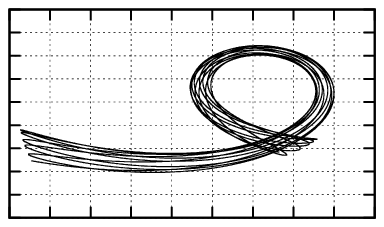}}%
    \gplfronttext
  \end{picture}%
\endgroup

%% file: Figure13d.tex
\begingroup
  \makeatletter
  \providecommand\color[2][]{%
    \GenericError{(gnuplot) \space\space\space\@spaces}{%
      Package color not loaded in conjunction with
      terminal option `colourtext'%
    }{See the gnuplot documentation for explanation.%
    }{Either use 'blacktext' in gnuplot or load the package
      color.sty in LaTeX.}%
    \renewcommand\color[2][]{}%
  }%
  \providecommand\includegraphics[2][]{%
    \GenericError{(gnuplot) \space\space\space\@spaces}{%
      Package graphicx or graphics not loaded%
    }{See the gnuplot documentation for explanation.%
    }{The gnuplot epslatex terminal needs graphicx.sty or graphics.sty.}%
    \renewcommand\includegraphics[2][]{}%
  }%
  \providecommand\rotatebox[2]{#2}%
  \@ifundefined{ifGPcolor}{%
    \newif\ifGPcolor
    \GPcolorfalse
  }{}%
  \@ifundefined{ifGPblacktext}{%
    \newif\ifGPblacktext
    \GPblacktexttrue
  }{}%
  \let\gplgaddtomacro\g@addto@macro
  \gdef\gplbacktext{}%
  \gdef\gplfronttext{}%
  \makeatother
  \ifGPblacktext
    \def\colorrgb#1{}%
    \def\colorgray#1{}%
  \else
    \ifGPcolor
      \def\colorrgb#1{\color[rgb]{#1}}%
      \def\colorgray#1{\color[gray]{#1}}%
      \expandafter\def\csname LTw\endcsname{\color{white}}%
      \expandafter\def\csname LTb\endcsname{\color{black}}%
      \expandafter\def\csname LTa\endcsname{\color{black}}%
      \expandafter\def\csname LT0\endcsname{\color[rgb]{1,0,0}}%
      \expandafter\def\csname LT1\endcsname{\color[rgb]{0,1,0}}%
      \expandafter\def\csname LT2\endcsname{\color[rgb]{0,0,1}}%
      \expandafter\def\csname LT3\endcsname{\color[rgb]{1,0,1}}%
      \expandafter\def\csname LT4\endcsname{\color[rgb]{0,1,1}}%
      \expandafter\def\csname LT5\endcsname{\color[rgb]{1,1,0}}%
      \expandafter\def\csname LT6\endcsname{\color[rgb]{0,0,0}}%
      \expandafter\def\csname LT7\endcsname{\color[rgb]{1,0.3,0}}%
      \expandafter\def\csname LT8\endcsname{\color[rgb]{0.5,0.5,0.5}}%
    \else
      \def\colorrgb#1{\color{black}}%
      \def\colorgray#1{\color[gray]{#1}}%
      \expandafter\def\csname LTw\endcsname{\color{white}}%
      \expandafter\def\csname LTb\endcsname{\color{black}}%
      \expandafter\def\csname LTa\endcsname{\color{black}}%
      \expandafter\def\csname LT0\endcsname{\color{black}}%
      \expandafter\def\csname LT1\endcsname{\color{black}}%
      \expandafter\def\csname LT2\endcsname{\color{black}}%
      \expandafter\def\csname LT3\endcsname{\color{black}}%
      \expandafter\def\csname LT4\endcsname{\color{black}}%
      \expandafter\def\csname LT5\endcsname{\color{black}}%
      \expandafter\def\csname LT6\endcsname{\color{black}}%
      \expandafter\def\csname LT7\endcsname{\color{black}}%
      \expandafter\def\csname LT8\endcsname{\color{black}}%
    \fi
  \fi
  \setlength{\unitlength}{0.0500bp}%
  \begin{picture}(3600.00,2520.00)%
    \gplgaddtomacro\gplbacktext{%
      \csname LTb\endcsname%
      \csname LTb\endcsname%
      \put(990,1620){\makebox(0,0)[r]{\strut{} 0}}%
      \csname LTb\endcsname%
      \csname LTb\endcsname%
      \put(990,1140){\makebox(0,0)[r]{\strut{} 0.2}}%
      \csname LTb\endcsname%
      \csname LTb\endcsname%
      \put(990,660){\makebox(0,0)[r]{\strut{} 0.4}}%
      \csname LTb\endcsname%
      \put(1122,440){\makebox(0,0){\strut{}-0.6}}%
      \csname LTb\endcsname%
      \csname LTb\endcsname%
      \put(1648,440){\makebox(0,0){\strut{}-0.4}}%
      \csname LTb\endcsname%
      \csname LTb\endcsname%
      \put(2174,440){\makebox(0,0){\strut{}-0.2}}%
      \csname LTb\endcsname%
      \csname LTb\endcsname%
      \put(2700,440){\makebox(0,0){\strut{} 0}}%
      \csname LTb\endcsname%
      \csname LTb\endcsname%
      \put(3226,440){\makebox(0,0){\strut{} 0.2}}%
      \put(220,1260){\rotatebox{90}{\makebox(0,0){\strut{}$z$ (m)}}}%
      \put(2174,110){\makebox(0,0){\strut{}$y$ (m)}}%
      \put(2174,2190){\makebox(0,0){\strut{}(d)}}%
    }%
    \gplgaddtomacro\gplfronttext{%
    }%
    \gplbacktext
    \put(0,0){\includegraphics{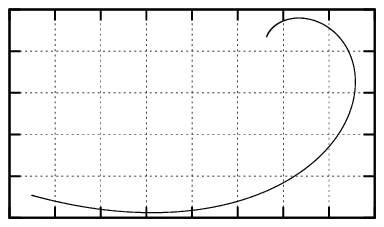}}%
    \put(2700,1620){\pscircle(0,0){0.25}}
    \gplfronttext
  \end{picture}%
\endgroup

%% file: Figure14a.tex
\begingroup
  \makeatletter
  \providecommand\color[2][]{%
    \GenericError{(gnuplot) \space\space\space\@spaces}{%
      Package color not loaded in conjunction with
      terminal option `colourtext'%
    }{See the gnuplot documentation for explanation.%
    }{Either use 'blacktext' in gnuplot or load the package
      color.sty in LaTeX.}%
    \renewcommand\color[2][]{}%
  }%
  \providecommand\includegraphics[2][]{%
    \GenericError{(gnuplot) \space\space\space\@spaces}{%
      Package graphicx or graphics not loaded%
    }{See the gnuplot documentation for explanation.%
    }{The gnuplot epslatex terminal needs graphicx.sty or graphics.sty.}%
    \renewcommand\includegraphics[2][]{}%
  }%
  \providecommand\rotatebox[2]{#2}%
  \@ifundefined{ifGPcolor}{%
    \newif\ifGPcolor
    \GPcolorfalse
  }{}%
  \@ifundefined{ifGPblacktext}{%
    \newif\ifGPblacktext
    \GPblacktexttrue
  }{}%
  \let\gplgaddtomacro\g@addto@macro
  \gdef\gplbacktext{}%
  \gdef\gplfronttext{}%
  \makeatother
  \ifGPblacktext
    \def\colorrgb#1{}%
    \def\colorgray#1{}%
  \else
    \ifGPcolor
      \def\colorrgb#1{\color[rgb]{#1}}%
      \def\colorgray#1{\color[gray]{#1}}%
      \expandafter\def\csname LTw\endcsname{\color{white}}%
      \expandafter\def\csname LTb\endcsname{\color{black}}%
      \expandafter\def\csname LTa\endcsname{\color{black}}%
      \expandafter\def\csname LT0\endcsname{\color[rgb]{1,0,0}}%
      \expandafter\def\csname LT1\endcsname{\color[rgb]{0,1,0}}%
      \expandafter\def\csname LT2\endcsname{\color[rgb]{0,0,1}}%
      \expandafter\def\csname LT3\endcsname{\color[rgb]{1,0,1}}%
      \expandafter\def\csname LT4\endcsname{\color[rgb]{0,1,1}}%
      \expandafter\def\csname LT5\endcsname{\color[rgb]{1,1,0}}%
      \expandafter\def\csname LT6\endcsname{\color[rgb]{0,0,0}}%
      \expandafter\def\csname LT7\endcsname{\color[rgb]{1,0.3,0}}%
      \expandafter\def\csname LT8\endcsname{\color[rgb]{0.5,0.5,0.5}}%
    \else
      \def\colorrgb#1{\color{black}}%
      \def\colorgray#1{\color[gray]{#1}}%
      \expandafter\def\csname LTw\endcsname{\color{white}}%
      \expandafter\def\csname LTb\endcsname{\color{black}}%
      \expandafter\def\csname LTa\endcsname{\color{black}}%
      \expandafter\def\csname LT0\endcsname{\color{black}}%
      \expandafter\def\csname LT1\endcsname{\color{black}}%
      \expandafter\def\csname LT2\endcsname{\color{black}}%
      \expandafter\def\csname LT3\endcsname{\color{black}}%
      \expandafter\def\csname LT4\endcsname{\color{black}}%
      \expandafter\def\csname LT5\endcsname{\color{black}}%
      \expandafter\def\csname LT6\endcsname{\color{black}}%
      \expandafter\def\csname LT7\endcsname{\color{black}}%
      \expandafter\def\csname LT8\endcsname{\color{black}}%
    \fi
  \fi
  \setlength{\unitlength}{0.0500bp}%
  \begin{picture}(3600.00,2520.00)%
    \gplgaddtomacro\gplbacktext{%
      \csname LTb\endcsname%
      \csname LTb\endcsname%
      \put(990,1727){\makebox(0,0)[r]{\strut{}-0.2}}%
      \csname LTb\endcsname%
      \csname LTb\endcsname%
      \put(990,1460){\makebox(0,0)[r]{\strut{} 0}}%
      \csname LTb\endcsname%
      \csname LTb\endcsname%
      \put(990,1193){\makebox(0,0)[r]{\strut{} 0.2}}%
      \csname LTb\endcsname%
      \csname LTb\endcsname%
      \put(990,927){\makebox(0,0)[r]{\strut{} 0.4}}%
      \csname LTb\endcsname%
      \csname LTb\endcsname%
      \csname LTb\endcsname%
      \csname LTb\endcsname%
      \put(1543,440){\makebox(0,0){\strut{}-0.4}}%
      \csname LTb\endcsname%
      \put(1964,440){\makebox(0,0){\strut{}-0.2}}%
      \csname LTb\endcsname%
      \put(2384,440){\makebox(0,0){\strut{} 0}}%
      \csname LTb\endcsname%
      \put(2805,440){\makebox(0,0){\strut{} 0.2}}%
      \csname LTb\endcsname%
      \put(220,1260){\rotatebox{90}{\makebox(0,0){\strut{}$z$ (m)}}}%
      \put(2174,110){\makebox(0,0){\strut{}$y$ (m)}}%
      \put(2174,2190){\makebox(0,0){\strut{}(a)}}%
    }%
    \gplgaddtomacro\gplfronttext{%
    }%
    \gplbacktext
    \put(0,0){\includegraphics{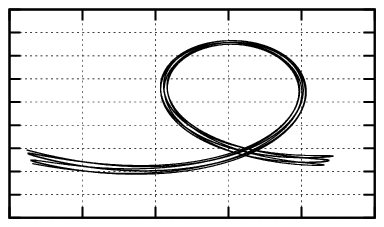}}%
    \gplfronttext
  \end{picture}%
\endgroup

%% file: Figure14b.tex
\begingroup
  \makeatletter
  \providecommand\color[2][]{%
    \GenericError{(gnuplot) \space\space\space\@spaces}{%
      Package color not loaded in conjunction with
      terminal option `colourtext'%
    }{See the gnuplot documentation for explanation.%
    }{Either use 'blacktext' in gnuplot or load the package
      color.sty in LaTeX.}%
    \renewcommand\color[2][]{}%
  }%
  \providecommand\includegraphics[2][]{%
    \GenericError{(gnuplot) \space\space\space\@spaces}{%
      Package graphicx or graphics not loaded%
    }{See the gnuplot documentation for explanation.%
    }{The gnuplot epslatex terminal needs graphicx.sty or graphics.sty.}%
    \renewcommand\includegraphics[2][]{}%
  }%
  \providecommand\rotatebox[2]{#2}%
  \@ifundefined{ifGPcolor}{%
    \newif\ifGPcolor
    \GPcolorfalse
  }{}%
  \@ifundefined{ifGPblacktext}{%
    \newif\ifGPblacktext
    \GPblacktexttrue
  }{}%
  \let\gplgaddtomacro\g@addto@macro
  \gdef\gplbacktext{}%
  \gdef\gplfronttext{}%
  \makeatother
  \ifGPblacktext
    \def\colorrgb#1{}%
    \def\colorgray#1{}%
  \else
    \ifGPcolor
      \def\colorrgb#1{\color[rgb]{#1}}%
      \def\colorgray#1{\color[gray]{#1}}%
      \expandafter\def\csname LTw\endcsname{\color{white}}%
      \expandafter\def\csname LTb\endcsname{\color{black}}%
      \expandafter\def\csname LTa\endcsname{\color{black}}%
      \expandafter\def\csname LT0\endcsname{\color[rgb]{1,0,0}}%
      \expandafter\def\csname LT1\endcsname{\color[rgb]{0,1,0}}%
      \expandafter\def\csname LT2\endcsname{\color[rgb]{0,0,1}}%
      \expandafter\def\csname LT3\endcsname{\color[rgb]{1,0,1}}%
      \expandafter\def\csname LT4\endcsname{\color[rgb]{0,1,1}}%
      \expandafter\def\csname LT5\endcsname{\color[rgb]{1,1,0}}%
      \expandafter\def\csname LT6\endcsname{\color[rgb]{0,0,0}}%
      \expandafter\def\csname LT7\endcsname{\color[rgb]{1,0.3,0}}%
      \expandafter\def\csname LT8\endcsname{\color[rgb]{0.5,0.5,0.5}}%
    \else
      \def\colorrgb#1{\color{black}}%
      \def\colorgray#1{\color[gray]{#1}}%
      \expandafter\def\csname LTw\endcsname{\color{white}}%
      \expandafter\def\csname LTb\endcsname{\color{black}}%
      \expandafter\def\csname LTa\endcsname{\color{black}}%
      \expandafter\def\csname LT0\endcsname{\color{black}}%
      \expandafter\def\csname LT1\endcsname{\color{black}}%
      \expandafter\def\csname LT2\endcsname{\color{black}}%
      \expandafter\def\csname LT3\endcsname{\color{black}}%
      \expandafter\def\csname LT4\endcsname{\color{black}}%
      \expandafter\def\csname LT5\endcsname{\color{black}}%
      \expandafter\def\csname LT6\endcsname{\color{black}}%
      \expandafter\def\csname LT7\endcsname{\color{black}}%
      \expandafter\def\csname LT8\endcsname{\color{black}}%
    \fi
  \fi
  \setlength{\unitlength}{0.0500bp}%
  \begin{picture}(3600.00,2520.00)%
    \gplgaddtomacro\gplbacktext{%
      \csname LTb\endcsname%
      \csname LTb\endcsname%
      \put(990,1710){\makebox(0,0)[r]{\strut{}-0.2}}%
      \csname LTb\endcsname%
      \csname LTb\endcsname%
      \put(990,1410){\makebox(0,0)[r]{\strut{} 0}}%
      \csname LTb\endcsname%
      \csname LTb\endcsname%
      \put(990,1110){\makebox(0,0)[r]{\strut{} 0.2}}%
      \csname LTb\endcsname%
      \csname LTb\endcsname%
      \put(990,810){\makebox(0,0)[r]{\strut{} 0.4}}%
      \csname LTb\endcsname%
      \csname LTb\endcsname%
      \csname LTb\endcsname%
      \put(1543,440){\makebox(0,0){\strut{}-0.4}}%
      \csname LTb\endcsname%
      \put(1964,440){\makebox(0,0){\strut{}-0.2}}%
      \csname LTb\endcsname%
      \put(2384,440){\makebox(0,0){\strut{} 0}}%
      \csname LTb\endcsname%
      \put(2805,440){\makebox(0,0){\strut{} 0.2}}%
      \csname LTb\endcsname%
      \put(220,1260){\rotatebox{90}{\makebox(0,0){\strut{}$z$ (m)}}}%
      \put(2174,110){\makebox(0,0){\strut{}$y$ (m)}}%
      \put(2174,2190){\makebox(0,0){\strut{}(b)}}%
    }%
    \gplgaddtomacro\gplfronttext{%
    }%
    \gplbacktext
    \put(0,0){\includegraphics{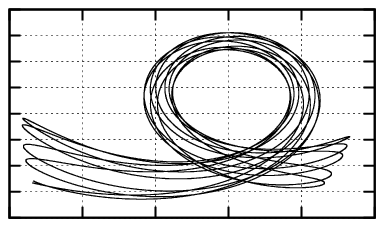}}%
    \gplfronttext
  \end{picture}%
\endgroup

%% file: Figure14c.tex
\begingroup
  \makeatletter
  \providecommand\color[2][]{%
    \GenericError{(gnuplot) \space\space\space\@spaces}{%
      Package color not loaded in conjunction with
      terminal option `colourtext'%
    }{See the gnuplot documentation for explanation.%
    }{Either use 'blacktext' in gnuplot or load the package
      color.sty in LaTeX.}%
    \renewcommand\color[2][]{}%
  }%
  \providecommand\includegraphics[2][]{%
    \GenericError{(gnuplot) \space\space\space\@spaces}{%
      Package graphicx or graphics not loaded%
    }{See the gnuplot documentation for explanation.%
    }{The gnuplot epslatex terminal needs graphicx.sty or graphics.sty.}%
    \renewcommand\includegraphics[2][]{}%
  }%
  \providecommand\rotatebox[2]{#2}%
  \@ifundefined{ifGPcolor}{%
    \newif\ifGPcolor
    \GPcolorfalse
  }{}%
  \@ifundefined{ifGPblacktext}{%
    \newif\ifGPblacktext
    \GPblacktexttrue
  }{}%
  \let\gplgaddtomacro\g@addto@macro
  \gdef\gplbacktext{}%
  \gdef\gplfronttext{}%
  \makeatother
  \ifGPblacktext
    \def\colorrgb#1{}%
    \def\colorgray#1{}%
  \else
    \ifGPcolor
      \def\colorrgb#1{\color[rgb]{#1}}%
      \def\colorgray#1{\color[gray]{#1}}%
      \expandafter\def\csname LTw\endcsname{\color{white}}%
      \expandafter\def\csname LTb\endcsname{\color{black}}%
      \expandafter\def\csname LTa\endcsname{\color{black}}%
      \expandafter\def\csname LT0\endcsname{\color[rgb]{1,0,0}}%
      \expandafter\def\csname LT1\endcsname{\color[rgb]{0,1,0}}%
      \expandafter\def\csname LT2\endcsname{\color[rgb]{0,0,1}}%
      \expandafter\def\csname LT3\endcsname{\color[rgb]{1,0,1}}%
      \expandafter\def\csname LT4\endcsname{\color[rgb]{0,1,1}}%
      \expandafter\def\csname LT5\endcsname{\color[rgb]{1,1,0}}%
      \expandafter\def\csname LT6\endcsname{\color[rgb]{0,0,0}}%
      \expandafter\def\csname LT7\endcsname{\color[rgb]{1,0.3,0}}%
      \expandafter\def\csname LT8\endcsname{\color[rgb]{0.5,0.5,0.5}}%
    \else
      \def\colorrgb#1{\color{black}}%
      \def\colorgray#1{\color[gray]{#1}}%
      \expandafter\def\csname LTw\endcsname{\color{white}}%
      \expandafter\def\csname LTb\endcsname{\color{black}}%
      \expandafter\def\csname LTa\endcsname{\color{black}}%
      \expandafter\def\csname LT0\endcsname{\color{black}}%
      \expandafter\def\csname LT1\endcsname{\color{black}}%
      \expandafter\def\csname LT2\endcsname{\color{black}}%
      \expandafter\def\csname LT3\endcsname{\color{black}}%
      \expandafter\def\csname LT4\endcsname{\color{black}}%
      \expandafter\def\csname LT5\endcsname{\color{black}}%
      \expandafter\def\csname LT6\endcsname{\color{black}}%
      \expandafter\def\csname LT7\endcsname{\color{black}}%
      \expandafter\def\csname LT8\endcsname{\color{black}}%
    \fi
  \fi
  \setlength{\unitlength}{0.0500bp}%
  \begin{picture}(3600.00,2520.00)%
    \gplgaddtomacro\gplbacktext{%
      \csname LTb\endcsname%
      \csname LTb\endcsname%
      \put(990,1710){\makebox(0,0)[r]{\strut{}-0.2}}%
      \csname LTb\endcsname%
      \csname LTb\endcsname%
      \put(990,1410){\makebox(0,0)[r]{\strut{} 0}}%
      \csname LTb\endcsname%
      \csname LTb\endcsname%
      \put(990,1110){\makebox(0,0)[r]{\strut{} 0.2}}%
      \csname LTb\endcsname%
      \csname LTb\endcsname%
      \put(990,810){\makebox(0,0)[r]{\strut{} 0.4}}%
      \csname LTb\endcsname%
      \csname LTb\endcsname%
      \csname LTb\endcsname%
      \put(1543,440){\makebox(0,0){\strut{}-0.4}}%
      \csname LTb\endcsname%
      \put(1964,440){\makebox(0,0){\strut{}-0.2}}%
      \csname LTb\endcsname%
      \put(2384,440){\makebox(0,0){\strut{} 0}}%
      \csname LTb\endcsname%
      \put(2805,440){\makebox(0,0){\strut{} 0.2}}%
      \csname LTb\endcsname%
      \put(220,1260){\rotatebox{90}{\makebox(0,0){\strut{}$z$ (m)}}}%
      \put(2174,110){\makebox(0,0){\strut{}$y$ (m)}}%
      \put(2174,2190){\makebox(0,0){\strut{}(c)}}%
    }%
    \gplgaddtomacro\gplfronttext{%
    }%
    \gplbacktext
    \put(0,0){\includegraphics{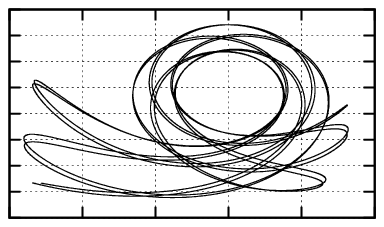}}%
    \gplfronttext
  \end{picture}%
\endgroup

%% file: Figure14d.tex
\begingroup
  \makeatletter
  \providecommand\color[2][]{%
    \GenericError{(gnuplot) \space\space\space\@spaces}{%
      Package color not loaded in conjunction with
      terminal option `colourtext'%
    }{See the gnuplot documentation for explanation.%
    }{Either use 'blacktext' in gnuplot or load the package
      color.sty in LaTeX.}%
    \renewcommand\color[2][]{}%
  }%
  \providecommand\includegraphics[2][]{%
    \GenericError{(gnuplot) \space\space\space\@spaces}{%
      Package graphicx or graphics not loaded%
    }{See the gnuplot documentation for explanation.%
    }{The gnuplot epslatex terminal needs graphicx.sty or graphics.sty.}%
    \renewcommand\includegraphics[2][]{}%
  }%
  \providecommand\rotatebox[2]{#2}%
  \@ifundefined{ifGPcolor}{%
    \newif\ifGPcolor
    \GPcolorfalse
  }{}%
  \@ifundefined{ifGPblacktext}{%
    \newif\ifGPblacktext
    \GPblacktexttrue
  }{}%
  \let\gplgaddtomacro\g@addto@macro
  \gdef\gplbacktext{}%
  \gdef\gplfronttext{}%
  \makeatother
  \ifGPblacktext
    \def\colorrgb#1{}%
    \def\colorgray#1{}%
  \else
    \ifGPcolor
      \def\colorrgb#1{\color[rgb]{#1}}%
      \def\colorgray#1{\color[gray]{#1}}%
      \expandafter\def\csname LTw\endcsname{\color{white}}%
      \expandafter\def\csname LTb\endcsname{\color{black}}%
      \expandafter\def\csname LTa\endcsname{\color{black}}%
      \expandafter\def\csname LT0\endcsname{\color[rgb]{1,0,0}}%
      \expandafter\def\csname LT1\endcsname{\color[rgb]{0,1,0}}%
      \expandafter\def\csname LT2\endcsname{\color[rgb]{0,0,1}}%
      \expandafter\def\csname LT3\endcsname{\color[rgb]{1,0,1}}%
      \expandafter\def\csname LT4\endcsname{\color[rgb]{0,1,1}}%
      \expandafter\def\csname LT5\endcsname{\color[rgb]{1,1,0}}%
      \expandafter\def\csname LT6\endcsname{\color[rgb]{0,0,0}}%
      \expandafter\def\csname LT7\endcsname{\color[rgb]{1,0.3,0}}%
      \expandafter\def\csname LT8\endcsname{\color[rgb]{0.5,0.5,0.5}}%
    \else
      \def\colorrgb#1{\color{black}}%
      \def\colorgray#1{\color[gray]{#1}}%
      \expandafter\def\csname LTw\endcsname{\color{white}}%
      \expandafter\def\csname LTb\endcsname{\color{black}}%
      \expandafter\def\csname LTa\endcsname{\color{black}}%
      \expandafter\def\csname LT0\endcsname{\color{black}}%
      \expandafter\def\csname LT1\endcsname{\color{black}}%
      \expandafter\def\csname LT2\endcsname{\color{black}}%
      \expandafter\def\csname LT3\endcsname{\color{black}}%
      \expandafter\def\csname LT4\endcsname{\color{black}}%
      \expandafter\def\csname LT5\endcsname{\color{black}}%
      \expandafter\def\csname LT6\endcsname{\color{black}}%
      \expandafter\def\csname LT7\endcsname{\color{black}}%
      \expandafter\def\csname LT8\endcsname{\color{black}}%
    \fi
  \fi
  \setlength{\unitlength}{0.0500bp}%
  \begin{picture}(3600.00,2520.00)%
    \gplgaddtomacro\gplbacktext{%
      \csname LTb\endcsname%
      \csname LTb\endcsname%
      \put(990,1710){\makebox(0,0)[r]{\strut{}-0.2}}%
      \csname LTb\endcsname%
      \csname LTb\endcsname%
      \put(990,1410){\makebox(0,0)[r]{\strut{} 0}}%
      \csname LTb\endcsname%
      \csname LTb\endcsname%
      \put(990,1110){\makebox(0,0)[r]{\strut{} 0.2}}%
      \csname LTb\endcsname%
      \csname LTb\endcsname%
      \put(990,810){\makebox(0,0)[r]{\strut{} 0.4}}%
      \csname LTb\endcsname%
      \csname LTb\endcsname%
      \csname LTb\endcsname%
      \put(1543,440){\makebox(0,0){\strut{}-0.4}}%
      \csname LTb\endcsname%
      \put(1964,440){\makebox(0,0){\strut{}-0.2}}%
      \csname LTb\endcsname%
      \put(2384,440){\makebox(0,0){\strut{} 0}}%
      \csname LTb\endcsname%
      \put(2805,440){\makebox(0,0){\strut{} 0.2}}%
      \csname LTb\endcsname%
      \put(220,1260){\rotatebox{90}{\makebox(0,0){\strut{}$z$ (m)}}}%
      \put(2174,110){\makebox(0,0){\strut{}$y$ (m)}}%
      \put(2174,2190){\makebox(0,0){\strut{}(d)}}%
    }%
    \gplgaddtomacro\gplfronttext{%
    }%
    \gplbacktext
    \put(0,0){\includegraphics{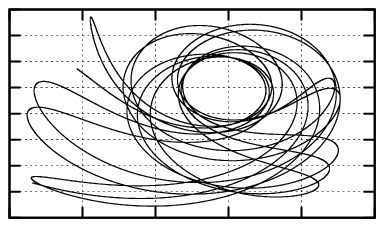}}%
    \gplfronttext
  \end{picture}%
\endgroup

%% file: Figure15a.tex
\begingroup
  \makeatletter
  \providecommand\color[2][]{%
    \GenericError{(gnuplot) \space\space\space\@spaces}{%
      Package color not loaded in conjunction with
      terminal option `colourtext'%
    }{See the gnuplot documentation for explanation.%
    }{Either use 'blacktext' in gnuplot or load the package
      color.sty in LaTeX.}%
    \renewcommand\color[2][]{}%
  }%
  \providecommand\includegraphics[2][]{%
    \GenericError{(gnuplot) \space\space\space\@spaces}{%
      Package graphicx or graphics not loaded%
    }{See the gnuplot documentation for explanation.%
    }{The gnuplot epslatex terminal needs graphicx.sty or graphics.sty.}%
    \renewcommand\includegraphics[2][]{}%
  }%
  \providecommand\rotatebox[2]{#2}%
  \@ifundefined{ifGPcolor}{%
    \newif\ifGPcolor
    \GPcolorfalse
  }{}%
  \@ifundefined{ifGPblacktext}{%
    \newif\ifGPblacktext
    \GPblacktexttrue
  }{}%
  \let\gplgaddtomacro\g@addto@macro
  \gdef\gplbacktext{}%
  \gdef\gplfronttext{}%
  \makeatother
  \ifGPblacktext
    \def\colorrgb#1{}%
    \def\colorgray#1{}%
  \else
    \ifGPcolor
      \def\colorrgb#1{\color[rgb]{#1}}%
      \def\colorgray#1{\color[gray]{#1}}%
      \expandafter\def\csname LTw\endcsname{\color{white}}%
      \expandafter\def\csname LTb\endcsname{\color{black}}%
      \expandafter\def\csname LTa\endcsname{\color{black}}%
      \expandafter\def\csname LT0\endcsname{\color[rgb]{1,0,0}}%
      \expandafter\def\csname LT1\endcsname{\color[rgb]{0,1,0}}%
      \expandafter\def\csname LT2\endcsname{\color[rgb]{0,0,1}}%
      \expandafter\def\csname LT3\endcsname{\color[rgb]{1,0,1}}%
      \expandafter\def\csname LT4\endcsname{\color[rgb]{0,1,1}}%
      \expandafter\def\csname LT5\endcsname{\color[rgb]{1,1,0}}%
      \expandafter\def\csname LT6\endcsname{\color[rgb]{0,0,0}}%
      \expandafter\def\csname LT7\endcsname{\color[rgb]{1,0.3,0}}%
      \expandafter\def\csname LT8\endcsname{\color[rgb]{0.5,0.5,0.5}}%
    \else
      \def\colorrgb#1{\color{black}}%
      \def\colorgray#1{\color[gray]{#1}}%
      \expandafter\def\csname LTw\endcsname{\color{white}}%
      \expandafter\def\csname LTb\endcsname{\color{black}}%
      \expandafter\def\csname LTa\endcsname{\color{black}}%
      \expandafter\def\csname LT0\endcsname{\color{black}}%
      \expandafter\def\csname LT1\endcsname{\color{black}}%
      \expandafter\def\csname LT2\endcsname{\color{black}}%
      \expandafter\def\csname LT3\endcsname{\color{black}}%
      \expandafter\def\csname LT4\endcsname{\color{black}}%
      \expandafter\def\csname LT5\endcsname{\color{black}}%
      \expandafter\def\csname LT6\endcsname{\color{black}}%
      \expandafter\def\csname LT7\endcsname{\color{black}}%
      \expandafter\def\csname LT8\endcsname{\color{black}}%
    \fi
  \fi
  \setlength{\unitlength}{0.0500bp}%
  \begin{picture}(3600.00,2520.00)%
    \gplgaddtomacro\gplbacktext{%
      \csname LTb\endcsname%
      \put(990,1860){\makebox(0,0)[r]{\strut{} 0}}%
      \put(990,1380){\makebox(0,0)[r]{\strut{} 0.2}}%
      \put(990,900){\makebox(0,0)[r]{\strut{} 0.4}}%
      \put(1473,440){\makebox(0,0){\strut{}-0.4}}%
      \put(2174,440){\makebox(0,0){\strut{} 0}}%
      \put(2875,440){\makebox(0,0){\strut{} 0.4}}%
      \put(220,1260){\rotatebox{90}{\makebox(0,0){\strut{}$z$ (m)}}}%
      \put(2174,110){\makebox(0,0){\strut{}$y$ (m)}}%
      \put(2174,2190){\makebox(0,0){\strut{}(a)}}%
    }%
    \gplgaddtomacro\gplfronttext{%
    }%
    \gplbacktext
    \put(0,0){\includegraphics{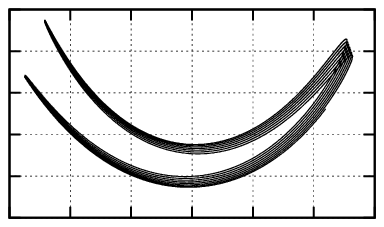}}%
    \gplfronttext
  \end{picture}%
\endgroup

%% file: Figure15b.tex
\begingroup
  \makeatletter
  \providecommand\color[2][]{%
    \GenericError{(gnuplot) \space\space\space\@spaces}{%
      Package color not loaded in conjunction with
      terminal option `colourtext'%
    }{See the gnuplot documentation for explanation.%
    }{Either use 'blacktext' in gnuplot or load the package
      color.sty in LaTeX.}%
    \renewcommand\color[2][]{}%
  }%
  \providecommand\includegraphics[2][]{%
    \GenericError{(gnuplot) \space\space\space\@spaces}{%
      Package graphicx or graphics not loaded%
    }{See the gnuplot documentation for explanation.%
    }{The gnuplot epslatex terminal needs graphicx.sty or graphics.sty.}%
    \renewcommand\includegraphics[2][]{}%
  }%
  \providecommand\rotatebox[2]{#2}%
  \@ifundefined{ifGPcolor}{%
    \newif\ifGPcolor
    \GPcolorfalse
  }{}%
  \@ifundefined{ifGPblacktext}{%
    \newif\ifGPblacktext
    \GPblacktexttrue
  }{}%
  \let\gplgaddtomacro\g@addto@macro
  \gdef\gplbacktext{}%
  \gdef\gplfronttext{}%
  \makeatother
  \ifGPblacktext
    \def\colorrgb#1{}%
    \def\colorgray#1{}%
  \else
    \ifGPcolor
      \def\colorrgb#1{\color[rgb]{#1}}%
      \def\colorgray#1{\color[gray]{#1}}%
      \expandafter\def\csname LTw\endcsname{\color{white}}%
      \expandafter\def\csname LTb\endcsname{\color{black}}%
      \expandafter\def\csname LTa\endcsname{\color{black}}%
      \expandafter\def\csname LT0\endcsname{\color[rgb]{1,0,0}}%
      \expandafter\def\csname LT1\endcsname{\color[rgb]{0,1,0}}%
      \expandafter\def\csname LT2\endcsname{\color[rgb]{0,0,1}}%
      \expandafter\def\csname LT3\endcsname{\color[rgb]{1,0,1}}%
      \expandafter\def\csname LT4\endcsname{\color[rgb]{0,1,1}}%
      \expandafter\def\csname LT5\endcsname{\color[rgb]{1,1,0}}%
      \expandafter\def\csname LT6\endcsname{\color[rgb]{0,0,0}}%
      \expandafter\def\csname LT7\endcsname{\color[rgb]{1,0.3,0}}%
      \expandafter\def\csname LT8\endcsname{\color[rgb]{0.5,0.5,0.5}}%
    \else
      \def\colorrgb#1{\color{black}}%
      \def\colorgray#1{\color[gray]{#1}}%
      \expandafter\def\csname LTw\endcsname{\color{white}}%
      \expandafter\def\csname LTb\endcsname{\color{black}}%
      \expandafter\def\csname LTa\endcsname{\color{black}}%
      \expandafter\def\csname LT0\endcsname{\color{black}}%
      \expandafter\def\csname LT1\endcsname{\color{black}}%
      \expandafter\def\csname LT2\endcsname{\color{black}}%
      \expandafter\def\csname LT3\endcsname{\color{black}}%
      \expandafter\def\csname LT4\endcsname{\color{black}}%
      \expandafter\def\csname LT5\endcsname{\color{black}}%
      \expandafter\def\csname LT6\endcsname{\color{black}}%
      \expandafter\def\csname LT7\endcsname{\color{black}}%
      \expandafter\def\csname LT8\endcsname{\color{black}}%
    \fi
  \fi
  \setlength{\unitlength}{0.0500bp}%
  \begin{picture}(3600.00,2520.00)%
    \gplgaddtomacro\gplbacktext{%
      \csname LTb\endcsname%
      \put(990,1660){\makebox(0,0)[r]{\strut{} 0}}%
      \put(990,1260){\makebox(0,0)[r]{\strut{} 0.2}}%
      \put(990,860){\makebox(0,0)[r]{\strut{} 0.4}}%
      \put(1473,440){\makebox(0,0){\strut{}-0.4}}%
      \put(2174,440){\makebox(0,0){\strut{} 0}}%
      \put(2875,440){\makebox(0,0){\strut{} 0.4}}%
      \put(220,1260){\rotatebox{90}{\makebox(0,0){\strut{}$z$ (m)}}}%
      \put(2174,110){\makebox(0,0){\strut{}$y$ (m)}}%
      \put(2174,2190){\makebox(0,0){\strut{}(b)}}%
    }%
    \gplgaddtomacro\gplfronttext{%
    }%
    \gplbacktext
    \put(0,0){\includegraphics{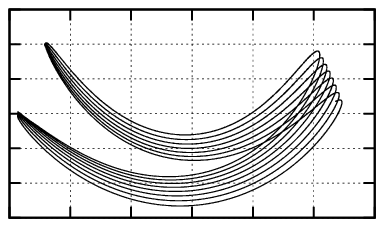}}%
    \gplfronttext
  \end{picture}%
\endgroup

%% file: Figure15c.tex
\begingroup
  \makeatletter
  \providecommand\color[2][]{%
    \GenericError{(gnuplot) \space\space\space\@spaces}{%
      Package color not loaded in conjunction with
      terminal option `colourtext'%
    }{See the gnuplot documentation for explanation.%
    }{Either use 'blacktext' in gnuplot or load the package
      color.sty in LaTeX.}%
    \renewcommand\color[2][]{}%
  }%
  \providecommand\includegraphics[2][]{%
    \GenericError{(gnuplot) \space\space\space\@spaces}{%
      Package graphicx or graphics not loaded%
    }{See the gnuplot documentation for explanation.%
    }{The gnuplot epslatex terminal needs graphicx.sty or graphics.sty.}%
    \renewcommand\includegraphics[2][]{}%
  }%
  \providecommand\rotatebox[2]{#2}%
  \@ifundefined{ifGPcolor}{%
    \newif\ifGPcolor
    \GPcolorfalse
  }{}%
  \@ifundefined{ifGPblacktext}{%
    \newif\ifGPblacktext
    \GPblacktexttrue
  }{}%
  \let\gplgaddtomacro\g@addto@macro
  \gdef\gplbacktext{}%
  \gdef\gplfronttext{}%
  \makeatother
  \ifGPblacktext
    \def\colorrgb#1{}%
    \def\colorgray#1{}%
  \else
    \ifGPcolor
      \def\colorrgb#1{\color[rgb]{#1}}%
      \def\colorgray#1{\color[gray]{#1}}%
      \expandafter\def\csname LTw\endcsname{\color{white}}%
      \expandafter\def\csname LTb\endcsname{\color{black}}%
      \expandafter\def\csname LTa\endcsname{\color{black}}%
      \expandafter\def\csname LT0\endcsname{\color[rgb]{1,0,0}}%
      \expandafter\def\csname LT1\endcsname{\color[rgb]{0,1,0}}%
      \expandafter\def\csname LT2\endcsname{\color[rgb]{0,0,1}}%
      \expandafter\def\csname LT3\endcsname{\color[rgb]{1,0,1}}%
      \expandafter\def\csname LT4\endcsname{\color[rgb]{0,1,1}}%
      \expandafter\def\csname LT5\endcsname{\color[rgb]{1,1,0}}%
      \expandafter\def\csname LT6\endcsname{\color[rgb]{0,0,0}}%
      \expandafter\def\csname LT7\endcsname{\color[rgb]{1,0.3,0}}%
      \expandafter\def\csname LT8\endcsname{\color[rgb]{0.5,0.5,0.5}}%
    \else
      \def\colorrgb#1{\color{black}}%
      \def\colorgray#1{\color[gray]{#1}}%
      \expandafter\def\csname LTw\endcsname{\color{white}}%
      \expandafter\def\csname LTb\endcsname{\color{black}}%
      \expandafter\def\csname LTa\endcsname{\color{black}}%
      \expandafter\def\csname LT0\endcsname{\color{black}}%
      \expandafter\def\csname LT1\endcsname{\color{black}}%
      \expandafter\def\csname LT2\endcsname{\color{black}}%
      \expandafter\def\csname LT3\endcsname{\color{black}}%
      \expandafter\def\csname LT4\endcsname{\color{black}}%
      \expandafter\def\csname LT5\endcsname{\color{black}}%
      \expandafter\def\csname LT6\endcsname{\color{black}}%
      \expandafter\def\csname LT7\endcsname{\color{black}}%
      \expandafter\def\csname LT8\endcsname{\color{black}}%
    \fi
  \fi
  \setlength{\unitlength}{0.0500bp}%
  \begin{picture}(3600.00,2520.00)%
    \gplgaddtomacro\gplbacktext{%
      \csname LTb\endcsname%
      \put(990,1689){\makebox(0,0)[r]{\strut{} 0}}%
      \put(990,1346){\makebox(0,0)[r]{\strut{} 0.2}}%
      \put(990,1003){\makebox(0,0)[r]{\strut{} 0.4}}%
      \put(990,660){\makebox(0,0)[r]{\strut{} 0.6}}%
      \put(1648,440){\makebox(0,0){\strut{}-0.4}}%
      \put(2349,440){\makebox(0,0){\strut{} 0}}%
      \put(3051,440){\makebox(0,0){\strut{} 0.4}}%
      \put(220,1260){\rotatebox{90}{\makebox(0,0){\strut{}$z$ (m)}}}%
      \put(2174,110){\makebox(0,0){\strut{}$y$ (m)}}%
      \put(2174,2190){\makebox(0,0){\strut{}(c)}}%
    }%
    \gplgaddtomacro\gplfronttext{%
    }%
    \gplbacktext
    \put(0,0){\includegraphics{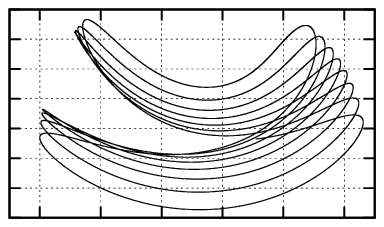}}%
    \gplfronttext
  \end{picture}%
\endgroup

%% file: Figure15d.tex
\begingroup
  \makeatletter
  \providecommand\color[2][]{%
    \GenericError{(gnuplot) \space\space\space\@spaces}{%
      Package color not loaded in conjunction with
      terminal option `colourtext'%
    }{See the gnuplot documentation for explanation.%
    }{Either use 'blacktext' in gnuplot or load the package
      color.sty in LaTeX.}%
    \renewcommand\color[2][]{}%
  }%
  \providecommand\includegraphics[2][]{%
    \GenericError{(gnuplot) \space\space\space\@spaces}{%
      Package graphicx or graphics not loaded%
    }{See the gnuplot documentation for explanation.%
    }{The gnuplot epslatex terminal needs graphicx.sty or graphics.sty.}%
    \renewcommand\includegraphics[2][]{}%
  }%
  \providecommand\rotatebox[2]{#2}%
  \@ifundefined{ifGPcolor}{%
    \newif\ifGPcolor
    \GPcolorfalse
  }{}%
  \@ifundefined{ifGPblacktext}{%
    \newif\ifGPblacktext
    \GPblacktexttrue
  }{}%
  \let\gplgaddtomacro\g@addto@macro
  \gdef\gplbacktext{}%
  \gdef\gplfronttext{}%
  \makeatother
  \ifGPblacktext
    \def\colorrgb#1{}%
    \def\colorgray#1{}%
  \else
    \ifGPcolor
      \def\colorrgb#1{\color[rgb]{#1}}%
      \def\colorgray#1{\color[gray]{#1}}%
      \expandafter\def\csname LTw\endcsname{\color{white}}%
      \expandafter\def\csname LTb\endcsname{\color{black}}%
      \expandafter\def\csname LTa\endcsname{\color{black}}%
      \expandafter\def\csname LT0\endcsname{\color[rgb]{1,0,0}}%
      \expandafter\def\csname LT1\endcsname{\color[rgb]{0,1,0}}%
      \expandafter\def\csname LT2\endcsname{\color[rgb]{0,0,1}}%
      \expandafter\def\csname LT3\endcsname{\color[rgb]{1,0,1}}%
      \expandafter\def\csname LT4\endcsname{\color[rgb]{0,1,1}}%
      \expandafter\def\csname LT5\endcsname{\color[rgb]{1,1,0}}%
      \expandafter\def\csname LT6\endcsname{\color[rgb]{0,0,0}}%
      \expandafter\def\csname LT7\endcsname{\color[rgb]{1,0.3,0}}%
      \expandafter\def\csname LT8\endcsname{\color[rgb]{0.5,0.5,0.5}}%
    \else
      \def\colorrgb#1{\color{black}}%
      \def\colorgray#1{\color[gray]{#1}}%
      \expandafter\def\csname LTw\endcsname{\color{white}}%
      \expandafter\def\csname LTb\endcsname{\color{black}}%
      \expandafter\def\csname LTa\endcsname{\color{black}}%
      \expandafter\def\csname LT0\endcsname{\color{black}}%
      \expandafter\def\csname LT1\endcsname{\color{black}}%
      \expandafter\def\csname LT2\endcsname{\color{black}}%
      \expandafter\def\csname LT3\endcsname{\color{black}}%
      \expandafter\def\csname LT4\endcsname{\color{black}}%
      \expandafter\def\csname LT5\endcsname{\color{black}}%
      \expandafter\def\csname LT6\endcsname{\color{black}}%
      \expandafter\def\csname LT7\endcsname{\color{black}}%
      \expandafter\def\csname LT8\endcsname{\color{black}}%
    \fi
  \fi
  \setlength{\unitlength}{0.0500bp}%
  \begin{picture}(3600.00,2520.00)%
    \gplgaddtomacro\gplbacktext{%
      \csname LTb\endcsname%
      \put(990,1860){\makebox(0,0)[r]{\strut{}-0.2}}%
      \put(990,1560){\makebox(0,0)[r]{\strut{} 0}}%
      \put(990,1260){\makebox(0,0)[r]{\strut{} 0.2}}%
      \put(990,960){\makebox(0,0)[r]{\strut{} 0.4}}%
      \put(990,660){\makebox(0,0)[r]{\strut{} 0.6}}%
      \put(1696,440){\makebox(0,0){\strut{}-0.4}}%
      \put(2461,440){\makebox(0,0){\strut{} 0}}%
      \put(3226,440){\makebox(0,0){\strut{} 0.4}}%
      \put(220,1260){\rotatebox{90}{\makebox(0,0){\strut{}$z$ (m)}}}%
      \put(2174,110){\makebox(0,0){\strut{}$y$ (m)}}%
      \put(2174,2190){\makebox(0,0){\strut{}(d)}}%
    }%
    \gplgaddtomacro\gplfronttext{%
    }%
    \gplbacktext
    \put(0,0){\includegraphics{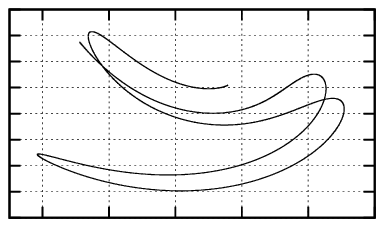}}%
    \gplfronttext
  \end{picture}%
\endgroup

%% file: Figure16a.tex
\begingroup
  \makeatletter
  \providecommand\color[2][]{%
    \GenericError{(gnuplot) \space\space\space\@spaces}{%
      Package color not loaded in conjunction with
      terminal option `colourtext'%
    }{See the gnuplot documentation for explanation.%
    }{Either use 'blacktext' in gnuplot or load the package
      color.sty in LaTeX.}%
    \renewcommand\color[2][]{}%
  }%
  \providecommand\includegraphics[2][]{%
    \GenericError{(gnuplot) \space\space\space\@spaces}{%
      Package graphicx or graphics not loaded%
    }{See the gnuplot documentation for explanation.%
    }{The gnuplot epslatex terminal needs graphicx.sty or graphics.sty.}%
    \renewcommand\includegraphics[2][]{}%
  }%
  \providecommand\rotatebox[2]{#2}%
  \@ifundefined{ifGPcolor}{%
    \newif\ifGPcolor
    \GPcolorfalse
  }{}%
  \@ifundefined{ifGPblacktext}{%
    \newif\ifGPblacktext
    \GPblacktexttrue
  }{}%
  \let\gplgaddtomacro\g@addto@macro
  \gdef\gplbacktext{}%
  \gdef\gplfronttext{}%
  \makeatother
  \ifGPblacktext
    \def\colorrgb#1{}%
    \def\colorgray#1{}%
  \else
    \ifGPcolor
      \def\colorrgb#1{\color[rgb]{#1}}%
      \def\colorgray#1{\color[gray]{#1}}%
      \expandafter\def\csname LTw\endcsname{\color{white}}%
      \expandafter\def\csname LTb\endcsname{\color{black}}%
      \expandafter\def\csname LTa\endcsname{\color{black}}%
      \expandafter\def\csname LT0\endcsname{\color[rgb]{1,0,0}}%
      \expandafter\def\csname LT1\endcsname{\color[rgb]{0,1,0}}%
      \expandafter\def\csname LT2\endcsname{\color[rgb]{0,0,1}}%
      \expandafter\def\csname LT3\endcsname{\color[rgb]{1,0,1}}%
      \expandafter\def\csname LT4\endcsname{\color[rgb]{0,1,1}}%
      \expandafter\def\csname LT5\endcsname{\color[rgb]{1,1,0}}%
      \expandafter\def\csname LT6\endcsname{\color[rgb]{0,0,0}}%
      \expandafter\def\csname LT7\endcsname{\color[rgb]{1,0.3,0}}%
      \expandafter\def\csname LT8\endcsname{\color[rgb]{0.5,0.5,0.5}}%
    \else
      \def\colorrgb#1{\color{black}}%
      \def\colorgray#1{\color[gray]{#1}}%
      \expandafter\def\csname LTw\endcsname{\color{white}}%
      \expandafter\def\csname LTb\endcsname{\color{black}}%
      \expandafter\def\csname LTa\endcsname{\color{black}}%
      \expandafter\def\csname LT0\endcsname{\color{black}}%
      \expandafter\def\csname LT1\endcsname{\color{black}}%
      \expandafter\def\csname LT2\endcsname{\color{black}}%
      \expandafter\def\csname LT3\endcsname{\color{black}}%
      \expandafter\def\csname LT4\endcsname{\color{black}}%
      \expandafter\def\csname LT5\endcsname{\color{black}}%
      \expandafter\def\csname LT6\endcsname{\color{black}}%
      \expandafter\def\csname LT7\endcsname{\color{black}}%
      \expandafter\def\csname LT8\endcsname{\color{black}}%
    \fi
  \fi
  \setlength{\unitlength}{0.0500bp}%
  \begin{picture}(3600.00,2520.00)%
    \gplgaddtomacro\gplbacktext{%
      \csname LTb\endcsname%
      \put(990,1660){\makebox(0,0)[r]{\strut{} 0}}%
      \put(990,1260){\makebox(0,0)[r]{\strut{} 0.2}}%
      \put(990,860){\makebox(0,0)[r]{\strut{} 0.4}}%
      \put(1473,440){\makebox(0,0){\strut{}-0.4}}%
      \put(2174,440){\makebox(0,0){\strut{} 0}}%
      \put(2875,440){\makebox(0,0){\strut{} 0.4}}%
      \put(220,1260){\rotatebox{90}{\makebox(0,0){\strut{}$z$ (m)}}}%
      \put(2174,110){\makebox(0,0){\strut{}$y$ (m)}}%
      \put(2174,2190){\makebox(0,0){\strut{}(a)}}%
    }%
    \gplgaddtomacro\gplfronttext{%
    }%
    \gplbacktext
    \put(0,0){\includegraphics{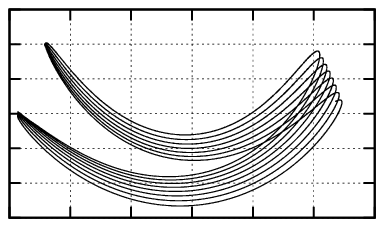}}%
    \gplfronttext
  \end{picture}%
\endgroup

%% file: Figure16b.tex
\begingroup
  \makeatletter
  \providecommand\color[2][]{%
    \GenericError{(gnuplot) \space\space\space\@spaces}{%
      Package color not loaded in conjunction with
      terminal option `colourtext'%
    }{See the gnuplot documentation for explanation.%
    }{Either use 'blacktext' in gnuplot or load the package
      color.sty in LaTeX.}%
    \renewcommand\color[2][]{}%
  }%
  \providecommand\includegraphics[2][]{%
    \GenericError{(gnuplot) \space\space\space\@spaces}{%
      Package graphicx or graphics not loaded%
    }{See the gnuplot documentation for explanation.%
    }{The gnuplot epslatex terminal needs graphicx.sty or graphics.sty.}%
    \renewcommand\includegraphics[2][]{}%
  }%
  \providecommand\rotatebox[2]{#2}%
  \@ifundefined{ifGPcolor}{%
    \newif\ifGPcolor
    \GPcolorfalse
  }{}%
  \@ifundefined{ifGPblacktext}{%
    \newif\ifGPblacktext
    \GPblacktexttrue
  }{}%
  \let\gplgaddtomacro\g@addto@macro
  \gdef\gplbacktext{}%
  \gdef\gplfronttext{}%
  \makeatother
  \ifGPblacktext
    \def\colorrgb#1{}%
    \def\colorgray#1{}%
  \else
    \ifGPcolor
      \def\colorrgb#1{\color[rgb]{#1}}%
      \def\colorgray#1{\color[gray]{#1}}%
      \expandafter\def\csname LTw\endcsname{\color{white}}%
      \expandafter\def\csname LTb\endcsname{\color{black}}%
      \expandafter\def\csname LTa\endcsname{\color{black}}%
      \expandafter\def\csname LT0\endcsname{\color[rgb]{1,0,0}}%
      \expandafter\def\csname LT1\endcsname{\color[rgb]{0,1,0}}%
      \expandafter\def\csname LT2\endcsname{\color[rgb]{0,0,1}}%
      \expandafter\def\csname LT3\endcsname{\color[rgb]{1,0,1}}%
      \expandafter\def\csname LT4\endcsname{\color[rgb]{0,1,1}}%
      \expandafter\def\csname LT5\endcsname{\color[rgb]{1,1,0}}%
      \expandafter\def\csname LT6\endcsname{\color[rgb]{0,0,0}}%
      \expandafter\def\csname LT7\endcsname{\color[rgb]{1,0.3,0}}%
      \expandafter\def\csname LT8\endcsname{\color[rgb]{0.5,0.5,0.5}}%
    \else
      \def\colorrgb#1{\color{black}}%
      \def\colorgray#1{\color[gray]{#1}}%
      \expandafter\def\csname LTw\endcsname{\color{white}}%
      \expandafter\def\csname LTb\endcsname{\color{black}}%
      \expandafter\def\csname LTa\endcsname{\color{black}}%
      \expandafter\def\csname LT0\endcsname{\color{black}}%
      \expandafter\def\csname LT1\endcsname{\color{black}}%
      \expandafter\def\csname LT2\endcsname{\color{black}}%
      \expandafter\def\csname LT3\endcsname{\color{black}}%
      \expandafter\def\csname LT4\endcsname{\color{black}}%
      \expandafter\def\csname LT5\endcsname{\color{black}}%
      \expandafter\def\csname LT6\endcsname{\color{black}}%
      \expandafter\def\csname LT7\endcsname{\color{black}}%
      \expandafter\def\csname LT8\endcsname{\color{black}}%
    \fi
  \fi
  \setlength{\unitlength}{0.0500bp}%
  \begin{picture}(3600.00,2520.00)%
    \gplgaddtomacro\gplbacktext{%
      \csname LTb\endcsname%
      \put(990,1660){\makebox(0,0)[r]{\strut{} 0}}%
      \put(990,1260){\makebox(0,0)[r]{\strut{} 0.2}}%
      \put(990,860){\makebox(0,0)[r]{\strut{} 0.4}}%
      \put(1473,440){\makebox(0,0){\strut{}-0.4}}%
      \put(2174,440){\makebox(0,0){\strut{} 0}}%
      \put(2875,440){\makebox(0,0){\strut{} 0.4}}%
      \put(220,1260){\rotatebox{90}{\makebox(0,0){\strut{}$z$ (m)}}}%
      \put(2174,110){\makebox(0,0){\strut{}$y$ (m)}}%
      \put(2174,2190){\makebox(0,0){\strut{}(b)}}%
    }%
    \gplgaddtomacro\gplfronttext{%
    }%
    \gplbacktext
    \put(0,0){\includegraphics{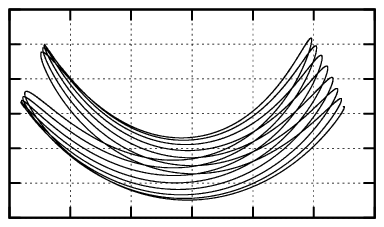}}%
    \gplfronttext
  \end{picture}%
\endgroup

%% file: Figure16c.tex
\begingroup
  \makeatletter
  \providecommand\color[2][]{%
    \GenericError{(gnuplot) \space\space\space\@spaces}{%
      Package color not loaded in conjunction with
      terminal option `colourtext'%
    }{See the gnuplot documentation for explanation.%
    }{Either use 'blacktext' in gnuplot or load the package
      color.sty in LaTeX.}%
    \renewcommand\color[2][]{}%
  }%
  \providecommand\includegraphics[2][]{%
    \GenericError{(gnuplot) \space\space\space\@spaces}{%
      Package graphicx or graphics not loaded%
    }{See the gnuplot documentation for explanation.%
    }{The gnuplot epslatex terminal needs graphicx.sty or graphics.sty.}%
    \renewcommand\includegraphics[2][]{}%
  }%
  \providecommand\rotatebox[2]{#2}%
  \@ifundefined{ifGPcolor}{%
    \newif\ifGPcolor
    \GPcolorfalse
  }{}%
  \@ifundefined{ifGPblacktext}{%
    \newif\ifGPblacktext
    \GPblacktexttrue
  }{}%
  \let\gplgaddtomacro\g@addto@macro
  \gdef\gplbacktext{}%
  \gdef\gplfronttext{}%
  \makeatother
  \ifGPblacktext
    \def\colorrgb#1{}%
    \def\colorgray#1{}%
  \else
    \ifGPcolor
      \def\colorrgb#1{\color[rgb]{#1}}%
      \def\colorgray#1{\color[gray]{#1}}%
      \expandafter\def\csname LTw\endcsname{\color{white}}%
      \expandafter\def\csname LTb\endcsname{\color{black}}%
      \expandafter\def\csname LTa\endcsname{\color{black}}%
      \expandafter\def\csname LT0\endcsname{\color[rgb]{1,0,0}}%
      \expandafter\def\csname LT1\endcsname{\color[rgb]{0,1,0}}%
      \expandafter\def\csname LT2\endcsname{\color[rgb]{0,0,1}}%
      \expandafter\def\csname LT3\endcsname{\color[rgb]{1,0,1}}%
      \expandafter\def\csname LT4\endcsname{\color[rgb]{0,1,1}}%
      \expandafter\def\csname LT5\endcsname{\color[rgb]{1,1,0}}%
      \expandafter\def\csname LT6\endcsname{\color[rgb]{0,0,0}}%
      \expandafter\def\csname LT7\endcsname{\color[rgb]{1,0.3,0}}%
      \expandafter\def\csname LT8\endcsname{\color[rgb]{0.5,0.5,0.5}}%
    \else
      \def\colorrgb#1{\color{black}}%
      \def\colorgray#1{\color[gray]{#1}}%
      \expandafter\def\csname LTw\endcsname{\color{white}}%
      \expandafter\def\csname LTb\endcsname{\color{black}}%
      \expandafter\def\csname LTa\endcsname{\color{black}}%
      \expandafter\def\csname LT0\endcsname{\color{black}}%
      \expandafter\def\csname LT1\endcsname{\color{black}}%
      \expandafter\def\csname LT2\endcsname{\color{black}}%
      \expandafter\def\csname LT3\endcsname{\color{black}}%
      \expandafter\def\csname LT4\endcsname{\color{black}}%
      \expandafter\def\csname LT5\endcsname{\color{black}}%
      \expandafter\def\csname LT6\endcsname{\color{black}}%
      \expandafter\def\csname LT7\endcsname{\color{black}}%
      \expandafter\def\csname LT8\endcsname{\color{black}}%
    \fi
  \fi
  \setlength{\unitlength}{0.0500bp}%
  \begin{picture}(3600.00,2520.00)%
    \gplgaddtomacro\gplbacktext{%
      \csname LTb\endcsname%
      \put(990,1660){\makebox(0,0)[r]{\strut{} 0}}%
      \put(990,1260){\makebox(0,0)[r]{\strut{} 0.2}}%
      \put(990,860){\makebox(0,0)[r]{\strut{} 0.4}}%
      \put(1473,440){\makebox(0,0){\strut{}-0.4}}%
      \put(2174,440){\makebox(0,0){\strut{} 0}}%
      \put(2875,440){\makebox(0,0){\strut{} 0.4}}%
      \put(220,1260){\rotatebox{90}{\makebox(0,0){\strut{}$z$ (m)}}}%
      \put(2174,110){\makebox(0,0){\strut{}$y$ (m)}}%
      \put(2174,2190){\makebox(0,0){\strut{}(c)}}%
    }%
    \gplgaddtomacro\gplfronttext{%
    }%
    \gplbacktext
    \put(0,0){\includegraphics{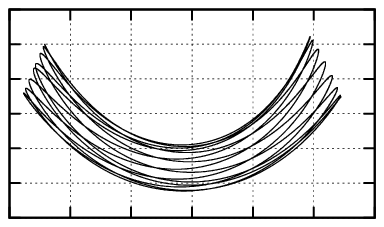}}%
    \gplfronttext
  \end{picture}%
\endgroup

%% file: Figure16d.tex
\begingroup
  \makeatletter
  \providecommand\color[2][]{%
    \GenericError{(gnuplot) \space\space\space\@spaces}{%
      Package color not loaded in conjunction with
      terminal option `colourtext'%
    }{See the gnuplot documentation for explanation.%
    }{Either use 'blacktext' in gnuplot or load the package
      color.sty in LaTeX.}%
    \renewcommand\color[2][]{}%
  }%
  \providecommand\includegraphics[2][]{%
    \GenericError{(gnuplot) \space\space\space\@spaces}{%
      Package graphicx or graphics not loaded%
    }{See the gnuplot documentation for explanation.%
    }{The gnuplot epslatex terminal needs graphicx.sty or graphics.sty.}%
    \renewcommand\includegraphics[2][]{}%
  }%
  \providecommand\rotatebox[2]{#2}%
  \@ifundefined{ifGPcolor}{%
    \newif\ifGPcolor
    \GPcolorfalse
  }{}%
  \@ifundefined{ifGPblacktext}{%
    \newif\ifGPblacktext
    \GPblacktexttrue
  }{}%
  \let\gplgaddtomacro\g@addto@macro
  \gdef\gplbacktext{}%
  \gdef\gplfronttext{}%
  \makeatother
  \ifGPblacktext
    \def\colorrgb#1{}%
    \def\colorgray#1{}%
  \else
    \ifGPcolor
      \def\colorrgb#1{\color[rgb]{#1}}%
      \def\colorgray#1{\color[gray]{#1}}%
      \expandafter\def\csname LTw\endcsname{\color{white}}%
      \expandafter\def\csname LTb\endcsname{\color{black}}%
      \expandafter\def\csname LTa\endcsname{\color{black}}%
      \expandafter\def\csname LT0\endcsname{\color[rgb]{1,0,0}}%
      \expandafter\def\csname LT1\endcsname{\color[rgb]{0,1,0}}%
      \expandafter\def\csname LT2\endcsname{\color[rgb]{0,0,1}}%
      \expandafter\def\csname LT3\endcsname{\color[rgb]{1,0,1}}%
      \expandafter\def\csname LT4\endcsname{\color[rgb]{0,1,1}}%
      \expandafter\def\csname LT5\endcsname{\color[rgb]{1,1,0}}%
      \expandafter\def\csname LT6\endcsname{\color[rgb]{0,0,0}}%
      \expandafter\def\csname LT7\endcsname{\color[rgb]{1,0.3,0}}%
      \expandafter\def\csname LT8\endcsname{\color[rgb]{0.5,0.5,0.5}}%
    \else
      \def\colorrgb#1{\color{black}}%
      \def\colorgray#1{\color[gray]{#1}}%
      \expandafter\def\csname LTw\endcsname{\color{white}}%
      \expandafter\def\csname LTb\endcsname{\color{black}}%
      \expandafter\def\csname LTa\endcsname{\color{black}}%
      \expandafter\def\csname LT0\endcsname{\color{black}}%
      \expandafter\def\csname LT1\endcsname{\color{black}}%
      \expandafter\def\csname LT2\endcsname{\color{black}}%
      \expandafter\def\csname LT3\endcsname{\color{black}}%
      \expandafter\def\csname LT4\endcsname{\color{black}}%
      \expandafter\def\csname LT5\endcsname{\color{black}}%
      \expandafter\def\csname LT6\endcsname{\color{black}}%
      \expandafter\def\csname LT7\endcsname{\color{black}}%
      \expandafter\def\csname LT8\endcsname{\color{black}}%
    \fi
  \fi
  \setlength{\unitlength}{0.0500bp}%
  \begin{picture}(3600.00,2520.00)%
    \gplgaddtomacro\gplbacktext{%
      \csname LTb\endcsname%
      \put(1122,1860){\makebox(0,0)[r]{\strut{} 0}}%
      \put(1122,1560){\makebox(0,0)[r]{\strut{} 0.1}}%
      \put(1122,1260){\makebox(0,0)[r]{\strut{} 0.2}}%
      \put(1122,960){\makebox(0,0)[r]{\strut{} 0.3}}%
      \put(1122,660){\makebox(0,0)[r]{\strut{} 0.4}}%
      \put(1583,440){\makebox(0,0){\strut{}-0.4}}%
      \put(2240,440){\makebox(0,0){\strut{} 0}}%
      \put(2897,440){\makebox(0,0){\strut{} 0.4}}%
      \put(220,1260){\rotatebox{90}{\makebox(0,0){\strut{}$z$ (m)}}}%
      \put(2240,110){\makebox(0,0){\strut{}$y$ (m)}}%
      \put(2240,2190){\makebox(0,0){\strut{}(d)}}%
    }%
    \gplgaddtomacro\gplfronttext{%
    }%
    \gplbacktext
    \put(0,0){\includegraphics{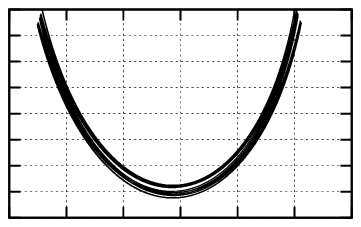}}%
    \gplfronttext
  \end{picture}%
\endgroup